\documentclass[nofootinbib,floats,aps,amsmath,showpacs]{revtex4}

\usepackage{simplewick}
\usepackage{graphicx}

\usepackage{amsmath}

\newcommand{\diag}{{\rm diag}}
\newcommand{\tr}{{\rm tr}\,}

\newcommand{\IN}{\,\rm in}
\newcommand{\OUT}{\,\rm out}
\newcommand{\I}{{\rm i}}
\newcommand{\e}{{\rm e}}

\newcommand{\str}{{\rm Str}\,}
\newcommand{\uB}{\underline{B}}
\newcommand{\uBt}{\underline{\tilde{B}}}
\newcommand{\tB}{\tilde{B}}
\newcommand{\tC}{\tilde{C}}

\begin{document}

\title{Periodic-orbit theory of universal level
correlations in quantum chaos}

\author{Sebastian M\"uller$^1$, Stefan Heusler$^2$,  Alexander
Altland$^3$, Petr Braun$^{4,5}$ and Fritz Haake$^4$ }
\address{$^1$Department of Mathematics, University of Bristol,
Bristol BS81TW, UK}

\address{ $^2$Institut f\"ur Didaktik der Physik, Universit\"at
  M\"unster, Wilhelm-Klemm Str.~10, 48149 M\"unster, Germany}

\address{$^3$Institut f\"ur Theoretische Physik, Universit\"at zu
  K\"oln, Z\"ulpicher Str.~77, 50937 K\"oln, Germany}

 \address{ $^{4}$Fachbereich Physik, Universit\"at Duisburg-Essen, 47048
 Duisburg, Germany}

 \address{ $^5$Institute of Physics, Saint-Petersburg University,
198504 Saint-Petersburg,
  Russia}

\email{sebastian.muller@bristol.ac.uk, stefanheusler@gmx.de, alexal@thp.uni-koeln.de,
petr.braun@uni-due.de, fritz.haake@uni-due.de}

\begin{abstract}
  Using Gutzwiller's semiclassical periodic-orbit theory we
  demonstrate universal behavior of the two-point correlator of the
  density of levels for quantum systems whose classical limit is fully
  chaotic.  We go beyond previous work in establishing the full
  correlator such that its Fourier transform, the spectral form
  factor, is determined for all times, below and above the Heisenberg
  time. We cover dynamics with and without time reversal invariance
  (from the orthogonal and unitary symmetry classes).
 A key step in our reasoning is to sum the periodic-orbit expansion
  in terms of a matrix integral,  like the one known from the sigma model
  of random-matrix theory.
\end{abstract}

  \maketitle

\section{Introduction}

Classical chaos is characterized by  sensitive dependence on
initial conditions, and by the fact that long trajectories
uniformly fill the available space. These classical features have
profound consequences in quantum mechanics: Many quantum
properties of chaotic systems become universal in the
semiclassical limit, i.e., they no longer depend on the system in
question but only on symmetries. For instance, as conjectured in
\cite{BGS} the statistics of energy eigenvalues  is universal:
Neighboring levels tend to repel each other, and the correlation
function of the level density $\rho(E)$,
\begin{equation}\label{defrealcorr}
R(\epsilon)=\frac{1}{\overline{\rho}^2}\left\langle \rho
\left(E+\frac{\epsilon}{2\pi\overline{\rho}}\right) \rho\left
(E-\frac{\epsilon}{2\pi\overline{\rho}}\right) \right\rangle -1\;,
\end{equation}
(where the brackets and overbars denote averages over the center
energy $E$) has the same form for all dynamics belonging to the
same symmetry class. The most important classes are those of
systems without any symmetries (unitary class; complex Hermitian
Hamiltonians) and systems whose sole symmetry is time-reversal
invariance (orthogonal class; real symmetric Hamiltonians).

If one accepts universal spectral correlations as a fact, a
phenomenological prediction for $R(\epsilon)$ can be obtained by
averaging over ensembles of systems sharing the same symmetries.
Modelling the corresponding Hamiltonians through matrices, one
arrives at random-matrix theory (RMT), as initially proposed by
Wigner and Dyson in the fifties \cite{Wigner} in the context of
atomic nuclei. RMT has since found broad applications in many
fields of physics. For the correlation function $R(\epsilon)$,
random-matrix averages yield the prediction
\cite{Haake,Efetov,GuhrWeidenmueller}
\begin{eqnarray}
\label{R} R(\epsilon)=\left\{
\begin{array}{cc}
-s(\epsilon)^2 & \mathrm{unitary} \\
-s(\epsilon)^2+s'(\epsilon)\left(\frac{1}{\pi}{\rm Si}(\epsilon)
-\frac{1}{2}{\rm sgn}(\epsilon)\right)& \mathrm{orthogonal}
\end{array}
\right.
\end{eqnarray}
with $s(\epsilon)=\frac{\sin\epsilon}{\epsilon}$ and ${\rm
  Si}(\epsilon)=\int_0^\epsilon d\epsilon' s(\epsilon')$.
It is convenient to access $R(\epsilon)$ through a complex
correlation function  $C(\epsilon)$ which is analytic in the
energy upper  half  plane and connected with the real correlator
as $R(\epsilon)=\lim_{\mbox{Im}\,
\epsilon\to+0}\mbox{Re}\,C(\epsilon)$. The correlator
$C(\epsilon)$ has the nice property of being retrievable by Borel
summation \cite{Sokal} from its asymptotic expansion
\begin{equation}
\label{RMTseries} C(\epsilon)\sim\sum_{n=2}^\infty(c_n+d_n
\e^{2\I\epsilon})\left(\frac{1}{\epsilon}\right)^n\;.
\end{equation}
The foregoing power series contains both non-oscillatory terms and
oscillatory terms  proportional to  $\e^{2 \I\epsilon}$; the
latter are responsible for the  singularity (discontinuity at
$\tau=1$ of the first or third derivative in the unitary and
orthogonal case, respectively) in the Fourier transform of the
spectral correlator $K(\tau)=\frac{1}{\pi}\int_{-\infty}^\infty
d\epsilon R(\epsilon)\e^{2\I \epsilon\tau}$ (the spectral form
factor). For the unitary class we have $c_2=-\frac{1}{2}$,
$d_2=\frac{1}{2}$ and all other coefficients vanish, whereas the
orthogonal class involves $c_2=-1$,
$c_n=\frac{(n-3)!(n-1)}{2\I^n}$ for $n\geq 3$, $d_2=d_3=0$ and
$d_n=\frac{(n-3)!(n-3)}{2\I^n}$ for $n\geq 4$.

This phenomenological RMT result leaves open the question {\it
why} chaotic systems behave universally, and that question we want
to address in the present paper. Taking a semiclassical approach
one can follow Gutzwiller \cite{Gutzwiller} and express the level
density as a sum over contributions of classical periodic orbits.
The correlation function then turns into a sum over pairs of
orbits.  Systematic contributions to that double sum are due to
pairs of orbits whose actions are sufficiently close for the
associated quantum amplitudes to interfere constructively.  The
task of finding correlations in quantum spectra is translated into
a classical one, namely to understand the correlations between
actions of periodic orbits \cite{Argaman}.

\begin{figure}
\begin{center}
  \includegraphics[scale=.4]{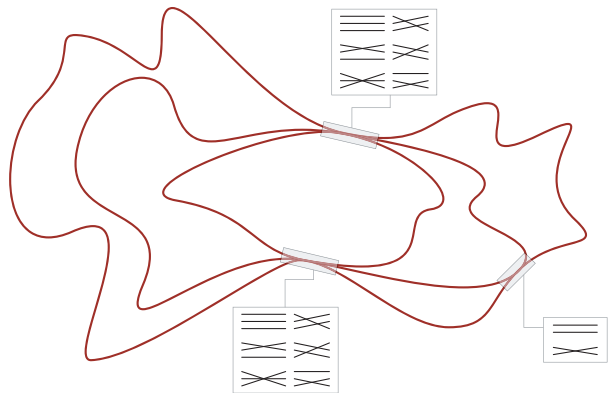}
\end{center}
\caption{Bunch of 72 (pseudo-)orbits differing in two 3-encounters
and one 2-encounter (a pseudo-orbit is  a set of several disjoined
orbits; see below). Different orbits not resolved except in
blowups of  encounters} \label{fig:bunch_of_72}
\end{figure}
The first step in this direction was taken by Berry \cite{Berry}
who showed that pairs of {\it identical} or {\it mutually
time-reversed} orbits explain the leading coefficient $c_2$.
Higher-order contributions $c_n$ with $n\geq 3$ are due to a still
not widely known feature of chaos, a certain ``bunching'' of
periodic orbits \cite{Bunches}.  A long orbit has many close
``self-encounters'' (see Fig.~\ref{fig:bunch_of_72}) where it
comes close to itself in phase space (possibly up to time
reversal, for dynamics with time reversal invariance).  Such an
orbit is just one of a closely packed ``bunch''. All orbits in a
bunch are nearly identical, except that the orbit pieces inside
the encounters are ``switched''.

\begin{figure}
\begin{center}
  \includegraphics[scale=.4] {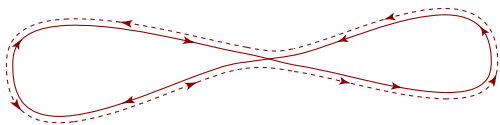}
\end{center}
\caption{Sieber-Richter pair} \label{fig:RS}
\end{figure}

Based on insight from disordered systems \cite{Disorder}, the
study of bunches of orbits differing in encounters was pioneered
by Sieber and Richter \cite{SR} who derived the next-to-leading
coefficient $c_3$ for time-reversal invariant systems from orbit
pairs differing in just one encounter where two stretches of an
orbit are close (see Fig.~\ref{fig:RS}). Full agreement with all
coefficients $c_n$ from RMT was established by the present authors
in \cite{TauSmallShort,TauSmallLong}.

In none of these works oscillatory contributions could be obtained
(Note however courageous forays by Keating \cite{HardyLittlewood}
and Bogomolny and Keating \cite{Bootstrap}), due to the fact that
Gutzwiller's formula for the level density is divergent. To
enforce convergence, one needs to allow for complex energies with
imaginary parts large compared to the mean level spacing, ${\rm Im
\,\epsilon
  \gg 1}$.  Oscillatory terms proportional to $\e^{2\I \epsilon}$ then
become exponentially small and cannot be resolved within the
conventional semiclassical approach.

In \cite{TauLargeShort} we proposed a way around this difficulty
that we here want to elaborate in detail.  The key idea is to
represent the correlation function through derivatives of a
generating function involving spectral determinants.  Two such
representations are available and entail different semiclassical
periodic-orbit expansions.  One of them recovers the
non-oscillatory part of the asymptotic expansion
(\ref{RMTseries}), essentially in equivalence to
\cite{SR,TauSmallShort,TauSmallLong}; the other representation
breaks new ground by giving the oscillatory part of
(\ref{RMTseries}).

The full random-matrix result also, and in fact most naturally,
arises within an alternative semiclassical approximation scheme
proposed by Berry and Keating in \cite{Resum}. That scheme
constrains the semiclassical periodic-orbit expansion of the
spectral determinants $\det(E-H)$ to be real and to converge for
real energy argument. Inserted into the generating function the
resulting ``Riemann-Siegel lookalike formula'' for $\det(E-H)$ was
shown in \cite{KM} to simultaneously produce both the
non-oscillatory and oscillatory parts of the expansion
(\ref{RMTseries}). Interestingly, the Riemann-Siegel lookalike
formula parallels an exact composition of the random-matrix
averaged generating function originating from a Weyl group
symmetry \cite{Zirnbauer}.

We want to stress that obtaining both the non-oscillatory and the
oscillatory contributions to $R(\epsilon)$ also implies a
derivation of level repulsion: By summing up the asymptotic series
(\ref{RMTseries}) we access the full $R(\epsilon)$ including its
behavior for small $\epsilon$,
$R(\epsilon)\propto|\epsilon|^\beta$ (with $\beta=2$ for the
unitary and $\beta=1$ for the orthogonal class), indicating a
suppression of small energy differences. To get access to small
$\epsilon$ the oscillatory contributions (varying on the scale of
the mean level spacing) are of crucial importance.

For our semiclassical approach to spectral statistics we have
drawn inspiration from a field-theoretical implementation of RMT,
the nonlinear sigma model. There, the ensemble averaged generating
function is ultimately transformed into an integral over matrices.
In a stationary-phase approximation to that integral, two
different saddles are found to respectively contribute
non-oscillatory and oscillatory terms. The relation between these
two contributions  bears close resemblance to semiclassics.
Moreover the topologically different structures of orbit bunches
in semiclassics turn out to be analogous to Feynman diagrams in a
perturbative treatment of the sigma model, with encounters
corresponding to vertices. That analogy helps to do the sums
required in semiclassics and to compact the semiclassical series
to a matrix integral like the one in the sigma model.
 This derivation has to be contrasted with a
previous field-theoretical approach to spectral universality, the
ballistic sigma model \cite{Ballistic,KoelleAlaaf}.

 The present
paper is organized as follows. We start by explaining the
generating-function approach (Section \ref{sec:semi_howto}), and
give the pertinent generalization of Berry's diagonal
approximation (Section \ref{sec:diagonal}).  We then evaluate the
contributions of bunches of periodic orbits to the generating
function and the correlation function (Section \ref{sec:off}) and
sum them up (Section \ref{sec:cancel}).  Our reasoning culminates
in the semiclassical construction of a sigma model lookalike
(Section \ref{sec:build_sigma}). For pedagogical reasons, Sections
\ref{sec:diagonal}-\ref{sec:build_sigma} are devoted to the
unitary symmetry class, the simplest case. The generalizations
necessary for time-reversal invariant dynamics are listed in
Section \ref{sec:orthogonal}.

The Appendix \ref{sec:contraction_rules} contains a derivation of
contraction rules needed for the semiclassical construction of the
matrix integral.  Further appendices contain
(\ref{sec:supersymmetric_sigma}) a brief introduction to the
supersymmetric sigma model, (\ref{sec:bosonic_sigma},
  \ref{sec:fermionic_sigma}) two replica versions of the
sigma model and their relation to the semiclassical theory,
(\ref{app_permutations}) a characterization of orbit topologies in
terms of permutations, and (\ref{sec:complex_corr}) the energy
average employed in the definition of the correlator and the
connection between the real and complex correlation functions.

\section{ Generating function}

\label{sec:semi_howto}

\subsection{Definition}\label{subsec:genfuncdef}

All approaches to resolve non-oscillatory and oscillatory
contributions start with representing
 the correlation
function through derivatives of a generating function.  To get
there we express the complex correlator $C(\epsilon)$ in terms of
Green functions $(E-H)^{-1}$ of the Hamiltonian $H$,
\begin{eqnarray}\label{defC}
C(\epsilon ^{+})=\frac{1}{2\pi ^{2}\overline{\rho}^2} \Big\langle
{\rm tr}\,\big(E+\frac{\epsilon^+}
{2\pi\overline{\rho}}-H\big)^{-1}\,{\rm tr}
\,\big(E-\frac{\epsilon^+}
{2\pi\overline{\rho}}-H\big)^{-1}\Big\rangle -\frac{1}{2}\,,
\end{eqnarray}
where the superscript $+$ denotes a positive imaginary part
$\I\eta$. The average will be understood as over the real center
energy $E$, with an averaging window large compared to the mean
level spacing (see Appendix \ref{sec:complex_corr}). The first
Green function in (\ref{defC}) is a retarded one due to the
positive imaginary part of the energy argument while the second is
an advanced one.

A suitable generating function involves four spectral determinants
$\Delta(E)=\det(E-H)$,
\begin{equation}
\label{Zdef} Z(\epsilon_A,\epsilon_B,\epsilon_C,\epsilon_D)=
\left\langle
\frac{\Delta\left(E+\frac{\epsilon_C}{2\pi\overline{\rho}}\right)
\Delta\left(E-\frac{\epsilon_D}{2\pi\overline{\rho}}\right)}
{\Delta\left(E+\frac{\epsilon_A}{2\pi\overline{\rho}}\right)
\Delta\left(E-\frac{\epsilon_B}{2\pi\overline{\rho}}\right)}
\right\rangle.
\end{equation}
The four parameters $\epsilon_A,\epsilon_B,\epsilon_C,\epsilon_D$
denote offsets from the center energy $E$, made dimensionless by
referral to the mean level spacing $1/\overline{\rho}$; they are
all taken to include a positive imaginary part $\I\eta$
\footnote{ Due to the imposed positivity of the imaginary parts of
the
  energy offsets the determinants involving $\epsilon_A$ and
  $\epsilon_C$ are associated with retarded Green functions. The minus
  signs in front of $\epsilon_B$ and $\epsilon_D$ in (\ref{Zdef}) let
  the respective spectral determinants be associated with advanced
  Green functions. Note that the present sign convention differs from
  the one in our letter \cite{TauLargeShort}. }. It is worth noting
the symmetry
\begin{equation}\label{Weyl}
Z(\epsilon_A,\epsilon_B,\epsilon_C,\epsilon_D)=
Z(\epsilon_A,\epsilon_B,-\epsilon_D,-\epsilon_C)
\end{equation}
which in the framework of the sigma-model treatment of RMT
reflects a Weyl symmetry \cite{Zirnbauer}. The normalization
$Z(\epsilon_A,\epsilon_B,\epsilon_A,\epsilon_B)=1$ also deserves
being mentioned.

The correlation function can be accessed from $Z$ in two different
ways. In the standard ``columnwise'' representation, we take two
derivatives w.r.t.  $\epsilon_A$ and $\epsilon_B$, and
subsequently identify all energy increments,
\begin{eqnarray}
\label{columnwise}\label{column}   -2\left.\frac{\partial^2
Z}{\partial\epsilon_A\partial\epsilon_B} \right|_{(\|)}
=\frac{1}{2(\pi\overline{\rho})^2} \left\langle {\rm
tr}\big(E+\frac{\epsilon_A}{2\pi\overline{\rho}}-H\big)^{-1} {\rm
tr}\big(E-\frac{\epsilon_B}{2\pi\overline{\rho}}-H\big)^{-1}\;
\frac{\Delta\big(E+\frac{\epsilon_C}{2\pi\overline{\rho}}\big)
\Delta\big(E-\frac{\epsilon_D}{2\pi\overline{\rho}}\big)}
{\Delta\big(E+\frac{\epsilon_A}{2\pi\overline{\rho}}\big)
\Delta\big(E-\frac{\epsilon_B}{2\pi\overline{\rho}}\big)}
\right\rangle\Bigg|_{(\|)}\nonumber\\
=\frac{1}{2(\pi\overline{\rho})^2} \Big\langle {\rm
tr}\big(E+\frac{\epsilon^+}{2\pi\overline{\rho}}-H\big)^{-1} {\rm
tr}\big(E-\frac{\epsilon^+}{2\pi\overline{\rho}}-H\big)^{-1}
\Big\rangle=C(\epsilon^+)+\frac{1}{2}\;;\\
(\|): \epsilon_A=\epsilon_B=\epsilon_C=\epsilon_D=\epsilon^+ \,.
\end{eqnarray}
Here we used the identity $\det=\exp\tr\ln$. The notation $(\|)$
signals that when equating $\epsilon_A,\epsilon_B,\epsilon_C$ and
$\epsilon_D$ and accounting for the signs in (\ref{Zdef}), we have
identified the four energy arguments in (\ref{Zdef}) in a
``columnwise'' way.

An alternative representation of the correlation function can be
obtained if we identify the energy arguments in a ``crosswise''
way (in view of the signs in (\ref{Zdef})),
\begin{eqnarray}\label{cross}
-2\left.\frac{\partial^2 Z}{\partial\epsilon_A\partial\epsilon_B}
\right|_{(\times)} =\frac{1}{2(\pi\overline{\rho})^2} \Big\langle
{\rm tr}\big(E+\frac{\epsilon^+}{2\pi\overline{\rho}}-H\big)^{-1}
{\rm tr}\big(E-\frac{\epsilon^+}{2\pi\overline{\rho}}-H\big)^{-1}
\frac{\Delta\left(E-\frac{\epsilon^-}{2\pi\overline{\rho}}\right)
\Delta\left(E+\frac{\epsilon^-}{2\pi\overline{\rho}}\right)}
{\Delta\left(E+\frac{\epsilon^+}{2\pi\overline{\rho}}\right)
\Delta\left(E-\frac{\epsilon^+}{2\pi\overline{\rho}}\right)}
\Big\rangle\nonumber\\
(\times): \qquad \epsilon_A=\epsilon_B=\epsilon^+\,, \,
\epsilon_C=\epsilon_D=-\epsilon^-\equiv-\epsilon+\I\,\eta\,.
\end{eqnarray}
In (\ref{cross}), the ratio of determinants\footnote {The above
ratio
  averages to $Z(\epsilon^+,\epsilon^+,-\epsilon^-,-\epsilon^-)$ which
  can also be written as
  $Z(\epsilon^+,\epsilon^+,-\epsilon^-,-\epsilon^-)
  =\langle\exp[2\pi\I\Delta N_\eta]\rangle$ with $\Delta N_\eta=
  N_\eta(E+\textstyle{\frac{\epsilon}{2\pi\bar\rho}})-
  N_\eta(E-\frac{\epsilon}{2\pi\bar\rho})$. Here $N_\eta$ denotes the
  level staircase smeared out over a range $\eta$, $N_\eta(E)=\sum_k
  \theta_\eta(E-E_k)$ with $\theta_\eta(x)=\pi^{-1}\mbox {Im}\ln(-x+\I
  \eta)$ the similarly smeared step function. The function
  $Z(\epsilon^+,\epsilon^+,-\epsilon^-,-\epsilon^-)$ could be used as
  a starting point for calculating the oscillatory part of the
  correlator $C(\epsilon)$, and then some similarity with the
  courageous foray of Bogomolny and Keating \cite{Bootstrap} mentioned
  in the Introduction would arise.  Indeed, replacing $N_\eta$ by the
  sum of its energy average and a fluctuating part, $N_\eta=\langle
  N_\eta\rangle+N_\eta^{\rm fluct}$, and treating $N_\eta^{\rm fluct}$
  in a Gaussian approximation we get $\langle\exp[2\pi\I\Delta
  N_\eta]\rangle\sim \e^{2\I\epsilon} \exp[-2\pi^2\langle(\Delta
  N_\eta^{\rm fluct})^2\rangle]$. The second factor contains the
  number variance in the exponent; the first factor signals that the
  oscillatory part of the two-point correlator becomes accessible.
  This rough argument anticipates the diagonal approximation below.}
becomes unity only in the limit $\eta\to 0$, i.e., one can access
$R(\epsilon)$ as
\begin{equation}
\label{cross2} {\rm Re}\,\lim_{\eta\to0}\Big(
-2\left.\frac{\partial^2 Z}{\partial\epsilon_A\partial\epsilon_B}
\right|_{(\times)}\Big)=R(\epsilon)+\frac{1}{2}\,.
\end{equation}

\subsection{Periodic-orbit representation}\label{sec:Gutzi}

 We are now prepared to semiclassically approximate the generating
function in terms of classical periodic orbits.  Our starting
point is Gutzwiller's formula for the trace of the Green function,
\begin{equation}\label{Gutzi}
{\rm tr}(E^+-H)^{-1}\sim-i\pi\overline{\rho}(E^+)
-\frac{\I}{\hbar}\sum_a T_a F_a {\rm e}^{\I S_a(E^+)/\hbar}\;.
\end{equation}
Herein, the first term involves the average level density
$\overline{\rho}$ given by Weyl's law as
$\overline{\rho}=\frac{\Omega}{(2\pi\hbar)^2}$ (for systems with
two degrees of freedom), with $\Omega$ the volume of the energy
shell.  The orbit $a$ is represented by its period $T_a$, action
$S_a$, and stability amplitude $F_a$ (defined to include the
Maslov phase factor but not the period)

\footnote{ We disregard  orbits which are multiple repetitions of
a shorter orbit since these are exponentially suppressed in the
limit of
 long  periods. As regards orbits
 incorporating multiple {\it approximate} traversals of
 short orbits, the respective corrections were shown to be
 immaterial for the  non-oscillatory part of the spectral
correlator (see \cite{TauSmallLong}) and can similarly be shown to
cancel in our four-determinant generating function $Z$.
Interestingly, such orbits do contribute to the averaged product
of {\it two} spectral determinants  \cite{Waltner}.}

To proceed to a semiclassical approximation of the spectral
determinant, we integrate and exponentiate in (\ref{Gutzi}) as
$\Delta(E^+)\propto\exp \Big(\int^{E^+}d E'{\rm
  tr}\frac{1}{E'-H}\Big)$. The resulting exponent inherits the
smoothed number $\overline{N}(E)$ of eigenvalues below $E$ and an
orbit sum from the trace formula (\ref{Gutzi}),
\begin{eqnarray}
\label{gutzwiller}  \Delta(E^+)\propto\exp
\Big(-\I\pi\overline{N}(E^+)-\sum_a F_a{\rm e}^{\I
S_a(E^+)/\hbar}\Big) ={\rm e}^{-\I\pi\overline{N}(E^+)}
\sum_AF_A(-1)^{n_A}{\rm e}^{\I S_A(E^+)/\hbar}\,.
\end{eqnarray}
The last member of the foregoing chain of equations has the
exponentiated orbit sum expanded into a sum over {\it
pseudo-orbits}, i.~e., non-ordered sets of periodic orbits $A$
\cite{Resum}. In that pseudo-orbit sum, $n_A$ is the number of
orbits inside $A$, $S_A$ the sum of their actions, and $F_A$ the
product of their stability amplitudes $F_a$.
The sum over $A$ includes the empty set, whose contribution is
unity. (Note that in the foregoing expressions we suppressed a
proportionality constant due to the lower limit of the
$E'$-integral. That factor is irrelevant since it cancels in the
generating function.

Eq.  (\ref{gutzwiller}) applies to energies with a positive
imaginary
part.  For 
energies with a negative imaginary part complex conjugation yields
\begin{eqnarray}
\label{gutzwiller_minus} \Delta(E^-) =\Delta(E^+)^*\propto\exp
\Big (\I\pi\overline{N}(E^-)-\sum_a F_a^* {\rm e}^{-\I
S_a(E^-)/\hbar} \Big) \nonumber\\= {\rm
e}^{\I\pi\overline{N}(E^-)} \sum_AF_A^*(-1)^{n_A}{\rm e}^{-\I
S_A(E^-)/\hbar}\;.
\end{eqnarray}
For inverse spectral determinants, the minus sign in the
periodic-orbit sum and hence the factor $(-1)^{n_A}$ in the
pseudo-orbit sum disappear,
\begin{eqnarray}
\label{det_inverse}
\Delta(E^+)^{-1}=\left(\Delta(E^-)^{-1}\right)^* \propto \exp\Big(
\I\pi\overline{N}(E^+)+\sum_a F_a {\rm e}^{\I S_a(E^+)/\hbar}\Big)\nonumber\\
= {\rm e}^{\I\pi\overline{N}(E^+)}\sum_AF_A{\rm e}^{\I
S_A(E^+)/\hbar} \,.
\end{eqnarray}

The four spectral determinants in $Z$  now yield semiclassical
asymptotics as a fourfold sum over pseudo-orbits,
\begin{eqnarray}
\label{Z1preexpansion} Z\sim Z^{(1)}&=&\left\langle {\rm
e}^{\I\pi\overline{N}
[E+\epsilon_A/(2\pi\overline{\rho})]}\sum_AF_A{\rm e}^{\I S_A
[E+\epsilon_A/(2\pi\overline{\rho})]/\hbar}\right. \nonumber\\
&&\times {\rm
e}^{-\I\pi\overline{N}[E-\epsilon_B/(2\pi\overline{\rho})]}
\sum_BF_B^*{\rm e}^{-\I
S_B[E-\epsilon_B/(2\pi\overline{\rho})]/\hbar} \nonumber\\
&&\times {\rm
e}^{-\I\pi\overline{N}\left[E+\epsilon_C/(2\pi\overline{\rho})
\right]} \sum_C F_C(-1)^{n_C}{\rm e}^{\I S_C
\left[E+\epsilon_C/(2\pi\overline{\rho})\right]/\hbar}
\nonumber\\&&\left.\times {\rm e}^{\I\pi\overline{N}
\left[E-\epsilon_D/(2\pi\overline{\rho})\right]} \sum_D
F_D^*(-1)^{n_D}{\rm e}^{-\I S_D
\left[E-\epsilon_D/(2\pi\overline{\rho})\right]/\hbar}\right\rangle\;,
\end{eqnarray}
where all offset variables
$\epsilon_A,\epsilon_B,\epsilon_C,\epsilon_D$ have positive
imaginary parts $\eta$.  Here the superscript 1 signals that
(\ref{Z1preexpansion}) is a semiclassical approximation based on
the Gutzwiller trace formula which requires
\begin{equation}\label{large_eta_prec}
\eta\gg1
\end{equation}
for convergence such that terms vanishing in that limit (such as
terms of the order $\e^{-\eta}$) will be lost. The complement
missed by $Z^{(1)}$ becomes important for real energies  (a
manifestation of the Stokes phenomenon of asymptotic analysis) and
will be introduced below.

Upon expanding as
$\overline{N}(E\pm\frac{\epsilon}{2\pi\overline{\rho}})
\sim\overline{N}(E)\pm\frac{\epsilon}{2\pi}$ and
$S(E\pm\frac{\epsilon}{2\pi\overline{\rho}})/\hbar \sim
S(E)\pm\frac{T\epsilon}{2\pi\hbar\overline{\rho}}=S(E)\pm
T\epsilon/T_H$ where
$T_H=2\pi\hbar\overline\rho=\frac{\Omega}{2\pi\hbar}$ is the
Heisenberg time, we simplify the quantity $Z^{(1)}$ to
\begin{eqnarray}
\label{Z1} Z^{(1)}&=& {\rm
e}^{\I(\epsilon_A+\epsilon_B-\epsilon_C-\epsilon_D)/2} \times
\Big\langle \sum_{A,B,C,D}F_A F_B^* F_C F_D^*(-1)^{n_C+n_D}\nonumber\\
&&\times{\rm e}^{\I(S_A(E)-S_B(E)+S_C(E)-S_D(E))/\hbar} {\rm
e}^{\I(T_A\epsilon_A+T_B\epsilon_B+T_C\epsilon_C+T_D\epsilon_D)/T_H}
\Big\rangle\;.
\end{eqnarray}

The foregoing sum over pseudo-orbit quadruplets combines classical
chaos with quantum interference due to the phase factor
${\e}^{\I\Delta
  S/\hbar}\equiv{\e}^{\I(S_A(E)-S_B(E)+S_C(E)-S_D(E))/\hbar}$.
For most quadruplets the interference is destructive since for
$\hbar\to 0$ the phase factor ${\e}^{\I\Delta S/\hbar}$ oscillates
rapidly as the energy varies and therefore vanishes after
averaging. For constructive interference, the action mismatch
$\Delta S$ must be small, at most of the order of $\hbar$; to
within such a quantum mismatch the cumulative action of the
pseudo-orbits $B\cup D$ must coincide with the cumulative action
of $A\cup C$.  Such action matching happens systematically only if
the orbits in $B\cup D$ either coincide with those in $A\cup C$
(and thus constitute the ``diagonal'' contributions, see Sect.
\ref{sec:diagonal}) or differ from them only by their connections
inside encounters (``off-diagonal'' contributions) as in
Figs.~\ref{fig:bunch_of_72},\ref{fig:RS}. Indeed, then, the
phenomenon of orbit bunching is crucial for spectral universality.

Now imagine $Z^{(1)}$ calculated from (\ref{Z1}) and then
analytically continued to real energy offsets, $\eta\to0^+$.  If
instead of $Z^{(1)}$ we had the full generating function we could
obtain both the non-oscillatory and the oscillatory contributions
to $C(\epsilon)$ by taking derivatives w.r.t.  $\epsilon_A$ and
$\epsilon_B$ and identifying the energy increments either in the
columnwise way (\ref{column}) or the crosswise one (\ref{cross}).
However, for $Z^{(1)}$ the columnwise procedure (\ref{column})
cannot yield oscillatory terms since it makes the (Weyl)
exponential
$\e^{\I(\epsilon_A+\epsilon_B-\epsilon_C-\epsilon_D)/2}$ collapse
to unity.  In contrast, with the crosswise representation
(\ref{cross}) the signs of  $\epsilon_C$ and $\epsilon_D$ are
flipped and the Weyl factor turns into $\e^{2\I\epsilon}$. Hence
the crosswise representation recovers only the oscillatory Fourier
component proportional to $\e^{2\I\epsilon}$.

The random-matrix result suggests that to get the {\it full
correlation function} one has to add both results
\begin{equation}
\label{addderivatives} R(\epsilon)=-{\rm Re}\left\{
2\frac{\partial^2 Z^{(1)}}{\partial\epsilon_A\partial\epsilon_B}
\Big|_{(\|)}
 +2\frac{\partial^2Z^{(1)}}{\partial\epsilon_A\partial\epsilon_B}
\Big|_{(\times)}+\frac{1}{2} \right\}\;.
\end{equation}
 Here the first and the third summand are due to (\ref{columnwise}) while
the second summand is due to (\ref{cross2})\footnote{Critical
readers may miss the $\frac{1}{2}$ from (\ref{cross2}). This
summand does not show up here since it is non-oscillatory and
cannot be retrieved from the crosswise representation.}. Of course
it is vital to justify this additivity independently from RMT, a
task to be attacked below.

The ''crosswise'' term $\frac{\partial^2
  Z^{(1)}}{\partial\epsilon_A\partial\epsilon_B}\big|_{(\times)}$
differs from the ''columnwise'' term $\frac{\partial^2
  Z^{(1)}}{\partial\epsilon_A\partial\epsilon_B}\big|_{(\|)}$ because
$\epsilon_C$ and $\epsilon_D$ are interchanged and flipped in
sign. This interchange and sign flip may as well be performed on
the level of the generating function, i.e., we can write
\begin{eqnarray}
\left.\frac{\partial^2
Z^{(1)}}{\partial\epsilon_A\partial\epsilon_B} \right|_{(\times)}
&=&\left.\frac{\partial^2
Z^{(2)}}{\partial\epsilon_A\partial\epsilon_B} \right|_{(||)}
\\
\label{Z2} Z^{(2)}(\epsilon_A,\epsilon_B,\epsilon_C,\epsilon_D)
&=&Z^{(1)}(\epsilon_A,\epsilon_B,-\epsilon_D,-\epsilon_C)
\end{eqnarray}

 Instead of adding the crosswise representation in
(\ref{addderivatives}) we can use $Z^{(1)}+Z^{(2)}$ as a
generating function, and get $R(\epsilon)$ in full through the
columnwise (or crosswise) representation alone. Therefore no extra
work is needed for the oscillatory part of the two-point
correlation function, once the ``standard'' part $Z^{(1)}$ of the
asymptotics of $Z$ is found.
 The resulting asymptotics of the
generating function
\begin{equation}
\label{Z1Z2} Z\sim Z^{(1)}+Z^{(2)}\,
\end{equation}
respects the Weyl symmetry and remains valid when  the arguments
are real ($\eta=0$).  We note that if one wants to evaluate
$Z^{(2)}$ using the Gutzwiller trace formula, the sign flip in
$\epsilon_C$ and $\epsilon_D$ means that the corresponding
arguments of $Z^{(2)}$ should be taken with large {\it negative}
imaginary parts.

\subsection{Riemann-Siegel lookalike}\label{subsec:RSlookalike}

As pointed out by Keating and M\"uller in \cite{KM}, the additive
structure (\ref{Z1Z2}) of the generating function most naturally
arises with the help of the Riemann-Siegel lookalike formula
\cite{Resum}.

The essential idea is to eliminate a drawback of the semiclassical
approximation (\ref{gutzwiller},\ref{gutzwiller_minus}) of the
spectral determinant, namely its failure to manifestly preserve
the reality of the energy eigenvalues, i.~e., the unitarity of the
quantum dynamics. In the limit $\eta\to 0$, $\Delta(E^+)$ and
$\Delta(E^-)$ must become real and identical to each other. That
property is not manifest in but may be imposed on Eqs.
(\ref{gutzwiller},\ref{gutzwiller_minus}). By imposing this
property, it was shown in \cite{Resum} that a duality arises
between ``long'' pseudo-orbits (sum of periods of contributing
orbits larger than half the Heisenberg time $T_H$) and ``short''
ones (cumulative period smaller than $T_H/2$); the overall
contribution of long orbits is the complex conjugate of the
contribution arising from short orbits.  The ensuing
Riemann-Siegel lookalike expresses the real spectral determinant
$\Delta(E)$ as a sum over pseudo-orbits with durations shorter
than $T_H/2$,
\begin{equation}
\label{resum} \Delta(E)\propto{\rm
e}^{-i\pi\overline{N}(E)}\!\!\sum_{A\; (T_A<T_H/2)}\!\!F_A
(-1)^{n_A}{\rm e}^{i S_A/\hbar} + {\rm c.c.}\,.
\end{equation}

To evaluate the generating function $Z$ one now uses the
Riemann-Siegel formula for the two spectral determinants in the
numerator.  For the determinants in the denominator we are not
allowed to use Riemann-Siegel as the non-zero imaginary parts with
appropriate signs in $\epsilon_A,\epsilon_B$ are crucial for the
definition of $Z$, and we just stick to Eq. (\ref{det_inverse}).
Since each determinant in the numerator becomes the sum of two
mutually complex conjugate terms, $Z$ has altogether four summands
which involve the four phases $\pm
\I\pi\left[\overline{N}\left(E+\epsilon_C/(2\pi\overline{\rho})\right)
  -\overline{N}\left(E-\epsilon_D/(2\pi\overline{\rho})\right)
\right]$ and $\pm
\I\,\pi\left[\overline{N}\left(E+\epsilon_C/(2\pi\overline{\rho})\right)
  +\overline{N}\left(E-\epsilon_D/(2\pi\overline{\rho})\right)\right]$.
The two latter contributions oscillate rapidly as functions of $E$
and thus vanish after averaging.

The two surviving additive terms yield the high-energy asymptotics
of the generating function $Z\sim Z^{(1)}+Z^{(2)}$ found in the
previous subsection, according to (\ref{Z1}, \ref{Z2}), except
that the sums over $C$ and $D$ are now restricted to pseudo-orbits
with durations shorter than $T_H/2$.  To make that difference
irrelevant it is convenient to
 evaluate the pseudo-orbit sum for
large imaginary parts $\eta$ of the dimensionless energy offsets
in the arguments of the action, $S_A(E)\to S_A(E+{\rm
i}\eta/2\pi\bar{\rho})$, see (\ref{large_eta_prec}). That artifice
makes sure that in the relevant integrals over orbit periods to be
encountered below the integrand falls to exponentially small
magnitude as the integration variable rises to near half the
Heisenberg time.  If all four energy offsets $\epsilon_{A,B,C,D}$
are thus provided with large positive imaginary parts, the two
additive pieces obtained through the Riemann-Siegel lookalike
become respectively
$Z^{(1)}(\epsilon_A,\epsilon_B,\epsilon_C,\epsilon_D)$ and
$Z^{(2)}(\epsilon_A,\epsilon_B,\epsilon_C^*,\epsilon_D^*)
=Z^{(1)}(\epsilon_A,\epsilon_B,-\epsilon_D^*,-\epsilon_C^*)$.
Complex conjugation of the last two arguments of $Z^{(2)}$ is
needed to enable its calculation  in terms of periodic orbits
according to (\ref{Z1}).  At the final stage we  use analyticity
to let $\eta$ go to zero; the sum of $Z^{(1)}$ and $Z^{(2)}$
yields then the above asymptotic representation (\ref{Z1Z2}).

To obtain a quantitative understanding of spectral statistics, we
now have to evaluate the quadruple sum over pseudo-orbits in
$Z^{(1)}$, taking into account orbits in $B\cup D$ that either
(diagonal contributions) coincide with or (off-diagonal
contributions) at most differ in encounters from the orbits in
$A\cup C$.

\section{Diagonal approximation}
\label{sec:diagonal}

In the diagonal approximation we effectively neglect all
correlations between orbits that are not identical. If we
represent the generating functions as a sum over $A,B,C,D$ as in
(\ref{Z1}) this means we should keep only contributions where the
orbits in $A\cup C$ are repeated inside $B\cup D$. However we can
save labor if we write the four spectral determinants entering our
generating function like the second member in (\ref{gutzwiller}),
i.~e., with the help of exponentiated orbit sums. We thus obtain
the equivalent representation of the generating function
\begin{equation}\label{Z1orb}
Z^{(1)}=\e^{\I\left( \epsilon _{A}+\epsilon _{B}-\epsilon
_{C}-\epsilon _{D}\right) /2}\left\langle
\prod\limits_{a}z_{a}\right\rangle
\end{equation}
wherein each periodic orbit participates with the factor
\begin{eqnarray}\label{orbitfactor}
z_{a} &=&\exp \left[
 F_{a}\e^{\frac{\I}{\hbar }S_{a}(E)}\underbrace{\left(
\e^{\I\frac{T_{a}}{T_{H}}\epsilon_{A}}
-\e^{\I\frac{T_{a}}{T_{H}}\epsilon_{C}}\right)}_{\equiv
f_{AC}^{a}} +F_{a}^{\ast }\e^{-\frac{\I}{\hbar
}S_{a}(E)}\underbrace{\left(
\e^{\I\frac{T_{a}}{T_{H}}\epsilon_{B}}
-\e^{\I\frac{T_{a}}{T_{H}}\epsilon_{D}}\right)}_{\equiv f_{BD}^a}
\right]
\end{eqnarray}
If we assume that contributions of different periodic orbits are
uncorrelated, $Z^{(1)}$ factorizes into a product of
energy-averaged orbit factors $z_{a}$,
\begin{equation}
Z_{\mathrm{diag}}^{(1)}=\e^{\I \left(
\epsilon_{A}+\epsilon_{B}-\epsilon _{C}-\epsilon _{D}\right)/2}
\prod\limits_{a}\left\langle z_{a}\right\rangle\,.
\end{equation}
Inasmuch as the stability coefficient $F_{a}\sim \e^{-\lambda
T_{a}/2}$ is small for long orbits we may expand the exponential
in $z_{a}$ and limit ourselves to the first nontrivial term,
$\left\langle z_{a}\right\rangle  \sim
1+|F_a|^2f_{AC}^{a}f_{BD}^a\approx
\e^{\,|F_a|^2f_{AC}^{a}f_{BD}^a}$; here terms 
proportional to $\e^{\pm\I\,S(E)/\hbar}$ and
$\e^{\pm2\I\,S(E)/\hbar}$ have been eliminated by the energy
average. We so obtain
\begin{eqnarray}
 \label{Zdiag}
Z_{\rm diag}^{(1)} &=&\e^{\I\left( \varepsilon _{A}+\varepsilon
_{B}-\varepsilon _{C}-\varepsilon _{D}\right) /2} \exp
\left(\sum_{a}|F_a|^2f_{AC}^{a}f_{BD}^a\right)\,.
\end{eqnarray}

This result allows for an intuitive interpretation: If we assume
that the non-identical orbits are uncorrelated the exponent in the
product of all orbit factors $z_a$, namely $\Phi =\sum_{a}\left[
  F_a\e^{\I S_{a}(E)/\hbar}f_{AC}^{a} +F_a^*\e^{-\I
    S_{a}(E)/\hbar}f_{BD}^a \right]$, effectively becomes a sum of many
independent random contributions; due to the central limit theorem
$\Phi $ must then obey Gaussian statistics such that $\langle
\e^{\Phi }\rangle =\e^{\langle \Phi^{2}\rangle /2}$ which just
yields (\ref{Zdiag}).

To bring our result to a more explicit final form, we need to
invoke ergodicity in the form of the sum rule of Hannay and Ozorio
de Almeida \cite{HOdA,Haake}: To sum a smooth function of the
 period $T$ over an ensemble of orbits weighed
with $|F_a|^2$, we may as well compute an integral over periods
weighed with $\frac{1}{T}$,
\begin{equation}
\label{HOdA} \sum_{a}|F_{a}|^{2}(\ldots)\approx
\int_{T_{0}}^{\infty } \frac{d T}{T}(\ldots)\,;
\end{equation}
here $T_0$ is some minimal period starting from which orbits
behave ergodically.  Employing the sum rule (\ref{HOdA}) we get
\[
\sum_{a}|F_a|^2f_{AC}^{a}f_{BD}^{a}\sim
\int_{T_{0}}^{\infty }\frac{d T}{T}\left(\e^{\I\frac{T}{%
T_{H}}\epsilon _{A}}-\e^{\I\frac{T}{T_{H}}\epsilon _{C}}\right)
\left( \e^{\I\frac{T}{T_{H}}\varepsilon _{B}}
-\e^{\I\frac{T}{T_{H}}\epsilon_{D}} \right)\,.
\]%
Since for $T\ll T_{H}$ the integrand is $\propto(T/T_{H})^{2}$ we
can set the lower integration limit to $0$, accepting an error of
the order $\left( T_{0}/T_{H}\right) ^{2} $.  Then using
$\int_{0}^{\infty }\frac{d x}{x}\left( \e^{\I ax}-\e^{\I
bx}\right) =\ln\frac{b}{a}$ we arrive at the diagonal part of the
generating function
\begin{equation}\label{Z1diag}
Z_{\rm diag}^{(1)}=\e^{\frac{\I}{2}\left(
\epsilon _{A}+\epsilon _{B}-\epsilon _{C}-\epsilon _{D}\right) }%
\frac{\left( \epsilon _{A}+\epsilon _{D}\right) \left( \epsilon
_{C}+\epsilon _{B}\right) }{\left( \epsilon _{A}+\epsilon
_{B}\right) \left( \epsilon _{C}+\epsilon _{D}\right) }.
\end{equation}

The diagonal approximation yields an estimate for the correlation
function $C(\epsilon)$. Taking into account both $Z_{\rm
diag}^{(1)}$ and its counterpart $Z_{\rm diag}^{(2)}$ defined in
Eq. (\ref{Z2}) we obtain
\begin{equation}
\label{Cdiag} C_{\rm diag}(\epsilon)= -2\frac{\partial^2
(Z^{(1)}_{\rm diag}+Z^{(2)}_{\rm diag})}
{\partial\epsilon_A\partial\epsilon_B}\big|_{(\|)}-\frac{1}{2} =
\frac{1}{2(\I\epsilon)^{2}}-\frac{\e^{2\I\epsilon}}{2(\I\epsilon)^{2}}
\,.
\end{equation}
This result already reproduces the full RMT prediction for the
unitary symmetry class under discussion. It remains to explain,
within our semiclassical approach, why all off-diagonal
contributions do in fact vanish.

\section{Off-diagonal contributions}

\label{sec:off}

Off-diagonal contributions arise from quadruplets of pseudo-orbits
where not all orbits of $A\cup C$ are simply repeated inside
$B\cup D$; some orbits of $A\cup C$ and $B\cup D$ differ in
encounters.  By studying these encounters in detail we shall show
that for systems without time-reversal invariance the off-diagonal
contributions cancel mutually in a nontrivial way.  This also sets
the stage for a generalization to time-reversal invariant dynamics
where this cancellation does not occur, and off-diagonal
quadruplets of pseudo-orbits give rise to terms of higher order in
$1/\epsilon$.

\subsection{Encounters and orbit bunches}
\label{subsec:enc_and_bunches}

Long periodic orbits often come close to themselves, and one
another, in phase space. We speak of an $l$-encounter whenever $l$
stretches of one or more periodic orbits in a chaotic system
approach in this way. Each stretch begins and ends at phase-space
points to be called ``entrance ports'' and ``exit ports''. These
ports can be uniquely defined by setting an upper bound for the
separation between the encountering stretches.  The parts of an
orbit in between encounter stretches will be called ``links''.

Now the crucial point is that the orbits and pseudo-orbits in
hyperbolic motion exist in bunches.  The members of a bunch differ
noticeably only in encounters where the ports are connected
differently by encounter stretches. In contrast the links are very
nearly identical. All members of such a bunch have almost
identical actions. If two of them are taken as $A\cup C$ and
$B\cup D$ in Eq. (\ref{Z1}) their contributions interfere
constructively and need to be taken into account for a full
understanding of spectral statistics.

The phenomenon of bunching is a consequence of hyperbolicity. For
a system with two degrees of freedom\footnote{In the present paper
we
  shall stick to two-dimensional systems for convenience but the
  generalization to systems with arbitrarily many degrees of freedom
  along the lines of \cite{EssenRegensburg,TauSmallLong} is
  straightforward.}, hyperbolicity means that deflections between
phase-space points in the three-dimensional energy shell can be
split into three components: one pointing along the flow, one
unstable component $u$, and one stable component $s$. As two
close-by phase-space points move, the unstable component of their
separation grows exponentially whereas the stable component
shrinks exponentially.  In the limit of large times the rate of
growth or shrinking is given by the Lyapunov exponent $\lambda$ as
\begin{eqnarray}
\label{su} u(t)\sim u(0){\rm e}^{\lambda t}\,,\qquad s(t)\sim
s(0){\rm e}^{-\lambda t}\;.
\end{eqnarray}

 As a consequence, given any two encounter stretches  that are long
  compared with the characteristic size of the system and close enough to
linearize the motion as in (\ref{su}) we can draw a piece of a
physical trajectory that starts close to the first stretch and
ends close to the second stretch.  Its separation from the first
stretch must point along the unstable direction at the respective
phase-space points (such that it approaches the first stretch for
large negative times) whereas the separation from the second
stretch must point along the stable direction (such that it
approaches that stretch for large positive times).  The new piece
of trajectory is thus located where the unstable manifold of one
encounter stretch of $A\cup C$ intersects the stable manifold of
the other stretch \cite{Braun}.  As a consequence we can reconnect
the ports of each encounter in any way we like while leaving the
links in between the encounter stretches practically unchanged.
(Note that subsequent to the construction just explained
exponentially small corrections to the encounter stretches and
links are needed to connect them to continuous orbits.)

\subsection{Structures  of pseudo-orbit  quadruplets}
\label{sec:structures}

\begin{figure} \begin{center} \includegraphics[scale=0.7]{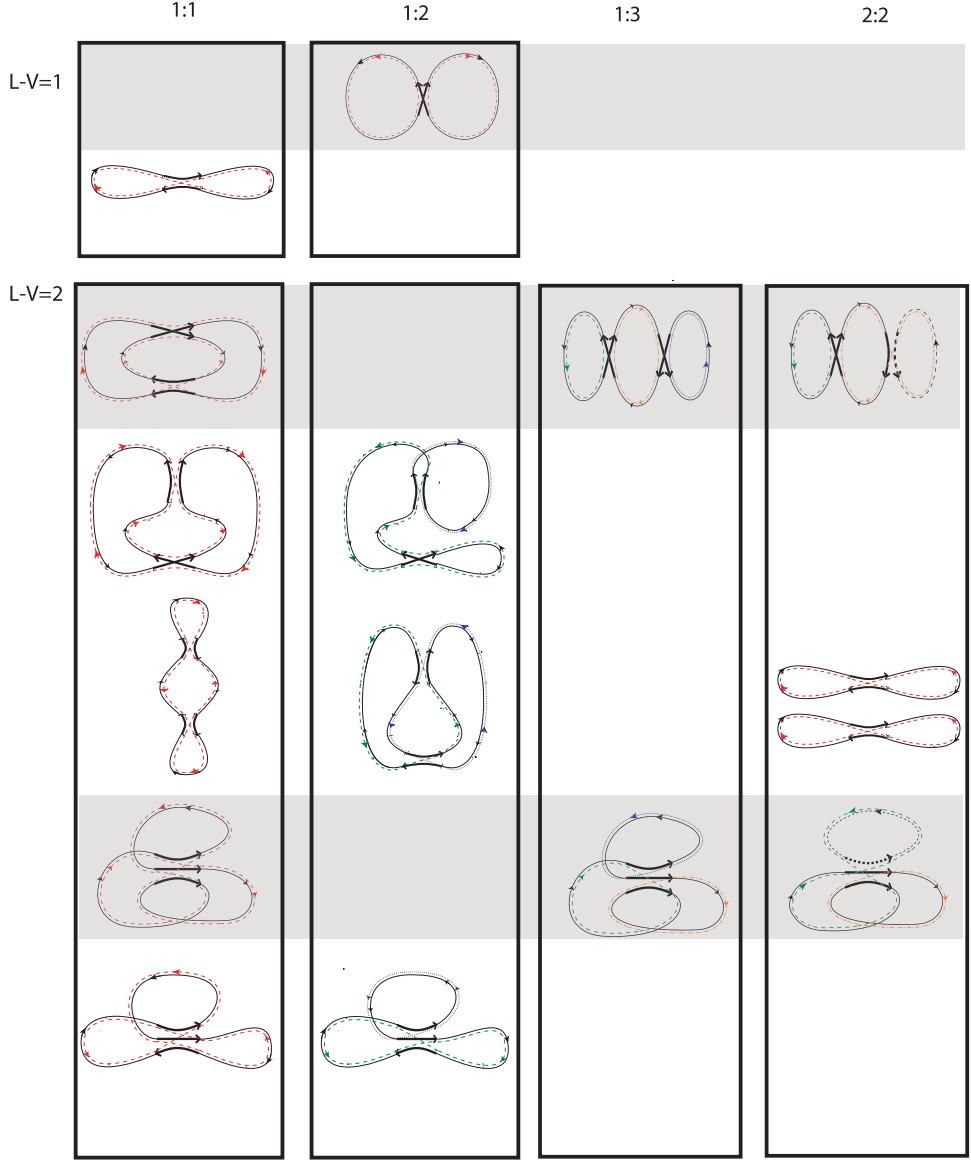}
 \end{center}
 \caption{Physically
    distinct diagrams contributing to the leading ($L-V=1$) and
    next-to-leading order ($L-V=2$) of the $\frac{1}{\epsilon}$
    expansion of $Z^{(1)}$. Only the ones with shaded backgrounds
    arise for the unitary class (only parallel encounters and single
    sense of traversal permitted, as indicated by arrows); they give
    mutually cancelling contributions in each order, both for the
    unitary and the orthogonal classes.  The numbers $n$ and $n'$ of
    orbits in respectively $A\cup C$  (full lines) and $B\cup D$ (dashed lines)
    are written above each column as $n:n'$.
   }
     \label{fig:gallery}
     \end{figure}

\begin{figure}
\begin{center}
\includegraphics[scale=0.5]{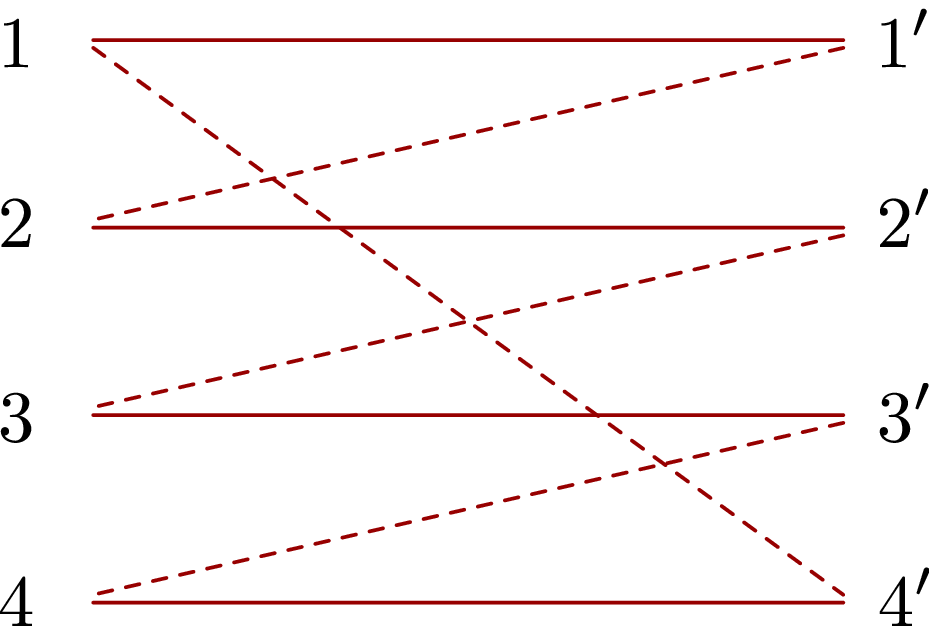}
\end{center}
\caption{Topological picture of an encounter where connections
  $1-1',2-2',3-3',4-4'$ are replaced by $1-4',2-1',3-2',4-3'$.}
\label{fig:connections}
\end{figure}

The pseudo-orbit quadruplets can have many different topologies
some of which are shown in Fig.~\ref{fig:gallery}. For their
classification  we introduce the notion of a {\it structure} by
which we mean a diagram like the ones  in Fig.~\ref{fig:gallery}
but with the encounters and their ports  numbered and each
periodic orbit  assigned to a particular pseudo-orbit. All
structures can be generated if we proceed as follows:
\begin{itemize}
\item[S1:] {\it Encounters} -- We consider any number $V$ of
encounters. For each of the encounters $\sigma=1,2,\ldots,V$, we
take into account all possible numbers of stretches
$l(\sigma)=2,3,4,\ldots$. The $l(\sigma)$ entrance and exit ports
of each encounter are numbered from $1$ to $l(\sigma)$  in such a
way that the $i$-th entrance port is connected  to the $i$-th exit
port before reconnection, and to the $(i-1)$-st exit port after
reconnection; the first entrance reconnects to the last exit. The
encounters can thus be depicted as in Fig. \ref{fig:connections}
where full and dashed lines denote the encounter stretches before
and after the reconnection.
 \item[S2:] {\it Links} --  We
connect the ports by links in all possible ways. Each link must
lead from an exit port to an entrance port of either the same or a
different encounter.
 \item[S3:] {\it Orbits} -- We
 include all possible distributions of the orbits among  the
pseudo-orbits $A,B,C,D$. The graph formed by all links and
encounter stretches before reconnection falls into a certain
number of disjoint parts each standing for a periodic orbit. Each
of these can be included in either $A$ or $C$. After changing
connections inside the encounters the graph consists of a
(possibly different) number of disjoint orbits which have to be
distributed between $B$ and $D$.
\end{itemize}

Thus equipped we shall split the sum over quadruplets in
(\ref{Z1}) into a sum over structures and  a sum over quadruplets
associated to each structure. The sum over quadruplets associated
to each structure will be done using ergodicity.  We shall
determine the likelihood for having encounters inside and between
periodic orbits; this likelihood depends on the length of the
links and the phase-space separation between the encounter
stretches. The sum can then be replaced by suitable integrals. The
subsequent summation over structures becomes a purely
combinatorial problem.

When doing the sums and integrals mentioned, we have to be aware
of  a certain overcounting.  Since our definition of a structure
involves an ordering of encounters and ports, each physical
quadruplet can be associated with a structure  in several
different ways. To begin with, the encounters of a given
quadruplet can be numbered in any one of $V!$ different orders.
Then, in each encounter any of its $l(\sigma)$ entrance ports can
be chosen as the first one. Note that as     soon as the first
entrance in an encounter is chosen, the numbers of all other
encounter ports of the quadruplet become uniquely defined by our
numbering scheme. In all, we get $V!\prod_\sigma l(\sigma)$
alternatives either leading to different structures or to the same
structure but with different numberings and thus to a different
set of the integration parameters. In both cases after summation
over structures and integration over the continuous parameters
 the same quadruplet is taken into account several times.
In order to compensate this overcounting, the contribution of each
structure needs to be divided by $V!\prod_\sigma l(\sigma)$.
Details can be found  in Appendix \ref{app_permutations} where a
formal definition of structures in terms of permutations is given.

\subsection{Classical properties of orbit bunches}

\begin{figure}
\begin{center}
  \includegraphics[scale=.4]{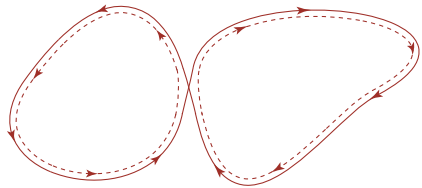}
\end{center}
\caption{Simplest (pseudo-)orbits.} \label{fig:ARS}
\end{figure}

We now follow \cite{TauSmallShort,TauSmallLong} to investigate the
classical properties associated to each structure, and then
evaluate their contribution to the generating function. We start
with the example of orbits differing in a 2-encounter (see Fig.
\ref{fig:ARS}).  First we need coordinates measuring the
phase-space separations between the two encounter stretches. We
therefore place a
 transversal Poincar\'e surface of section at an arbitrary position inside the
encounter and consider the two points where the encounter
stretches pierce through that section.  The separation of the
piercings can be decomposed into a stable component $s$ and an
unstable component $u$ \cite{Braun,Spehner,TurekRichter}. These
coordinates determine the difference between the action of the
original orbit and the cumulative action of the two partner orbits
as $\Delta S=su$ (see also \cite{SR,Mueller}). Moreover, if we
define the encounter as the region with phase-space separations
$|s|,|u|<c$ (where $c$ is arbitrary but classically small) the
time until the end of the encounter can be estimated as
$\frac{1}{\lambda}\ln\frac{c}{|u|}$ and the time since the
beginning of the encounter is obtained as
$\frac{1}{\lambda}\ln\frac{c}{|s|}$.  Summation thus yields the
overall duration\footnote{
  We note that for general hyperbolic systems (where
  Eq. (\ref{su}) is only valid in the limit of large times)
  our formula for $t_{\rm enc}$ is just an approximation.
  The treatment can be made more rigorous if one avoids this
  approximation, by following the lines of \cite{EssenRegensburg}
  and Appendix B of \cite{TauSmallLong}.
}
 of the encounter $t_{\rm
  enc}\sim\frac{1}{\lambda}\ln\frac{c^2}{|su|}$ \cite{TauSmallLong}.
For the semiclassically relevant encounters we need $\Delta
S=su\sim\hbar$, i.e., the encounter duration is of the order of
the Ehrenfest time $T_E=\frac{1}{\lambda}\ln\frac{c^2}{\hbar}$.
Encounters are thus considerably shorter than the orbits, whose
periods must be of the order of the Heisenberg time $T_H\propto
\hbar^{-1}$.

The  likelihood of finding encounters in a periodic orbit can be
determined using ergodicity, i.e., the fact that long trajectories
and (ensembles of) long periodic orbits uniformly fill the energy
shell \cite{Equidistribution}. Ergodicity implies that the
probability for finding two piercings through a Poincar\'e section
with separations $t,s,u$ in differential intervals $d t,d s,d u$
is uniform and reads $\frac{d t d s
  d u}{\Omega}$, with $\Omega$ the volume of the energy shell.
Statistical independence of the two piercings is assumed here;
that assumption is an allowable one since for orbits with a period
of the order $T_H$ the links between the encounter stretches
typically have durations much longer than  the classical timescale
needed for the decay of correlations.

Now integration of the probability density $1/\Omega$ yields the
number density $w_T(s,u)$ for finding, inside an orbit of period
$T$, 2-encounters with stable and unstable separations $s$ and
$u$.  That density was determined in
\cite{TauSmallLong,MuellerPhD} as
\begin{equation}\label{wT}
w_T(s,u)=\frac{T\int d t}{\Omega t_{\rm enc}(s,u)}\;.
\end{equation}
Here the factor $T$ is the period of the orbit (originating from
integration over the time of the first piercing). $\int d t$ is an
integral over the time of the second piercing reckonned from the
first one, with the following restriction: The time separation
between the two piercings must be at least $t_{\rm enc}$ so that
in between the piercings the orbit can go through the rest of the
first encounter stretch, a link of positive duration, and the
beginning of the second stretch. This reasoning leads to $\int d
t=T-2t_{\rm enc}$, but we prefer to just write $\int d t$.

Now let us consider more complicated structures with, say, $V$
different encounters inside or between, say, $n$ orbits of periods
$T_1,T_2,\ldots,T_n$. The encounters are labelled by an index
$\sigma$ running from 1 to $V$ and the $\sigma$th encounter may
have $l(\sigma)$ stretches. Drawing a Poincar\'e section in each
encounter we face altogether $L=\sum_\sigma l(\sigma)$ piercings,
$l(\sigma)$ ones for the $\sigma$th encounter. The separations
between the piercings belonging to one encounter are characterized
by $l(\sigma)-1$ stable and unstable coordinates $s_{\sigma
m},u_{\sigma m}$ with $m=1,2,\ldots,l(\sigma)-1$ (see
\cite{TauSmallLong} for the precise definition). These coordinates
determine the action difference as $\Delta
S=\sum_{\sigma,m}s_{\sigma m}u_{\sigma m}$ and the duration of the
$\sigma$-th encounter as
$ t_{{\rm
enc},\sigma}=\frac{1}{\lambda}\ln\frac{c^2}{\max_m|s_{\sigma
m}|\max_{m'}|u_{\sigma m'}|}$
\cite{TauSmallLong}. By following the same steps as for (\ref{wT})
the number density of encounters is generalized to
\begin{equation}
\label{wTgeneral} w_{T_1,T_2,\ldots T_n}(s,u)=\frac{\prod_{i=1}^n
T_i\int d^{L-n}t}{\prod_\sigma \Omega^{l(\sigma)-1}t_{{\rm
enc},\sigma}}
\end{equation}
with $s=\{s_{\sigma m}\}$, $u=\{u_{\sigma m}\}$. Here the orbit
periods $T_i$ arise from integrating over the time of the first
piercing in every orbit, and the integral $\int d^{L-n} t$ goes
over the possible times of the remaining piercings reckoned from
the first one.

\subsection{Contribution of orbit bunches to $Z$}

\label{sec:eight}

We can now investigate the impact orbit bunches have on spectral
statistics. We thus evaluate the generating function $Z^{(1)}$ as
given by the fourfold pseudo-orbit sum in (\ref{Z1}), taking into
account quadruplets with $A\cup C$ and $B\cup D$ differing in any
number of encounters. As before $A\cup C$ and $B\cup D$ may also
coincide in any number of component orbits, and therefore
$Z^{(1)}_{\rm diag}$ still arises as a {\it factor}. We can thus
rewrite (\ref{Z1}) as
\begin{equation}
\label{multiply} Z^{(1)}=Z_{\rm diag}^{(1)}(1+Z_{\rm
off}^{(1)})\,,
\end{equation}
with $Z_{\rm off}^{(1)}$ picking up non-empty pseudo-orbits $A\cup
C$ and $B\cup D$ that differ in encounters while no longer
comprising identically repeated orbits,
\begin{eqnarray}  \label{off}\nonumber
Z_{\mathrm{off}}^{(1)}&=&\sum_{A,B,C,D\atop{\mathrm{diff.\;in\;
enc.}}} \hspace{-.4cm}\Big\langle F_{A} F_{B}^* F_{C}
F_{D}^*\,(-1)^{n_{C}+n_{D}}\e^{\I\Delta S/\hbar} \e
^{\I(T_{A}\epsilon_A+T_{B}\epsilon_B+T_{C}\epsilon_C+T_{D}
\epsilon_D)/T_H}\Big\rangle\\
&\sim&\sum_{A,B,C,D\atop{\mathrm{diff.\;in\;enc.}}}
\hspace{-.4cm}\Big\langle |F_{A}|^2 |F_{C}|^2
\,(-1)^{n_{C}+n_{D}}\e^{\I\Delta S/\hbar} \e
^{\I(T_{A}\epsilon_A+T_{B}\epsilon_B+T_{C}\epsilon_C+T_{D}
\epsilon_D)/T_H}\Big\rangle \,.
\end{eqnarray}

The off-diagonal part $Z_{\rm off}^{(1)}$ of the generating
function can now be written as a sum over structures, and then
over all quadruplets of pseudo-orbits pertaining to each
structure. We thus obtain
\begin{eqnarray}\label{master1}
Z^{(1)}_{\rm off} =\sum_{\rm structures}\sum_{(A,B,C,D)\atop
\in\,{\rm structure}} \frac{(-1)^{n_C+n_D}}{V!\prod_\sigma
l(\sigma)}\nonumber\\
\times \Big\langle |F_{A}|^2 |F_{C}|^2 \,\e^{\I\Delta S/\hbar} \e
^{\I(T_{A}\epsilon_A+T_{B}\epsilon_B
     +T_{C}\epsilon_C+T_{D}\epsilon_D)/T_H}
\Big\rangle\,,
\end{eqnarray}
where as explained above we divided out the number of possible
orderings of encounters and stretches, $V!\prod_\sigma l(\sigma)$,
since otherwise each quadruplet would be counted multiple times.
 The sum over
pseudo-orbits permitted by a structure can be done by summing over
all choices for the $n$ component orbits  $p_1,p_2,\ldots,p_n$ of
$A\cup C$ and integrating over the $L-V=\sum_\sigma (l(\sigma)-1)$
pairs of stable and unstable coordinates characterizing the $V$
encounters, with the number density (\ref{wTgeneral}) as weight,
\begin{eqnarray}\label{masteranderthalb}
Z^{(1)}_{\rm off} &=&\sum_{\rm
structures}\frac{(-1)^{n_C+n_D}}{V!\prod_\sigma l(\sigma)}
\sum_{p_1,p_2,\ldots,p_n} \prod_{i=1}^n |F_{p_i}|^2\Big\langle\int
d^{L-V}s\,d^{L-V}u
\\ \nonumber
&&\hspace{1 cm}\times\,w_{T_1,T_2,\ldots T_n}(s,u) \;
\e^{\I\sum_{\sigma,m}s_{\sigma m}u_{\sigma m}/\hbar}
\e^{\I(T_{A}\epsilon_A+T_{B}\epsilon_B
      +T_{C}\epsilon_C+T_{D}\epsilon_D)/T_H}
\Big\rangle\,.
\end{eqnarray}
Each of the $n$ orbit sums can now be done using the sum rule of
Hannay and Ozorio de Almeida,
$\sum_{p_i}|F_{p_i}|^2\to\int_{T_0}^\infty\frac{d T}{T}$.
Importing the density $w$ from (\ref{wTgeneral}) we see the orbit
periods in $w$ cancel against those brought in by the sum rule and
face an $L$-fold time integral, over the $n$ periods and $L-n$
piercing times.

To simplify the foregoing expression for $Z_{\rm off}^{(1)}$ we
change the $L$-fold time integral just mentioned to an integral
over the $L$ link durations. The Jacobian of that transformation
is unity. We thus obtain
\begin{eqnarray}\label{master2}
Z^{(1)}_{\rm off} =\sum_{\rm
structures}\frac{(-1)^{n_C+n_D}}{V!\prod_\sigma l(\sigma)}
\left\langle\left( \prod_{\sigma=1}^V \int
d^{l(\sigma)-1}s_\sigma\,
d^{l(\sigma)-1}u_\sigma\frac{T_H^{l(\sigma)}}
{\Omega^{l(\sigma)-1}t_{{\rm enc},\sigma}}
\e^{\I\sum_{m=1}^{l_\sigma -1}s_{\sigma m}u_{\sigma
m}/\hbar}\right)\right.
\\ \left.\nonumber
\times\int_0^\infty d^Lt_{\rm link}\frac{1}{T_H^L}
\e^{\I(T_{A}\epsilon_A+T_{B}\epsilon_B
      +T_{C}\epsilon_C+T_{D}\epsilon_D)/T_H}
\right\rangle\,,
\end{eqnarray}
where we inserted compensating powers of the Heisenberg time.  The
summand coming from a structure in (\ref{master2}) then becomes a
product over all links times a product over all encounters. It is
easy to evaluate the pertinent factors.

Each link appears precisely once in $A\cup C$ as well as in $B\cup
D$. The integral over its duration picks up the integrand
$\frac{1}{T_H}\e^{\I(\epsilon_{A\;{\rm or}\;C}+\epsilon_{B\;{\rm
      or}\;D})t_{\rm link}/T_H}$ and contributes
\begin{equation}\label{link_factor}
\frac{\I}{\epsilon_{A\;{\rm or}\;C}+\epsilon_{B\;{\rm or}\;D}}
\qquad \qquad\qquad
 \qquad {\rm link\;factor}\,;
\end{equation}
clearly, the index alternatives have to be decided according to
which pseudo-orbit the link pertains to before reconnection ($A$
or $C$) and after reconnection ($B$ or $D$).

We now turn to the factor coming from an $l$-encounter.  In the
original pseudo-orbit $A\cup C$, each of the $l$ stretches may
belong to either $A$ or $C$. We denote by $l_A$ the number of
stretches belonging to $A$ and by $l_C$ the number of stretches
belonging to $C$, with $l_A+l_C=l$.  When switching connections
inside the encounter these stretches are replaced by $l$ new
stretches differently connecting the same ports as the old
stretches. These new stretches belong to either $B$ or $D$, and
the corresponding numbers are denoted by $l_B$ and $l_D$ with
$l_B+l_D=l$.  The encounter now yields a contribution $l_A t_{\rm
enc}$ to the duration of pseudo-orbit $A$; analogously for $B$,
$C$ and $D$.  Hence the encounter draws the part
$\e^{\I(l_A\epsilon_A+l_B\epsilon_B
  +l_C\epsilon_C+l_D\epsilon_D)t_{\rm enc}/T_H}$ from the phase factor
$\e^{\I(T_{A}\epsilon_A+T_{B}\epsilon_B
  +T_{C}\epsilon_C+T_{D}\epsilon_D)/T_H}$ in (\ref{master1}).
Collecting everything from (\ref{master2}) pertaining to the
encounter under consideration we get the dimensionless term
\begin{equation}
T_H^l\int\frac{d^{l-1}sd^{l-1}u}{\Omega^{l-1}t_{\rm enc}(s,u)} \,
\e^{\,\I\sum_m s_mu_m/\hbar}\,
\e^{\,\I(l_A\epsilon_A+l_B\epsilon_B+l_C\epsilon_C+l_D\epsilon_D)t_{\rm
enc}(s,u)/T_H}\,.
\end{equation}
For the calculation of the remaining integral we refer to
\cite{TauSmallLong,Transport} and here but briefly sketch the
argument. The key is to Taylor expand the second exponential and
show that only the term linear in $t_{\rm enc}$ survives in the
limit $\hbar\to 0$. In that surviving term the encounter duration
$t_{\rm
  enc}$ cancels and the factor
$\I(l_A\epsilon_A+l_B\epsilon_B+l_C\epsilon_C+l_D\epsilon_D)/T_H$
can be pulled out from the integral. Using
$T_H=\frac{\Omega}{2\pi\hbar}$ and
$\big(\frac{T_H}{\Omega}\big)^{l-1}\int d^{l-1}s\,d^{l-1}u
\,\e^{\,\I\sum_m s_mu_m/\hbar} =\big(\int\!\frac{d sd
u}{2\pi}\,\e^{\,\I
  su}\big)^{l-1}\to 1$ we get
\begin{eqnarray}\label{encounter_factor}
\I(l_A\epsilon_A+l_B\epsilon_B+l_C\epsilon_C+l_D\epsilon_D)
\qquad\qquad\qquad {\rm encounter\; factor}\,.
\end{eqnarray}
The sum over structures (\ref{master2}) thus simplifies to
\begin{eqnarray}\label{master3}
Z^{(1)}_{\rm off}=\sum_{\rm structures} \frac{1}{V!\prod_\sigma
l(\sigma)}(-1)^{n_C+n_D} \frac{\prod_{\rm
enc}\I(l_A\epsilon_A+l_B\epsilon_B
                       +l_C\epsilon_C+l_D\epsilon_D)}
     {\prod_{\rm links}(-\I(\epsilon_{A\;{\rm or}\;C}
                           +\epsilon_{B\;{\rm or}\;D}))}\,.
\end{eqnarray}

A further simplification can be achieved after inspecting more
closely the encounter contributions.  For an $l$-encounter of some
structure we include the overcounting factor $\frac{1}{l}$ and
face the contribution
$\frac{1}{l}\I(l_A\epsilon_A+l_B\epsilon_B+l_C\epsilon_C
+l_D\epsilon_D)$. Each of the $l$ structures obtained from the
given one by cyclic renumbering of the stretches of the selected
encounter contributes identically, and altogether these $l$
structures contribute
$\I(l_A\epsilon_A+l_B\epsilon_B+l_C\epsilon_C +l_D\epsilon_D)$. We
now claim that the same result is obtained if we choose to assign
to each structure the encounter factor $\I(\epsilon_{A\;{\rm
or}\;C}+\epsilon_{B\;{\rm or}\;D})$ where the index alternatives
must be decided according to the placement of the {\it first
entrance port};  e.g., for a structure where the entrance port of
the first encounter stretch belongs to the pseudo-orbit $A$ before
reconnection and to $D$ afterwards we would write
$\I(\epsilon_A+\epsilon_D)$. Taking a sum over our group of $l$
equivalent structures we  then obtain $\sum \epsilon_{A\;{\rm
    or}\;C}=l_A\epsilon_A+l_C\epsilon_C$ since the first entrance
belongs to $A$ and $C$ in $l_A$ and $l_C$ members of the group,
respectively. Similarly we have $\sum\epsilon_{B\;{\rm
    or}\;D}=l_B\epsilon_B+l_D\epsilon_D$.  Consequently, the encounter
contribution
$\frac{1}{l}\I(l_A\epsilon_A+l_B\epsilon_B+l_C\epsilon_C
+l_D\epsilon_D)$ can be replaced by $\I(\epsilon_{A\;{\rm
    or}\;C}+\epsilon_{B\;{\rm or}\;D})$ since after summation over the
group of equivalent structures associated with the encounter the
two expressions give the same result.  Eq. (\ref{master3}) can
thus be replaced by the simpler formula in which the numerator
refers to the first entrance port of each encounter,
\begin{eqnarray}\label{master}
Z^{(1)}_{\rm off}=\sum_{\rm structures}\frac{1}{V!}(-1)^{n_C+n_D}
\frac{\prod_{\rm enc}\I(\epsilon_{A\;{\rm or}\;C}
                       +\epsilon_{B\;{\rm or}\;D})}
{\prod_{\rm links}(-\I(\epsilon_{A\;{\rm or}\;C}
                       +\epsilon_{B\;{\rm or}\;D}))}\,.
\end{eqnarray}
A symmetry between link and encounter factors thus results and
entails an interesting property of the foregoing ``master
formula'' for $Z^{(1)}_{\rm off}$: Apart from a factor $-1$, each
encounter factor is cancelled by the link factor for the link
ending at the first entrance port of that encounter. Indeed, the
first entrance port and the link ending there belong to the same
pseudo-orbit before the reconnection, either $A$ or $C$, and
likewise after the reconnection, either $B$ or $D$. After that
cancellation, the numerator of the $\epsilon$-dependent fraction
becomes $(-1)^V$ and only the denominator remains
$\epsilon$-dependent.  That property will be important for proving
the cancellation of all encounter contributions in
Sect.~\ref{sec:cancel}.

According to (\ref{master}) the contribution of a structure to the
generating function is of the order $\frac{1}{\epsilon^{L-V}}$.
The sum over structures therefore generates a power series in
$\frac{1}{\epsilon}$. Low-order diagrams can easily be drawn; for
instance, Fig.~\ref{fig:gallery} depicts  all diagrams
contributing in the leading two orders, $\frac{1}{\epsilon}$ and
$\frac{1}{\epsilon^2}$, both for the unitary and orthogonal class
 (apart from a further diagram containing two copies of Fig. \ref{fig:ARS}
and a diagram containing one copy of Fig. \ref{fig:ARS} and one
Sieber-Richter pair).

\subsection{Example: Structures for single 2-encounter}

\label{sec:example}

\begin{figure}
\begin{center}
    \includegraphics[scale=0.33]{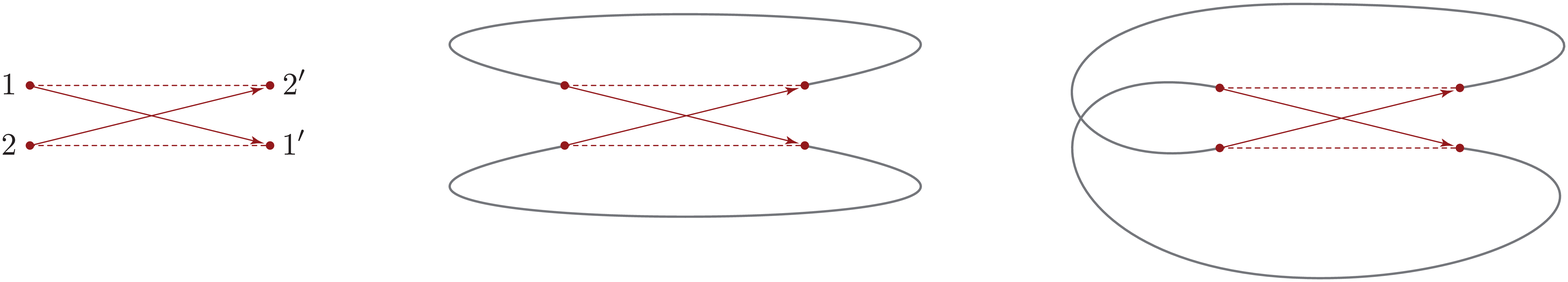}
\end{center}
\caption{(a) 2-encounter with ports labelled, exit ports primed;
(b,
  c) two ways of connecting ports with links. Pre-reconnection
  encounter stretches full lines and post-reconnection ones dashed}
\label{fig_antisr2}
\end{figure}

It is well to illustrate the formula (\ref{master}) for the simple
example of a single 2-encounter with two entrance and exit ports,
depicted in Fig.~\ref{fig_antisr2} (a); the full and dashed lines
respectively show the encounter stretches before and after
reconnection.  We can connect the ports by links (stage S2) in
$L!=2$ ways; the two resulting diagrams are shown in parts (b) and
(c) of Fig.~\ref{fig_antisr2} .

We first deal with Fig. ~\ref{fig_antisr2} (b) where we face a
single pre-reconnection orbit $\gamma_0$ incorporating all ports
and links. After the reconnection the graph disconnects into two
orbits $\gamma_1$ and $\gamma_2$ each of which contains one
entrance and one exit port as well as one link; the labels 1 and 2
signal the number of the respective entrance port.

At the stage S3 we generate structures by dividing the orbits
among the pseudo-orbits. Since we deal with three orbits there are
$2^3$ distinct such allotments; four of these are
\begin{eqnarray}\label{liststruct}
1.&\;\;&A=\{\gamma_0\},\;B=\{\gamma_1,\gamma_2\},\; C=D=\emptyset,\nonumber\\
2.&\;\;&A=\{\gamma_0\},\;D=\{\gamma_1,\gamma_2\},\; C=B=\emptyset,\nonumber\\
3.&\;\;&A=\{\gamma_0\},\;B=\{\gamma_1\},\;\; D=\{\gamma_2\},\;\;C=\emptyset,\nonumber\\
4.&\;\;&A=\{\gamma_0\},\;D=\{\gamma_2\},\; \;B=\{\gamma_1\},\;\;
C=\emptyset,
\end{eqnarray}
and four similar structures have $A$ interchanged with $C$.
Consider the first structure in the foregoing list. The entrance
port $1$ belongs to $\gamma_0\in A$ before, and to $\gamma_1\in B$
after reconnection.  Therefore the encounter factor in the
numerator reads $\I\left(\epsilon_A+\epsilon_B\right)$. Both links
belong to $A$ before and $B$ after reconnection; hence the
denominator is
$\left[-\I\left(\epsilon_A+\epsilon_B\right)\right]^2$. The
factors in (\ref{master}) are calculated using $V=1,n_C=n_D=0$;
the structure thus yields $Z_1=\frac{\I}{\epsilon_A+\epsilon_B}$.
The contribution of the second structure differs by the
replacement $B\to D$; there are two orbits in $D$ and none in $C$
such that the sign is unchanged and we get
$Z_{2}=\frac{\I}{\epsilon_A+\epsilon_D}$.  In the third structure
the first entrance port belongs to $A$ before and $B$ after
reconnection. The two links are in $A$ before reconnection; after
reconnection one of them goes to $B$, another one to $D$. The sign
factor is negative since $n_C=0,n_D=1$. Hence we get
$Z_{3}=-\frac{\I(\epsilon_A+\epsilon_B)}
{(\epsilon_A+\epsilon_B)(\epsilon_A+\epsilon_D)}
=-\frac{\I}{\epsilon_A+\epsilon_D}$.  Similarly, the fourth
structure gives $Z_{4}=-\frac{\I(\epsilon_A+\epsilon_D)}
{(\epsilon_A+\epsilon_B)(\epsilon_A+\epsilon_D)}
=-\frac{\I}{\epsilon_A+\epsilon_B}$.  Obviously, $Z_{3}$ cancels
with $Z_{2}$ and $Z_{4}$ with $Z_1$.

The reader is invited to repeat the calculations for the remaining
four structures and again find full cancellation. Finally, a
patient reader will check that Fig.~\ref{fig_antisr2} (c) leads to
8 more structures whose contributions coincide with those of
Fig.~\ref{fig_antisr2} (b) and also cancel pairwise.

\section{Cancellation of off-diagonal contributions}

\label{sec:cancel}

We now show that for systems without time-reversal invariance all
off-diagonal contributions mutually cancel. Our proof is based on
two cancellation mechanisms: Each structure is cancelled either by
a structure in which the orbits are {\it partitioned differently
into
  pseudo-orbits} or in which {\it one link has been removed}.

\begin{figure}[!ht]
\begin{center}
\includegraphics[ scale=0.6]{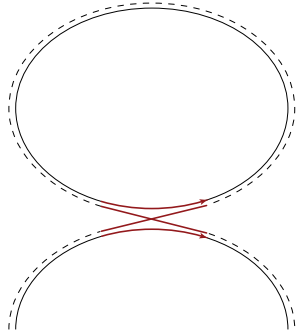}
  \caption{Case 1 of cancellation: Structure involving orbit with
    single encounter stretch}
\label{fig:cancel1}
\end{center}
\end{figure}

\paragraph*{Different partition:}
To start we consider structures where one of orbits (say, in $A$)
only contains {\it one encounter stretch with two ports and a
link}, as in Fig.~\ref{fig:cancel1}, and {\it  the stretch is the
first in its encounter}. The ports and the link then also have to
belong to the same partner orbit (included, say, in $B$).  Then
according to Eq. (\ref{master}) the contribution of the structure
contains a factor $\I(\epsilon_A+\epsilon_B)$ from the entrance
port of the stretch and a factor
$\frac{1}{-\I(\epsilon_A+\epsilon_B)}$ from the link; these link
and stretch factors multiply to $-1$.

Now we compare with a structure where the orbit from
Fig.~\ref{fig:cancel1} is included in $C$ rather than $A$.  Then
the above factors are replaced by $\I(\epsilon_C+\epsilon_B)$ and
$\frac{1}{-\I(\epsilon_C+\epsilon_B)}$, and the product is again
$-1$. However, there is an extra factor $-1$ because the number of
orbits in $C$ is increased by one. Hence the two structures
considered have cancelling contributions.

The same argument goes through if $B$ is replaced by $D$, or if
the orbit in Fig.~\ref{fig:cancel1} is a partner orbit; in the
latter case its inclusion in $B$ and $D$ yields mutually
cancelling contributions. In all cases we are free to drop from
Eq. (\ref{master}) all structures where an orbit includes just one
link and one encounter stretch which is the first in its
encounter.

\begin{figure}[!ht]
\begin{center}
\includegraphics[ scale=0.4]{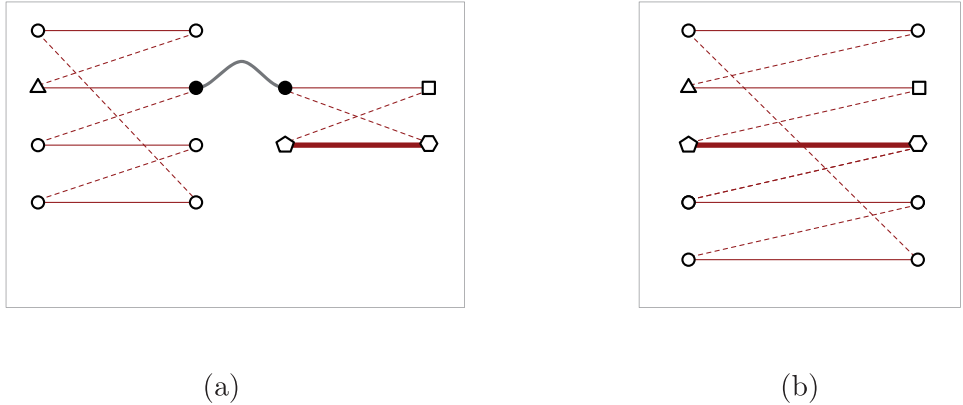}
  \caption{Case 2 of cancellation: Structure involving 2-encounter and
    $l$-encounter cancels against structure with the two encounters
    merged to an $(l+1)$-encounter; only the link disappearing in the
    merger is drawn. Reference stretches in both structures thick}
    \label{fig:cancel2}
\end{center}
\end{figure}

\paragraph*{Shrunken links:}

To get rid of the remaining contributions we consider an arbitrary
structure involving a 2-encounter, and there focus specifically on
the {\it link ending at the first entrance port of the
2-encounter}.  Then where does that link start?  The only
possibility that we still have to consider is that it starts at
the exit port of a {\it different} encounter (with an arbitrary
number $l$ of stretches).  All other possibilities have been dealt
with above: if the link would start at the exit port of the same
stretch the corresponding original orbit would just include one
stretch and one link; if the link would start at the exit of the
second stretch, there would be a partner orbit with only one
stretch and one link. Hence we must have a situation as depicted
schematically in Fig.~\ref{fig:cancel2}a for $l=4$.

Now let us consider a structure that looks topologically the same
as Fig.~\ref{fig:cancel2}a but has the link just considered shrunk
away.  In that structure the two ports connected by the link in
question (the black ports in the picture) are ignored, leaving
altogether $l+1$ entrance ports and $l+1$ exit ports. These ports
have to be connected as in Fig.~\ref{fig:cancel2}a, with
connections in the original orbits following the full lines and
connections in the partner orbits following the dashed lines. If
we leave out the black ports, this leads to connections as in
Fig.~\ref{fig:cancel2}b.  We conclude that when the link between
the 2-encounter and the $l$-encounter is removed these two
encounters merge into a single  ($l+1$)-encounter. The
contribution of the new structure is proportional to the old one.
Just the factor $-\frac{1}{\I(\epsilon_{A\;{\rm
      or}\;C}+\epsilon_{B\;{\rm or}\;D})}$ from the link and the
corresponding factor $\I(\epsilon_{A\;{\rm
or}\;C}+\epsilon_{B\;{\rm
    or}\;D})$ from the removed entrance port are gone, and a change of
sign is again incurred. Hence the two structures depicted have a
chance to cancel.

What remains to be done is a little bit of book-keeping (e.g., to
account for the numbering of encounters, and for the possibility
that removing links from different structures yields the same
result).  To do so, we single out one stretch in each structure as
a reference; that stretch must not be the first stretch in its
encounter. The number of eligible stretches (i.e. the number of
all stretches $L$ minus the number of encounters $V$) will be
denoted by $n$.  We now sum over all structures {\it and} over all
possibilities of choosing references stretches inside a structure;
to avoid an obvious overcounting we modify the summand of Eq.
(\ref{master}) to include an additional divisor $n$.

Now for each contribution where the reference stretch belongs to a
2-encounter we can find exactly one contribution with opposite
sign where the reference stretch belongs to an $l$-encounter with
$l>2$.  This is done by (i) merging the 2-encounter with a
different encounter as described above, (ii) renumbering the
encounters to make up for the loss of one encounter (i.e. decrease
by 1 the labels of all encounters which are larger than the label
of the removed encounter), and (iii) choosing the encounter
stretch following the merged one (e.g.,  the fourth encounter
stretch in Fig.~\ref{fig:cancel2}b) as our new reference stretch.
The last step guarantees that the reference stretch ends up in any
position but the first in its encounter, as demanded above. It is
clear that the new contribution has the same $n$ as the initial
one (as both $L$ and $V$ have been reduced by one), hence the
additional divisor $1/n$ weighing every contribution remains
unchanged.

 Each contribution where the reference stretch belongs to an $l$-encounter can be accessed by following this procedure.
This becomes clear if we take a contribution as in Fig.
\ref{fig:cancel2}b and then undo the steps described above to
arrive at Fig. \ref{fig:cancel2}a. The same contribution as in
Fig. \ref{fig:cancel2}b results from $V$ different contributions
as in Fig. \ref{fig:cancel2}a since the removed 2-encounter may
have an arbitrary index between 1 and $V$. However this degeneracy
is compensated by the change of the factorial in the denominator
of Eq. (\ref{master}). Since the remaining factors are the same
but the sign is different, we thus see that every contribution
where the reference stretch belongs to an $l$-encounter exactly
cancels with  $V$ ``parent'' contributions with the reference
stretch in the 2-encounter.

 To summarize, we have seen that all off-diagonal contributions cancel due to two mechanisms:
For quadruplets where at least one orbit is limited to just one
encounter stretch  (which is the first in its encounter) and one
link, the structures where this orbit is assigned to different
pseudo-orbits mutually cancel. For quadruplets where no orbit is
of this type we singled out a reference stretch. We then
distinguished between contributions where this reference stretch
is part of a 2-encounter or an $l$-encounter with $l>2$. By the
construction of Fig. \ref{fig:cancel2} we were able to show that
these two types of contributions mutually cancel. Hence for
systems without time-reversal invariance we have $Z_{\rm
off}^{(1)}=0$ and only the diagonal term remains.

\section{Semiclassical construction of a sigma model}
\label{sec:build_sigma}

An alternative proof for the cancellation of all off-diagonal
contributions can be based on field-theoretical techniques. We
shall show that the whole series (\ref{master}) can be summed to
become an integral over a certain matrix manifold.
 That matrix integral
turns out to coincide with the one known from the $\sigma$-model
in RMT, and gives the same result as the diagonal approximation.

\subsection{Matrix elements for ports and contraction lines for links}

 We recall that structures of pseudo-orbit quadruplets were defined by
(i) fixing the numbers of encounters and stretches involved (the
stretches being ordered according to the way they are connected in
$A\cup C$ and $B\cup D$), (ii) connecting the encounter ports by
links, and (iii) distributing the resulting pre-reconnection
orbits over the pseudo-orbits $A$ and $C$, and the
post-reconnection orbits between $B$ and $D$.

Now we encode each structure in (\ref{master}) by a sequence of
alternating symbols $B_{kj}$ and $\tilde B_{jk}$ respectively
standing for entrance and exit ports. We start with the first
entrance port of the first encounter, continue with the first exit
port of the first encounter, then list all further ports of that
encounter, then of all further encounters. The indices of
$B,\tilde B$ indicate the affiliation to the pseudo-orbits.
Namely, the second index of $B$ and the first index of $\tilde B$
determine whether before reconnection the port belongs to $A$
(case $j=1$) or to C ($j=2$). Similarly $k$ shows whether the port
belongs after reconnection to $B$ (if $k=1$) or to $D$ ($k=2$).
Due to the two possible values for both $k$ and $j$ we altogether
confront four $B_{kj}$'s and four $\tilde{B}_{jk}$'s.

Our conventions for the numbering of ports imply that the ports
corresponding to a $B$ and the subsequent $\tilde{B}$ (entrance
and exit ports with the same number) are connected by an encounter
stretch of an original orbit. Hence they must belong to the same
original pseudo-orbit and have the same index $j$, as in
$B_{k'j}\tilde{B}_{jk}$. Similarly the ports associated to a
$\tilde{B}$ and the subsequent $B$ (representing an exit port and
the entrance port with the following number) must be connected by
an encounter stretch of a partner orbit. Consequently they belong
to the same partner pseudo-orbit and share the same index $k$, as
in $\tilde{B}_{jk}B_{kj''}$. The same applies for the last
$\tilde{B}$ and the first $B$ in each encounter.  In any case
neighboring indices of successive port symbols must coincide. For
a 2-encounter with two entrance and two exit ports as in
Fig.~\ref{fig_antisr2} (a) we thus get the four-symbol sequence
\begin{equation}
\label{Bexample} B_{k_{1},j_{1}}\tilde{B}_{j_{1},k_{2}}
B_{k_{2},j_{2}}\tilde{B}_{j_{2},k_{1}}\,,
\end{equation}
just like in a trace of a matrix product; note that the
sub-indices of the indices $j,k$ signal port or encounter stretch
labels within the encounter.  For a structure with arbitrarily
many encounters indexed by $\sigma=1,2,\ldots V$ and stretches
inside each encounter indexed by $i=1,2,\ldots l(\sigma)$ the
sequence of ports gives rise to the product
\begin{equation}
\label{general} \prod_{\sigma=1}^V\left(
B_{k_{\sigma1},j_{\sigma1}} \tilde{B}_{j_{\sigma1},k_{\sigma2}}
B_{k_{\sigma2},j_{\sigma2}} \ldots \tilde{B}_{j_{\sigma
l(\sigma)},k_{\sigma1}} \right)\,.
\end{equation}
Note that we do not imply summation over repeated indices either
here or anywhere else in the paper. The indices $j$ and $k$ now
carry two subindices each, $\sigma=1,2,\ldots,V$ to label the
encounter and $i=1,2,\ldots,l(\sigma)$ to label the stretches
within an encounter. Such ``proliferation'' of subindices
notwithstanding there are still only four different $B_{kj}$'s and
four $\tilde{B}_{jk}$'s. Whenever possible below we shall suppress
the ``address labels'' on $k$ and $j$.  If we take $B_{kj},\tilde
B_{jk}$ to be elements of $2\times 2$ matrices
$B=\left({B_{11}\atop B_{21}}{B_{12}\atop B_{22}}\right)$ and
similarly for $\tilde{B}$ we may consider the expression above as
one of the summands in the expansion of the trace product
$\prod_{\sigma=1}^V {\rm tr}\,\big({B}\tilde{B}\big)^{l(\sigma)}$.

The links of a structure will be indicated by drawing
``contraction lines'' between the respective exit and entrance
ports.  Each such line connects one entrance and one exit port of
some orbit {\it and} its partner (recall that links do not change
when shifting connections in encounters). The indices $j,k$ of $B$
and $\tilde{B}$ indicating accommodation in original and partner
pseudo-orbits therefore have to coincide.

{\it Example:}  In our above example of a product of four symbols
the ports can be connected by links in two ways:
\begin{equation}\label{ex2enc}
\contraction[4.5mm]{}{B}{{}_{k_1j_1}\tilde B_{j_1k_2}B_{k_2j_2}}{\tilde B}%
\contraction[2.5mm]{B_{k_1j_1}}{\tilde B}{{}_{j_1k_2}}{\tilde B}
B_{k_1j_1}\tilde B_{j_1k_2}B_{k_2j_1}\tilde B_{j_1k_1}
 \hspace{-1cm} \quad
\hspace{2cm}%
{\rm or} \hspace{2cm}%
\contraction[2.0mm]{}{B}{{}_{k_1j_1}}{\tilde B}%
\contraction[2.0mm]{B_{k_1j_1}\tilde B_{j_1k_1}}{B}{{}_{k_1j_2}}{\tilde B}%
 B_{k_1j_1}\tilde B_{j_1k_1}B_{k_1j_2}\tilde B_{j_2k_1}
 \,;
\end{equation}
note the agreement of indices of the connected ports. We can even
complement these diagrams to topological pictures of the orbits.
If we draw lower lines to indicate how the ports are connected
inside the encounter we obtain
\begin{eqnarray} \label{ex2encextended}
\bcontraction{}{B}{{}_{k_1j_1}}{\tilde B}%
\bcontraction{B_{k_1j_1}\tilde B_{k_1j_1}}{\!B}{{}_{k_1j_1}}{\tilde B}%
\contraction[4.5mm]{}{B}{{}_{k_1j_1}\tilde B_{k_1j_1}B_{k_1j_1}}{\tilde B}%
\contraction[2.5mm]{B_{k_1j_1}}{\tilde B}{{}_{j_1k_2}}{B}%
 B_{k_1j_1}\tilde B_{j_1k_2}\,B_{k_2j_1}\tilde B_{j_1k_1}
 \hspace{.3cm}{\rm and} \hspace{.3cm}%
\bcontraction{B_{k_1j_1}}{\tilde B}{{}_{j_1k_1}}{B}%
\bcontraction[2.5mm]{}{B}{{}_{k_1j_1}\tilde B_{k_1j_1}B_{k_1j_1}}{\tilde B}%
\contraction[4.5mm]{}{B}{{}_{k_1j_1}\tilde B_{k_1j_1}B_{k_1j_1}}{\tilde B}%
\contraction[2.5mm]{B_{k_1j_1}}{\tilde B}{{}_{k_1j_1}}{B}%
 B_{k_1j_1}\tilde B_{j_1k_2}\,B_{k_2j_1}\tilde
 B_{j_1k_1}\nonumber\\
  {\rm or} \nonumber\\%
\bcontraction{}{B}{{}_{k_1j_1}}{\tilde B}%
\bcontraction{b_{k_1j_1}\tilde B_{k_1j_11}}{B}{{}_{k_1j_1}}{\tilde B}%
\contraction{}{B}{{}_{k_1j_1}}{\tilde B}%
\contraction{B_{k_1j_1}\tilde B_{k_1j_1}}{B}{{}_{k_1j_1}}{\tilde
B}
  B_{k_1j_1}\tilde B_{j_1k_1}B_{k_1j_2}\tilde B_{j_2k_1}%
\hspace{.3cm} {\rm and}%
 \hspace{.3cm}
\bcontraction{B_{k_1j_1}}{\tilde B}{{}_{k_1j_1}}{B}%
\bcontraction[3mm]{}{B}{{}_{k_1j_1}\tilde B_{k_1j_1}B_{k_1j_1}}{\tilde B}%
\contraction{}{B}{{}_{k_1j_1}}{\tilde B}%
\contraction{B_{k_1j_1}\tilde B_{k_1j_1}}{B}{{}_{k_1j_1}}{\tilde
B}
  B_{k_1j_1}\tilde B_{j_1k_1}B_{k_1j_2}\tilde B_{j_2k_1}%
 \, .
\end{eqnarray}
Here the upper pair represents the structures in the list
(\ref{liststruct}) and pertains to Fig.~\ref{fig_antisr2}. The
first picture shows the single orbit before reconnection (with
encounter stretches connecting the ports as
$\bcontraction{}{B}{}{\tilde{B}}B\tilde{B}$). The second picture
depicts the two orbits after reconnection (with encounter
stretches connecting the ports as
$\bcontraction{}{\tilde{B}}{}{B}\tilde{B}B$). Analogously the
lower pair represents the structures pertaining to
Fig.~\ref{fig_antisr2} (c).

{\it Diagrammatic rules:} We can now translate the previous
diagrammatic rules for the contribution of a structure in
(\ref{master}) to the new language of contracted sequences of
matrix elements:

(i) The ports $B_{kj}$ and $\tilde{B}_{j'k'}$  connected by a link
 must have the same subscripts $j=j'$ and $k=k'$, and the
contribution from that link in (\ref{master}) equals
$[-\I(\epsilon_{A\,{\rm or}\,C}+\epsilon_{B\,{\rm or}\,D})]^{-1}$.
We can thus write the factor provided by a contraction between
$B_{kj},\tilde B_{j'k'}$ as
$[-\I(\epsilon_j+\epsilon_k')]^{-1}\delta_{kk'}\delta_{jj'}$ where
$\epsilon_j$ stands for $\epsilon_A$ ($j=1$) or $\epsilon_C$
($j=2$); similarly $\epsilon_k'$ equals $\epsilon_B$ for $k=1$ and
$\epsilon_D$ for $k=2$. (The primed energy will always refer to
the post-reconnection pseudo-orbits $B$ or $D$.)

(ii) The indices in an encounter factor $\I\left(\epsilon_{A\,{\rm
or}\,C}+\epsilon_{B\,{\rm or}\,D}\right)$ in (\ref{master}) are
determined by the pseudo-orbits containing the first entrance port
of the encounter. In (\ref{general}) the first entrance port of
the $\sigma$th encounter is denoted by
$B_{k_{\sigma1},j_{\sigma1}}$ such that the respective encounter
factor can be rewritten as
$\I(\epsilon_{j_{\sigma1}}+\epsilon_{k_{\sigma1}}')$.

(iii) Finally Eq. (\ref{master}) implies that there should be a
factor $(-1)^{n_C+n_D}$ depending on the numbers of orbits
included in the pseudo-orbits $C$ and $D$.

\subsection{Wick's theorem and link summation}
\label{subsec:Wick}

We now want to sum over all structures, which in particular
involves summation over all possible ways to draw links alias all
contractions.  A useful tool for that summation is provided by
Wick's theorem. Consider the integral
\begin{equation}\label{tobecontracted}
\int d b d b^* \, \e^{\I\epsilon bb^*}f(b,b^*)
\end{equation}
over the complex plane with
$d bd b^*\equiv \frac{d{\rm Re}b\,d{\rm Im}b}{\pi}$ and $f(b,b^*)$
a product involving an equal number of $b$'s and $b^*$'s, and
${\rm Im} \epsilon>0$ to ensure convergence.
 Wick's theorem states (i) that such Gaussian integrals can be written as
 sums over ``diagrams'' where
each $b$ in $f(b,b^*)$ is connected to one $b^*$ by a
``contraction line'', and (ii) that for complex $b,b^*$ these
diagrams can be evaluated by performing the Gaussian integral over
contracted elements as if the rest of the integrand were absent.
The contraction lines can then be eliminated as in
\begin{equation}
\label{simplecontr} \int d bd b^*\; \contraction{}{b}{b}{}%
\,b\,b^*g(b,b^*)\e^{\I\epsilon
bb^*}\equiv\frac{1}{-\I\epsilon}\int d bd b^*
\,g(b,b^*)\e^{\I\epsilon bb^*}\,.
\end{equation}%
 Using this contraction rule  one can stepwise remove
contraction lines and the $b$'s and $b^*$'s connected by these
lines and thereby in each step earn a factor
$-\frac{1}{\I\epsilon}$.  For complex $b$ and $b^*$ Wick's theorem
is just a fancy way of writing down the integral $\int d b d b^*
\e^{\I\epsilon bb^*}(bb^*)^n=\frac{n!}{(-\I\epsilon)^{n+1}}$; the
factorial gives the number of possible ways to draw contraction
lines.

But remarkably, Wick's theorem
 also holds, up to a sign in Eq.~(\ref{simplecontr}) ,
if we replace $b$ and $b^*$ by Grassmann variables $\eta$ and
$\eta^*$. Grassmann variables are formally defined as
anticommuting, $\eta\eta^*=-\eta^*\eta$, and with integrals $\int
d\eta =\int d\eta^*=0$, $\int d\eta \eta=\int d\eta^*\eta^*=1$
\cite{Haake}. While $\eta$ and $\eta^*$ should in general be seen
as independent variables, for the purposes of this paper  it is
convenient to imagine $\eta^*$ to be the complex conjugate of
$\eta$ and define the complex conjugate of $\eta^*$ to be
$(\eta^*)^*=-\eta$. Imagining
$\exp\left(\I\epsilon\eta\eta^*\right)$ Taylor-expanded we obtain
\begin{equation}\label{fermicontr}
 \int d\eta d\eta^*\; \contraction{}{\eta}{\eta}{{}^*}%
\,\eta\,\eta^*\e^{\I\epsilon \eta\eta^*}\equiv\int d\eta d\eta^*\;
\,\eta\,\eta^*\e^{\I\epsilon
\eta\eta^*}=-1=\frac{1}{\I\epsilon}\int d\eta d\eta^*
\e^{\I\epsilon \eta\eta^*}\,.
\end{equation}
This is the analogue of (\ref{simplecontr}) with $g=1$ which  is
the only case of interest since all powers of the Grassmann
variables higher than 1 vanish. Below we shall apply contraction
rules of the foregoing types to integrals over an arbitrary number
of pairs of commuting and anti-commuting variables. Following
traditions of the physical literature we shall refer to the
anti-commuting variables as ``Fermionic'' as opposed to the
commuting ``Bosonic'' ones.

We now want to use Wick's theorem to represent sums over all
possible ways of drawing contraction lines between $B_{kj}$'s and
$\tilde{B}_{jk}$'s, for general products of the form
(\ref{general}). This can be done if we declare these symbols 
to be either complex Bosonic or Fermionic variables and make the
variables $B_{kj}$ and $\tilde{B}_{jk}$ between which we want to
draw contraction lines mutually conjugate up to a sign, i.e., we
set $\tilde{B}_{jk}\equiv \pm B_{kj}^*$. (The choice of the
Bosonic or Fermionic character of the variables and of the sign
will be made later.) Then all possible contractions are generated
by taking the corresponding expression without contraction lines,
and integrating with a Gaussian weight that has a term
proportional to $B_{kj}B_{kj}^*$ in the exponent. The prefactor of
$B_{kj}B_{kj}^*$ should be chosen as $\I(\epsilon_j+\epsilon_k')$.
Then Wick's theorem yields the desired factor $[-\I(\epsilon_j
+\epsilon'_k)]^{-1}$ for each link belonging to the pseudo-orbits
$j$ and $k$ (up to a sign factor arising for Fermionic variables).
If we also include the encounter factor
$\I(\epsilon_{j_{\sigma1}}+\epsilon'_{k_{\sigma1}})$ we are led to
expressions of the type
\begin{equation}
\label{uptosign} \pm \int d[B,\tilde{B}]\,
\e^{\pm\sum_{j,k=1,2}\I(\epsilon_j+\epsilon_k')B_{kj}B_{kj}^*}
\left\{\prod_{\sigma=1}^V \left(
\pm\I\big(\epsilon_{j_{\sigma1}}+\epsilon'_{k_{\sigma1}}\big)
B_{k_{\sigma1},j_{\sigma1}} \tilde{B}_{j_{\sigma1},k_{\sigma2}}
B_{k_{\sigma2},j_{\sigma2}}\ldots \tilde{B}_{j_{\sigma
l(\sigma)},k_{\sigma1}} \right)\right\}
\end{equation}
where the integration measure involves the product of all
independent differentials,
\begin{equation}
d[B,\tilde{B}]=\frac{1}{\pi^2}d \left({\rm Re} B_{11}\right)d
\left({\rm Im} B_{11}\right)d\left({\rm Re}
B_{22}\right)d\left({\rm Im} B_{22}\right)d B_{12}d B_{12}^*d
B_{21}d B_{21}^*\,.
\end{equation}
All (convergent) expressions of this type contain all relevant
structures with the appropriate link and encounter factors.

\subsection{Signs}\label{signs}

The most delicate task is to fix the signs in (\ref{uptosign}) and
the Bosonic vs. Fermionic nature of the variables so as to produce
the sign factor $(-1)^{n_C+n_D}$ in the sum over structures
(\ref{master}). We shall make the following choices which are
inspired by the supersymmetric sigma model of RMT (see
\cite{Haake} and   Appendix \ref{susy_sigma} ) but will here be
justified purely on semiclassical grounds: We take $B_{11},B_{22}$
Bosonic and $B_{12},B_{21}$ Fermionic,  and let the supermatrices
$B$ and $\tilde{B}$ be related as
\begin{equation}
\tilde{B}=\sigma_zB^\dagger= \left({B_{11}^*\atop
B_{12}^*}{B_{21}^*\atop-B_{22}^*}\right)\;
\end{equation}
where $\sigma_z=\left({1\atop 0}{0\atop -1}\right)$ and the
adjoint of a supermatrix is
$B^\dagger= \left({B_{11}^*\atop -B_{12}^*}{B_{21}^*\atop
B_{22}^*}\right)$. Then we pick the signs in the Gaussian and the
product of $B$'s and $\tilde{B}$'s of (\ref{uptosign}) such that
these terms can be written in terms of supertraces $\str
X\equiv\mathrm{tr}\ \sigma_z X$. We also add an overall minus
sign.

For the {\it Gaussian integral} we thus write
\begin{eqnarray}
\label{signgauss}  -\int d[B,\tilde{B}]
\exp\Big(\I(\epsilon_1+\epsilon'_1)B_{11}B_{11}^*
-\I(\epsilon_1+\epsilon'_2)B_{21}B_{21}^*
+\I(\epsilon_2+\epsilon'_1)B_{12}B_{12}^*
+\I(\epsilon_2+\epsilon'_2)B_{22}B_{22}^*
\Big)\{\ldots\}\nonumber\\
 = -\int
d[B,\tilde{B}]\exp\left(\I{\str}\hat{\epsilon}'B\tilde{B}
+\I{\str}\hat{\epsilon}\tilde{B}B\right)\{\ldots\} \equiv
\langle\!\langle\{\ldots\}\rangle\!\rangle\,;
\end{eqnarray}
here the curly brackets $\{\ldots\}$ have the same content as in
(\ref{uptosign}); to compact the notation we  employ the matrices
\begin{equation}\label{epsilon_matrices}
\hat\epsilon=\diag(\epsilon_1,\epsilon_2)\,, \qquad
\hat\epsilon'=\diag(\epsilon'_1,\epsilon'_2)\;;
\end{equation}
moreover, in the last member of the foregoing chain of equations
we abbreviate the $B$-integral with the Gaussian weight (including
the negative sign factor) by double angular brackets
$\langle\!\langle\ldots\rangle\!\rangle$.  Since the Gaussian is
not normalized, $\langle\!\langle\ldots\rangle\!\rangle$ is not an
average, i.e., $\langle\!\langle1\rangle\!\rangle\neq 1$. Instead
we have
\begin{eqnarray}\label{btodiag}
\langle\!\langle 1 \rangle\!\rangle=-\int d[B,\tilde{B}]\exp\Big(
\I(\epsilon_1+\epsilon_1')B_{11}B_{11}^*
-\I(\epsilon_1+\epsilon_2')B_{21}B_{21}^*
+\I(\epsilon_2+\epsilon_1')B_{12}B_{12}^*
+\I(\epsilon_2+\epsilon_2')B_{22}B_{22}^*\Big)
\nonumber\\
=\frac{(\epsilon_1+\epsilon'_2)(\epsilon_2+\epsilon'_1)}
{(\epsilon_1+\epsilon'_1)(\epsilon_2+\epsilon'_2)}
=\frac{(\epsilon_A+\epsilon_D)(\epsilon_C+\epsilon_B)}
{(\epsilon_A+\epsilon_B)(\epsilon_C+\epsilon_D)}\,,
\end{eqnarray}
reminiscent of the diagonal part $Z^{(1)}_{\mathrm{diag}}$ of
$Z^{(1)}$.  The factor $\langle\!\langle1\rangle\!\rangle$ remains
after all contractions are performed and thus all $B$, $\tilde{B}$
removed. Reassuringly it accounts, once multiplied with  $\e^{\I
(\epsilon_A+\epsilon_B-\epsilon_C-\epsilon_D)/2}$, for all
diagonal quadruplets.

When writing the {\it product of $B$'s and $\tilde{B}$'s} in
(\ref{uptosign}) in terms of a supertrace, we still need to keep
the indices fixed (note that no summation over port labels is
implied!). We thus employ projection matrices $P_1=\diag(1,0)$,
$P_2=\diag(0,1)$ and write  (\ref{uptosign}) as
\begin{equation}
\left\langle\!\!\!\left\langle \prod_{\sigma=1}^V
\I(\epsilon_{j_{\sigma1}}+\epsilon'_{k_{\sigma1}})\, {\rm Str}\,
P_{k_{\sigma1}}B P_{j_{\sigma1}}\tilde{B}P_{k_{\sigma2}} \ldots
P_{j_{\sigma l(\sigma)}}\tilde{B}
\right\rangle\!\!\!\right\rangle\;.
\end{equation}
Note that just like for a trace, it is possible to cyclically
permute the supermatrices appearing in this supertrace.

If we also install the weight $\frac{1}{V!}$ of a structure and
the Weyl factor, we obtain
\begin{eqnarray}
\label{proj} Z^{(1)}(V,\{l(\sigma),j_{\sigma i},k_{\sigma i}\})
=\frac{\e^{\I(\epsilon_A+\epsilon_B-\epsilon_C-\epsilon_D)/2}}{V!}\nonumber\\
\times\left\langle\!\!\!\left\langle \prod_{\sigma=1}^V
\I(\epsilon_{j_{\sigma1}}+\epsilon'_{k_{\sigma1}})\, {\rm Str}\,
P_{k_{\sigma1}}B P_{j_{\sigma1}}\tilde{B}P_{k_{\sigma2}} \ldots
P_{j_{\sigma l(\sigma)}}\tilde{B} \right\rangle\!\!\!\right\rangle
\end{eqnarray}
which gives the cumulative contribution to $Z^{(1)}=Z_{\rm
  diag}^{(1)}\big(1+Z_{\rm off}^{(1)})$ of all structures with $V$
encounters, each having a fixed number of stretches
$l(\sigma),\sigma=1\ldots V$ and a fixed distribution of ports
$\{j_{\sigma i},k_{\sigma i}\}$ among the pseudo-orbits; each
allowed way of drawing contraction lines makes for one structure.
The purely diagonal additive term is due to $V=0$.

The only thing still to be shown for a proof of this statement is
that our conventions incorporate the correct sign $(-1)^{n_C+n_D}$
for each structure depending on the number of orbits included in
$C$ and $D$. To fill this gap we reexpress the contraction rules
(\ref{simplecontr}) and (\ref{fermicontr}) in terms of $B$ and
$\tilde{B}$. Let us consider one way of drawing contraction lines
in Eq.~(\ref{proj}), and moreover select a particular contraction,
either between ports within the same encounters (inside the same
supertrace) or from different encounters. Then we can prove two
identities connecting the contribution of a structure with that of
another structure containing two ports and one link less (see
\ref{sec:contraction_rules} for details of the derivation),
\begin{eqnarray}
\label{rule1}  \left\langle\!\!\!\left\langle%
\contraction{\str (P_k}{B}{P_jY) \str (XP_{j'}}{\tilde{B}}%
\str (P_k B P_jY) \str (XP_{j'} \tilde{B}P_{k'})%
\ldots\right\rangle\!\!\!\right\rangle
&=&-\frac{\delta_{jj'}\delta_{kk'}}{\I(\epsilon_j+\epsilon'_k)}
\big\langle\!\big\langle{\str}(P_kXP_jY)\ldots
\big\rangle\!\big\rangle\;, \\%
\label{rule2}  \left\langle\!\!\!\left\langle%
\contraction{ \str\,(P_k}{B}{P_j UP_{j'}}{\tilde{B}}%
 \str\,(P_k B P_j UP_{j'} \tilde{B}P_{k'}V)%
  \ldots\right\rangle\!\!\!\right\rangle
&=&-\frac{\delta_{jj'}\delta_{kk'}}{\I(\epsilon_j+\epsilon'_k)}
\left\langle\!\left\langle{\str}(P_jU){\str}(P_kV)\ldots
\right\rangle\!\right\rangle\;.
\end{eqnarray}
In these formulae, contraction lines other than the selected one
are not drawn and are assumed to coincide in the left and right
hand sides. In terms of periodic orbits, the rules (\ref{rule1},\
\ref{rule2}) respectively account for the two ports connected by a
link belonging to either different encounters or to the same one.
The symbols $X,Y,U,V$ represent matrix products as in (\ref{proj})
with $Y=\tilde{B}\ldots \tilde{B}$, $X= B\ldots B$; on the other
hand, $U=\tilde B\ldots B,\quad V=B\ldots\tilde{B}$.  The terms
$-\delta_{jj'}\delta_{kk'}\frac{1}{\I(\epsilon_j+\epsilon'_k)}$
are consistent with the diagrammatic rules stated above: Links can
only connect ports associated to the same original pseudo-orbit
$j=j'$ and the same partner pseudo-orbit $k=k'$, and contribute
factors $-\frac{1}{\I(\epsilon_j+\epsilon'_k)}$. The contraction
rules can be used to stepwise simplify our expressions for the
structure contributions. Each step removes two matrices or, in
other words, a link connecting two ports from a periodic-orbit
structure, without changing other links.

At first sight, the rules (\ref{rule1},\ref{rule2}) do not seem to
yield any sign factors. But we must pay special attention to steps
where the contraction line pertaining to the {\it last} link of an
orbit is removed, after removal of all other contraction lines of
some ``parent" orbit in previous steps. The $B$ and $\tilde{B}$ so
connected represent the only ports of the remnant orbit.  That
orbit is still periodic, and therefore the two ports are connected
not only by the link but also by an encounter stretch. The two
matrices $B$'s and $\tilde{B}$'s then have just one projection
matrix in between. If the orbit in question is an original one,
the matrices follow each other like $B P_j \tilde{B}$. We thus
have to use contraction rule (\ref{rule2}) with $U =1$. Then the
first supertrace on the r.h.s. turns into $\str P_j$ which equals
1 if $j=1$ (the orbit belongs to pseudo-orbit $A$) and $-1$ if
$j=2$ (the orbit belongs to $C$). As desired a sign factor $-1$
thus arises for each orbit included in $C$.

An analogous result holds for partner orbits. Imagine that we
remove the last contraction line associated to a partner orbit.
Then the ports connected by this line must also be connected by a
partner encounter stretch. The rule (\ref{rule2}) then applies
with  $V=1$. The second supertrace on the r.h.s. thus equals $\str
P_k$ and yields $-1$ if $k=2$, i.e., if the orbit belongs to $D$.%

To conclude, by turning $B$ and $\tilde{B}$ into supermatrices, we
have successfully incorporated the sign factor $(-1)^{n_C+n_D}$.
Other ways to deal with that sign factor (the replica trick) are
discussed in Appendices \ref{sec:bosonic_sigma} and
\ref{sec:fermionic_sigma}.

\subsection{Towards a sigma model }

To determine $Z^{(1)}$ we sum over all structures of pseudo-orbit
quadruplets.  In Eq. (\ref{proj}) we had already collected all
ways of connecting ports with links.  Now we sum over all
possibilities of assigning ports to pseudo-orbits, i.~e., over the
indices $j_{\sigma i}$, $k_{\sigma i}$ in (\ref{proj}).  This
means that we have to drop the projection matrices. To keep the
factors $\I(\epsilon_{\sigma1}+\epsilon'_{\sigma1})$ connected to
the indices in the supertrace we use the diagonal offset matrices
 $\hat\epsilon,\hat\epsilon'$ and write
\begin{equation}
\label{summedoverind} Z^{(1)}(V,\{l(\sigma)\})=
\frac{\e^{\I(\epsilon_A+\epsilon_B-\epsilon_C-\epsilon_D)/2}}{V!}
\left\langle\!\!\!\left\langle \prod_{\sigma=1}^V
\I\,\str\left(\hat{\epsilon}(\tilde{B}B)^{l(\sigma)}+\hat{\epsilon}'(B\tilde{B})^{l(\sigma)}
\right) \right\rangle\!\!\!\right\rangle\;.
\end{equation}
Here the term $\str\hat{\epsilon}'(B\tilde{B})^{l(\sigma)}$
contains the factor $\epsilon'_{k_{\sigma 1}}$ from (\ref{proj})
while $\str\hat{\epsilon}(\tilde{B}B)^{l(\sigma)}$ includes
$\epsilon_{j_{\sigma 1}}$. It remains to sum over the number $V$
of encounters and over their sizes $l(\sigma)$,
$\sigma=1,2,\ldots,V$. We let $V$ run from 0 to $\infty$, the term
$V=0$ taking care of the purely diagonal contribution, and obtain
\begin{eqnarray}
\label{nonsummed}
Z^{(1)}&=&\e^{\I\,(\epsilon_A+\epsilon_B-\epsilon_C-\epsilon_D)/2}
\sum_{V=0}^\infty\frac{1}{V!} \left\langle\!\!\!\left\langle
\left[\I\,\str\,\sum_{l=2}^\infty
\Big(\hat{\epsilon}(\tilde{B}B)^l+\hat{\epsilon}'(B\tilde{B})^l\Big)
\right]^V \right\rangle\!\!\!\right\rangle
\nonumber\\
&=&\e^{\I\,(\epsilon_A+\epsilon_B-\epsilon_C-\epsilon_D)/2}
\left\langle\!\!\!\left\langle
\exp\left[\I\,\str\,\sum_{l=2}^\infty
\left(\hat{\epsilon}(\tilde{B}B)^l+\hat{\epsilon}'(B\tilde{B})^l\right)
\right]\right\rangle\!\!\!\right\rangle.
\end{eqnarray}

Now recall that (\ref{nonsummed}) was derived using a
semiclassical approximation, which converges only with large
imaginary parts $\eta$ for all energy offsets. In the limit
$\eta\gg 1$ the $B$-integral draws overwhelmingly dominant
contributions from near $B=0$ whereas  other $B$ only make for
exponentially small corrections of order $e^{-\eta}$. Therefore in
the important integration region we have $||B\tilde B||\ll 1$ and
the geometric series in the foregoing exponent can be replaced by
its sum.  Upon explicitly writing the Gaussian integral we get the
compact result
\begin{eqnarray}\label{SUSYZ}
Z^{(1)}(\epsilon_A,\epsilon_B,\epsilon_C,\epsilon_D)\sim
-\e^{\I\,(\epsilon_A+\epsilon_B-\epsilon_C-\epsilon_D)/2}\nonumber\\
\times\int_{\cal L} d[B,\tilde{B}]\exp\left[\I\;\str\! \left(
\hat{\epsilon}\,\frac{\tilde{B}B}{1-\tilde{B}B}
+\hat{\epsilon}'\,\frac{B\tilde{B}}{1-B\tilde{B}} \right)\right]\\
\nonumber = -\int_{\cal L}
d[B,\tilde{B}]\exp\left[\frac{\I}{2}\;\str\! \left(
\hat{\epsilon}\,\frac{1+\tilde{B}B}{1-\tilde{B}B}
+\hat{\epsilon}'\,\frac{1+B\tilde{B}}{1-B\tilde{B}}
\right)\right],\qquad \eta\gg 1.
\end{eqnarray}
The integration domain ${\cal L}$  must include the vicinity of
the point $B=\tilde B=0$ and obey $1-B\tilde B >0$, to exclude
regions where the integrand is exponentially large; otherwise it
is arbitrary. Under these limitations and $\eta\gg 1$ the
 asymptotic $1/\epsilon$-expansion of the integral (\ref{SUSYZ})
coinciding with the semiclassical series (\ref{summedoverind}) is
created by  the saddle point of the exponent at $B=\tilde B=0$.

If we  evaluate (\ref{SUSYZ}) in the limit $\eta\gg 1$ (see
 Appendix \ref{susy_sigma}
) we obtain the same result $Z^{(1)}$ as in the diagonal
approximation, Eq (\ref{Z1diag}). Then the same $Z^{(2)}$ as in
the diagonal approximation is obtained using the Riemann-Siegel
relation $Z^{(2)}(\epsilon_A,\epsilon_B,\epsilon_C,\epsilon_D)
=Z^{(1)}(\epsilon_A,\epsilon_B,-\epsilon_D,-\epsilon_C)$. We see
once again that all off-diagonal terms in the generating function
cancel.


\subsection{Comparison with the RMT sigma model}
\label{subsec:RMTsigma}

In the so called rational parametrization of the zero-dimensional
sigma model of RMT  (see, e.g.,   Appendix \ref{susy_sigma} ) the
{\it exact} generating function $Z$ is given by the integral
\begin{equation}\label{RMTZunitary}
Z(\epsilon_A,\epsilon_B,\epsilon_C,\epsilon_D)= -\int_{{\cal L}_0}
d[B,\tilde{B}]\exp\left[\frac{\I}{2}\;\str\! \left(
\hat{\epsilon}\,\frac{1+\tilde{B}B}{1-\tilde{B}B}
+\hat{\epsilon}'\,\frac{1+B\tilde{B}}{1-B\tilde{B}}
\right)\right]\,,
\end{equation}
identical in appearance with (\ref{SUSYZ}) and with the
integration measure and the structure of the supermatrices the
same as there. However there are two important differences: First,
the offset variables $\epsilon_{A,B,C,D}$ are all allowed to be
real. Second, the complex Bosonic integration variables are
assigned specific integration domains, the open unit disk
$|B_{11}|<1$ and the full complex plane  $ |B_{22}|<\infty$. It is
easy to show that one of the eigenvalues of $B\tilde B$ is then
between 0 and 1 while the other one can take any negative value.
Consequently $1-B\tilde B>0$ and the integration domain ${\cal
L}_0 $ is in fact a special case of $\cal L$ in (\ref{SUSYZ}).

Now suppose that instead of calculating the integral exactly we
consider its high-energy asymptotics. We then get a sum of
contributions of the two existing stationary points of the
exponent. The first stationary point is at $B=0$; its contribution
coincides with our semiclassical result $Z^{(1)}$.

The second stationary ``point'' associated with the so called
Andreev-Altshuler saddle  corresponds to
$B_{11}=0,B_{22}\to\infty$ (see Appendix \ref{susy_sigma} ). For
$\eta\gg 1$ its contribution is exponentially small compared to
the first one and can be dropped, but it is of the same order for
small or zero $\eta$. Now importantly this contribution is related
to $Z^{(1)}$ by
$Z^{(2)}(\epsilon_A,\epsilon_B,\epsilon_C,\epsilon_D)
=Z^{(1)}(\epsilon_A,\epsilon_B,-\epsilon_D,-\epsilon_C)$ (see
 Appendix \ref{susy_sigma}). This is the same relation as between the two summands in our
semiclassical approximation of the generating function, based on
the Riemann-Siegel lookalike formula and thus unitarity. As
anticipated the generating function and the two-point correlation
function of fully chaotic dynamics without time-reversal
invariance thus agree with those derived from the Gaussian Unitary
Ensemble of RMT.

The following extension of the comparison with the RMT sigma model
requires some more knowledge of that model. Previously uninitiated
readers may want to first consult with introductions provided in
 Appendix \ref{susy_sigma} and in \cite{Haake} or with the
pioneering work of {\it K. Efetov} \cite{Efetov} .

{\it $Q$-integral and geometrical interpretation:} We first
compact the RMT matrix integral (\ref{RMTZunitary}) to
\begin{equation}\label{Qint}
Z=\int d Q \,\e^{\frac{1}{2}\str Q\left({\hat{\epsilon}'\atop 0}
{0 \atop -\hat{\epsilon}}\right)}\,;
\end{equation}
here the $4\times 4$ supermatrix $Q=\I\,T\Lambda^{(1)} T^{-1}$ is
related to a diagonal matrix (called ``standard saddle'')
$\Lambda^{(1)}=\diag(1,1,-1,-1)$ by conjugation with the
supermatrix $T=\left({1 \atop \tilde{B}}{B \atop 1}\right)$.  The
matrices $Q$ and $\Lambda^{(1)}$ obey the nonlinear constraints
$-Q^2=\left(\Lambda^{(1)}\right)^2=1$. The integration domain for
the Bosonic elements in $Q$ can be shown to be the two-sphere
$S_2$ and (one of the disjoint halves of) the two dimensional
hyperboloid $H_2$. In particular, the complex variable $B_{22}$ in
the foregoing rational parametrization can be identified as a
stereographic projection variable for the sphere $S_2$. The point
$B_{22}=0$ corresponds to the ``south pole'' while the antipodal
``north pole'' is approached as $B_{22}\to \infty$.  The complex
variable $B_{11}$, on the other hand, is associated with $H_2$.

The foregoing rational parametrization of the manifold of matrices
$Q$ is just one of many. In fact, the manifold in question can be
characterized without specifying any ``coordinates'', and we would
like to just state that  characterization here, referring to
 Appendix \ref{susy_sigma}
 of the present paper for some details.
 For a geometrical interpretation of the matrix integral
it is well to reorder the lines and columns in the $4 \times 4$
matrices $Q,T,\Lambda$ as $1234\to 1324$, so as to be able to
employ $2\times 2$ blocks as
$Q=\left({Q_{BB}
    \atop Q_{FB}} {Q_{BF}\atop Q_{FF}}\right)$; the Bose-Bose block
$Q_{BB}$ and the Fermi-Fermi Block $Q_{FF}$ contain only commuting
entries while all anticommuting entries are assembled in the
off-diagonal blocks $Q_{BF}, Q_{FB}$. In that ``Bose-Fermi
notation'' (the previously used ordering is known as the
advanced-retarded notation) \cite{Haake} the integration domain
for the blocks $Q_{BB},Q_{FF}$ in the sigma model integral become
\begin{equation}\label{sphere_and_hyperboloid}
Q_{BB}\in H_2\equiv \mathrm{U}(1,1)/\mathrm{U}(1)\times
\mathrm{U}(1),\qquad Q_{FF}\in S_2\equiv
\mathrm{U}(2)/\mathrm{U}(1)\times \mathrm{U}(1),
\end{equation}
where $\mathrm{U}(2)$ is the 2-dimensional unitary group, and
$\textrm{U}(1,1)$ is the group of 2-dimensional pseudo-unitary
matrices, i.e. the group of matrices obeying $U^\dagger \sigma_z
U=\sigma_z$, where $\sigma_z={\rm diag}(1,-1)$. Geometrically,
$H_2$ and $S_2$ define the two-dimensional hyperboloid and the
two-sphere mentioned before.

 {\it Andreev-Altshuler saddle point \cite{YurkevichLerner,KamenevMezard,AndreevAltshuler}:}
We now comment on another specific parametrization of the RMT
sigma model connected to the contributions $Z^{(2)}$ in our
semiclassical approach. This parametrization is obtained from the
above rational one by the transformation $Q=RQ'R$ with $R$ a
permutation matrix interchanging rows and columns as $1234\to
1432$ and obeying $R^2=1$. The pertinent Jacobian is unity and the
integration manifold is unchanged. The exact generating function
can thus be written as the ``$Q'$-integral''
\begin{equation}\label{Q'int}
Z=\int d Q' \,\e^{\frac{1}{2}\str Q'R\left({\hat{\epsilon}'\atop
0} {0 \atop -\hat{\epsilon}}\right)R}
\end{equation}
and the rational parametrization chosen for $T$ in
$Q'=T\Lambda^{(1)}T^{-1}$. We have thus recovered the Weyl
symmetry,
$Z(\epsilon_A,\epsilon_B,\epsilon_C,\epsilon_D)=Z(\epsilon_A,\epsilon_B,-\epsilon_D,-\epsilon_C)$,
since conjugation of the matrix of energy offsets with $R$ just
amounts to $\epsilon_C\leftrightarrow -\epsilon_D$. On the other
hand, it is most illuminating to check that the ``point''
$B=\tilde{B}=0$ for $Q'$ corresponds to $Q=R\Lambda^{(1)}R\equiv
\Lambda^{(2)}$, the so called Andreev-Altshuler saddle. The two
diagonal ``saddles'' are not within reach of one another through
the rational parametrization with any finite $B,\tilde{B}$.
Rather, $\Lambda^{(2)}$ can only be approached as
$T\Lambda^{(1)}T^{-1}$ for $B,\tilde{B}\to \infty$ and vice versa.
In particular, south and north pole of the sphere $S_2$ are
swapped in the $Q'$-integral relative to the $Q$-integral. In that
sense one may say that the $Q$-integral (\ref{Qint}) rationally
parametrizes $Q$ w.r.t. the standard saddle while the
$Q'$-integral (\ref{Q'int}) rationally parametrizes $Q$ w.r.t. the
Andreev-Altshuler saddle. We would like to emphasize that either
variant of the rational parametrization yields, upon exact
evaluation of the superintegral, the full generating function; the
 inaccessibility of the antipodal saddle does not matter since all
of its neighborhood is accounted for.

As already stated, our semiclassical construction of the
$[B,\tilde{B}]$-integral corresponds to choosing the rational
parametrization according to the $Q$-integral in (\ref{Qint}),
since that choice for the RMT sigma model integral does yield
$Z^{(1)}$ in a Gaussian approximation around $B=\tilde{B}=0$. Had
we done our semiclassical analysis for
$Z^{(2)}(\epsilon_A,\epsilon_B,\epsilon_C,\epsilon_D)$, we would
have ended up with a $[B,\tilde{B}]$-integral corresponding, in
the RMT context, to the alternative rational parametrization in
the $Q'$-integral in (\ref{Q'int}): The Gaussian approximation
about $B=\tilde{B}=0$ in (\ref{Q'int}) does yield
$Z^{(2)}(\epsilon_A,\epsilon_B,\epsilon_C,\epsilon_D)$, as is
easily checked. In other words, the asymptotic contribution which
stems from the infinitely remote integration point in the
parametrization using the standard saddle, becomes the
contribution of the stationary point $B=\tilde B=0$ in the
Andreev-Altshuler parametrization and vice versa.

\section{Time-reversal invariant systems}

\label{sec:orthogonal}

We now propose to generalize the above to time-reversal invariant
systems. For them, the possibility to revert orbits (and their
pieces) in time makes room for more constructive interference. We
shall see that the contributions of orbits differing in encounters
no longer cancel but rather generate terms of arbitrarily high
orders in $1/\epsilon$. The summation of all relevant pseudo-orbit
quadruplets can again be done with the help of a sigma-model type
matrix integral.

\subsection{Diagonal approximation and off-diagonal corrections}

For time-reversal invariant systems the orbits of $(A,C)$ can not
only be repeated in $(B,D)$ but may also be reversed in time and
only then included in $(B,D)$.  Thus if we again formulate the
{\it diagonal approximation} in terms of an exponentiated double
sum over orbits (\ref{Zdiag}), the relevant contributions to the
exponent involve pairs of orbits which are either identical or
mutually time-reversed. This doubling of orbit pairs leads to an
additional factor 2 in the exponent, and the resulting term
$\frac{(\epsilon_A+\epsilon_D)(\epsilon_C+\epsilon_B)}
{(\epsilon_A+\epsilon_B)(\epsilon_C+\epsilon_D)}$ in Eq.
(\ref{Z1diag}) needs to be squared.

Moreover, we now have to consider {\it encounters} where some of
the stretches are almost mutually time-reversed rather than close
in phase space. For instance, Sieber-Richter pairs as depicted in
Fig.~\ref{fig:RS} contain one encounter with two almost
time-reversed stretches. If we change connections inside this
encounter, the two links outside the encounter still look almost
the same in configuration space, but one of them must be traversed
with opposite sense of motion, and this is possible only if the
dynamics is  time reversal invariant.

Since time reversal changes the sides where a stretch enters or
leaves an encounter, it becomes cumbersome to work with the notion
of entrance and exit ports.  Instead we arbitrarily refer to the
ports on one side of each encounter as ``left'' ports and those on
the opposite side as ``right'' ports. Encounter stretches may lead
either from left to right or from right to left. The links may
connect left and left, right and right, or left and right ports.

Our formal definition of structures is  then changed as follows:
 S1: When numbering ports we now make sure that in $A\cup C$
the $i$-th left port is connected to the $i$-th right port,
whereas in $B\cup D$ it is connected to the $(i-1)$-st right port.
 S2: When connecting ports through links we have to allow for
arbitrary connections between ports on either side of the
encounters.  S3: Apart from dividing the orbits before
reconnection between the pseudo-orbits $A$ and $C$ and the orbits
after reconnection among $B$ and $D$ we have to choose the sense
of traversal of every orbit.

Again, several structures can describe the same quadruplet. In
particular, we decide arbitrarily which side of each encounter is
chosen as left or right.  For $V$ encounters there are $2^V$
equivalent possibilities of labelling the sides. To avoid
overcounting, this factor needs to be divided out.

The rest of our argument proceeds in analogy to Section
\ref{sec:off}. Our semiclassical {\it generating function}  turns
into
\begin{eqnarray}
Z_{\mathrm{diag}}^{(1)}&=&\mathrm{e}^{\,\I\,
(\epsilon_{A}+\epsilon_{B}-\epsilon_{C}-\epsilon_{D})/2}
\left(\frac{(\epsilon_{A}+\epsilon_{D})(\epsilon_{C}+\epsilon
_{B})}{(\epsilon_{A}+\epsilon_{B})(\epsilon_{C}+\epsilon_{D})}\right)^2
\nonumber\\
Z^{(1)}_{\rm off}&=&\sum_{\rm structures}\frac{1}{2^V
V!}(-1)^{n_C+n_D}\frac{\prod_{\rm enc}\I\, (\epsilon_{A\;{\rm
or}\;C}+\epsilon_{B\;{\rm or}\;D})} {\prod_{\rm
links}(-\I\,(\epsilon_{A\;{\rm or}\;C}+\epsilon_{B\;{\rm
or}\;D}))} \label{masterorth}
\end{eqnarray}
where the subscripts in the encounter factor depend on the first
left port.

We note that some of these structures already appear for systems
without time-reversal invariance, or coincide with such structures
up to the directions of motion in some of the orbits. One can show
that the proof of section \ref{sec:cancel} generalizes
accordingly, i.e., all structures trivially related to those in
the unitary case mutually cancel.

Let us use the semiclassical expansion (\ref{masterorth}) to
calculate the {\it leading-order oscillatory contribution} to the
correlation function. The oscillatory term is obtained by
application of $\partial^2/\partial \epsilon_A\partial\epsilon_B$
to
\begin{eqnarray}
&&Z^{(2)}\left( \epsilon _{A},\epsilon _{B},\epsilon _{C},\epsilon
_{D}\right)  =Z^{(1)}\left( \epsilon _{A},\epsilon
_{B},-\epsilon _{D},-\epsilon _{C}\right) \\
&=&\e^{\frac{\I}{2}\left( \epsilon _{A}+\epsilon _{B}+\epsilon
_{C}+\epsilon _{D}\right) }
 \left[ \frac{\left( \epsilon _{A}-\epsilon _{C}\right)
\left( \epsilon _{B}-\epsilon _{D}\right) }{\left( \epsilon
_{A}+\epsilon _{B}\right) \left( \epsilon _{C}+\epsilon
_{D}\right) }\right] ^{2}\left(1+Z_{\mathrm{off}}^{(1)}\left(
\epsilon _{A},\epsilon _{B},-\epsilon _{D},-\epsilon
_{C}\right)\right)\nonumber
\end{eqnarray}%
and then  taking the limit
$\epsilon_A,\epsilon_B,\epsilon_C,\epsilon_D\to\epsilon$. The
factor $\left[ \left( \epsilon _{A}-\epsilon _{C}\right) \left(
\epsilon_{B}-\epsilon _{D}\right) \right] ^{2}$ tends to zero in
this limit even after taking the derivatives. Therefore the
oscillatory contribution to the correlator is purely off-diagonal.
It must be connected to summands in $Z_{\mathrm{off}}^{(1)}\left(
\epsilon_{A},\epsilon _{B},-\epsilon _{D},-\epsilon _{C}\right) $
which tend to infinity like $\left[ \left( \epsilon
    _{A}-\epsilon _{C}\right) \left( \epsilon _{B}-\epsilon
    _{D}\right) %
\right] ^{-1}$,  since multiplication with such terms and
differentiation w.r.t. $\epsilon_A$ and $\epsilon_B$ is needed
 to eliminate the factor
 $\left[ \left( \epsilon _{A}-\epsilon _{C}\right) \left( \epsilon
    _{B}-\epsilon _{D}\right) \right] ^{2}$ .
 The respective contributions in $Z_{\mathrm{off}}^{(1)}\left(
  \epsilon _{A},\epsilon _{B},\epsilon _{C},\epsilon _{D}\right) $
behave like $\left[ \left( \epsilon _{A}+\epsilon _{D}\right)
\left(
    \epsilon _{C}+\epsilon _{B}\right) \right] ^{-1}$.
Since inverse powers of the energy offsets  can only originate
from the link factors, such terms can only be due to quadruplets
that contain a link belonging to the pseudo-orbit $A$ before and
to $D$ after reconnection, and another link belonging to $C$
before and to $B$ after the reconnection.  In particular this
means that the corresponding quadruplets must contain at least two
orbits, both before and after reconnection.

In the lowest order these are the quadruplets depicted in column
2:2 of Fig. \ref{fig:gallery}.
 (We omitted quadruplets containing two
copies of Fig. \ref{fig:ARS} or one copy of Fig. \ref{fig:ARS} and
one Sieber-Richter pair. These quadruplets are irrelevant since
the contribution of a disconnected diagram can be written as the
product of its components, and the factor due to Fig.
\ref{fig:ARS} vanishes by the argument in section
\ref{sec:example}.)
 Among them the only diagram that
explicitly requires time-reversal invariance is the one consisting
of two Sieber-Richter pairs (the middle diagram in  column 2:2 of
Fig. \ref{fig:gallery}). In order to satisfy the above conditions
on membership in pseudo-orbits we only need to consider the
situation where one of these Sieber-Richter pairs
 involves one orbit included in $A$ and one orbit included in $D$,
whereas the other pair involves one orbit included in $B$ and one
orbit included in $C$. There are $2^4=16$ ways to choose
directions of motion in each of the four orbits involved, and two
choices for considering either of the two encounters as the first,
leaving altogether 32 equivalent structures. Their total
contribution to $Z^{(1)}_{\mathrm {off}}\left( \epsilon
  _{A},\epsilon _{B},\epsilon _{C},\epsilon _{D}\right)$ is
\[
32\times \frac{1}{2^2 2!}\;\frac{\left[ \I\left( \epsilon _{A}+\epsilon _{D}\right) %
\right] \left[ \I\left( \epsilon _{C}+\epsilon _{B}\right) \right] }{%
\left[ -\I\left( \epsilon _{A}+\epsilon _{D}\right) \right] ^{2}%
\left[ -\I\left( \epsilon _{C}+\epsilon _{B}\right) \right] ^{2}}=-%
\frac{4}{\left( \epsilon _{A}+\epsilon _{D}\right) \left( \epsilon
_{C}+\epsilon _{B}\right) };
\]
The leading oscillatory contribution to the correlator thus reads
\begin{eqnarray}
C_{\rm double-SR }\left( \epsilon \right) &=&-2\frac{\partial
^{2}}{\partial \epsilon _{A}\partial \epsilon _{B}}\left[ \left( -%
\frac{4}{\left( \epsilon _{A}-\epsilon _{C}\right) \left( \epsilon
_{B}-\epsilon _{D}\right) }\right) \right. \nonumber \\
&&\times \left. \e^{\frac{\I}{2}\left( \epsilon _{A}+\epsilon
_{B}+\epsilon _{C}+\epsilon _{D}\right) }\frac{\left( \epsilon
_{A}-\epsilon _{C}\right) ^{2}\left( \epsilon _{B}-\epsilon
_{D}\right) ^{2}}{\left( \epsilon _{A}+\epsilon _{B}\right)
^{2}\left( \epsilon _{C}+\epsilon _{D}\right) ^{2}}\right]
_{\parallel } =\frac{1}{2\epsilon ^{4}}\e^{\I 2\epsilon }
\end{eqnarray}%
 which coincides with the predictions of RMT.

\subsection{Semiclassical construction of a sigma model}

To go beyond the leading order, we again write our semiclassical
results in terms of a matrix integral. The derivation of this
matrix integral is very similar to the unitary case. First we
encode every structure by a sequence of symbols, each symbol
standing for a port, and links depicted by contraction lines.
Second, we replace every symbol by a Bosonic or Fermionic variable
and integrate the product with a suitably chosen Gaussian weight.
The contribution of every structure will then coincide with the
corresponding contractions in the resulting Gaussian integral; the
sum over all possible link connections will be provided by the
integral itself. Finally, summing over all encounter sets we get
an integral representation of the generating function known from
the zero dimensional sigma model of RMT.

\subsubsection{Matrix elements for ports and contraction lines for links}

We again write down the sequence of ports starting with the first
left port of the first encounter, continuing with the first right
port, etc. The left ports will be denoted by $B_{\nu k,\mu j}$ and
the right ports by $\tilde{B}_{\mu j,\nu k}$. The first and the
second pair of indices of $B$ respectively refer to the properties
of the port before and after reconnection, and vice versa for
$\tilde B$. As in the unitary case the Latin index $j=1,2$
indicates whether the port belongs to the original pseudo-orbit
$A$ ($j=1$), or $C$ ($j=2$). The index $k$ reveals whether  after
reconnection the port belongs to the partner pseudo-orbit $B$
($k=1$) or $D$ ($k=2$).

 The Greek indices $\mu,\nu$ 
account for the directions of motion through the port and the
attached encounter stretches, both in the pertinent original and
partner orbits: If $\mu=1$ the direction is from left to right in
the original orbit; if $\mu=2$ it is from right to left. The index
$\nu=1,2$ indicates the same for the partner orbits. The symbols
$B_{\nu k,\mu j},\tilde{B}_{\mu j,\nu k}$ can be considered as
elements of $4\times 4$ matrices $B,\tilde B$; to simplify the
notation we shall often substitute a capital Latin letter for the
composite index like $J=(\mu,j)$.

 As for systems without time-reversal invariance the ports are ordered
 such that an element of $B$ and the immediately following $\tilde{B}$
 represent ports that are connected by an encounter stretch of an
 original orbit. This means that they have to belong to the same
 original pseudo-orbit, and that the direction of motion (left to
 right or right to left) must be the same as well.  Hence their
 indices $J=(\mu,j)$ have to coincide. Similarly each $\tilde{B}$ and
 the immediately following $B$ represent ports that are connected by a
 stretch of a partner orbit, and the corresponding subscripts
 $K=(\nu,k)$ coincide. (When a $\tilde{B}$ represents the last right
 port in its encounter, its subscripts $\nu,k$ have to coincide with
 those of the first $B$.) Thus the arrangement of indices $J$ and $K$
 is the same as in a product of matrices and we obtain a product of
 the same form as in Eq. (\ref{general}) but with $j$ and $k$ replaced
 by  $J,K$.

\begin{figure}[!ht]
\begin{center}
\includegraphics[ scale=0.3]{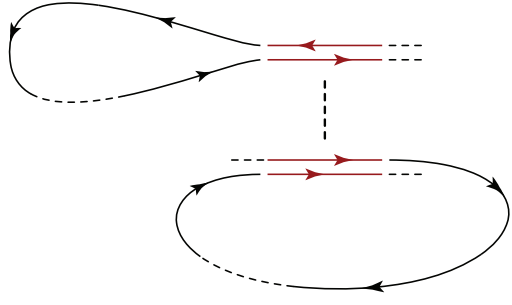}
  \caption{In the presence of time reversal invariance a link can connect
ports on the same side as well as on opposite sides of an
encounter. The linked encounter stretches are traversed in
antiparallel or parallel direction,  correspondingly.}
\label{fig:sidesofenc}
\end{center}
\end{figure}

 We now build in the links and depict them by ``contraction lines''
 above the symbol sequences.
 Since for time-reversal invariant systems links can be drawn between ports
 on either side of the encounters, the contraction lines
 can now connect not only $B$ to
 $\tilde B$ but also $B$ to $B$ and $\tilde B$ to $\tilde B$. Any
 ports connected by a link must belong to the same pseudo-orbit,
 before and after  reconnection and hence their Latin indices must
 coincide, as in the unitary case. The Greek direction indices require
 new reasoning. A contraction between a $\tilde{B}$ and a $B$ stands
 for a link connecting a right port to a left port. In this case, if
 an orbit {\it leaves} one encounter at the right port it must {\it
 enter} the other encounter stretch at the left port. The direction
 of motion (left to right) is thus the same for both encounter
 stretches.  The same applies for a direction of motion from right to
 left, and for original and partner orbits alike. Hence for
 contractions between $\tilde{B}$ and $B$ the Greek subscripts
 indicating the directions of motion must coincide,
 $\mu_1=\mu_2,\nu_1=\nu_2$. On the other hand, in the contractions %
 $\contraction{}{B}{\ldots}{ B} B\ldots B$
 $\contraction{}{\tilde B}{\ldots}{\tilde B}%
 \tilde B\ldots \tilde B$ the connected ports lie on the same side,
 and hence their directions of motion must be opposite. These possibilities
 are demonstrated in Fig.~\ref{fig:sidesofenc}.
  We shall use a
 bar over $\nu,\mu$ to indicate that the port direction of the motion
 is flipped like $ 1\rightleftharpoons 2$ such that $\bar
 \mu=3-\mu$. Again denoting the energy offsets as
$\epsilon_1=\epsilon_A,\,\epsilon_2=\epsilon_C,\,\epsilon_1'
=\epsilon_B,\,\epsilon_2'=\epsilon_D$ we have the link
contribution
\begin{eqnarray}\label{btobtilde} &\bullet&\qquad
[-\I(\epsilon_{j}+\epsilon'_{k})]^{-1}\delta_{kk'}\delta_{jj'}
\delta_{\mu\mu'}\delta_{\nu\nu'} \qquad \mbox{for}
\quad\contraction{}{B}{{}_{\nu k,\mu j}}{\tilde{B}}
B_{\nu k,\mu j}\tilde B_{\mu'j',\nu'k'}\,,\\
\label{btildetobtilde}  &\bullet&\qquad
[-\I(\epsilon_{j}+\epsilon'_{k})]^{-1}\delta_{kk'}\delta_{jj'}
\delta_{\bar\mu\mu'}\delta_{\bar\nu\nu'} \qquad\mbox{for}\quad
\contraction{}{B}{{}_{\nu k,\mu j}}{ B} B_{\nu k,\mu
j}B_{\nu'k',\mu'j'}\quad \,\mbox{and} \quad
\contraction{}{B}{{}_{\mu j,\nu k}}{ B} \tilde B_{\mu j,\nu
k}\tilde B_{\mu' j',\nu'k''} \,.
\end{eqnarray}
The contribution of an encounter  is defined by its first left
port and can be written as  $\I(\epsilon_{j_{\sigma
1}}+\epsilon'_{k_{\sigma 1}})$. And again, there will be a factor
$(-1)^{n_C+n_D}$ depending on the numbers of orbits included in
$C$ and $D$.

\subsubsection{From symbol sequence to Gaussian integral}

 In order to represent additive terms in (\ref{masterorth}) by
contractions in a Gaussian integral we interpret the symbols
$B_{\nu k,\mu j},\tilde B_{\mu'j',\nu'k'}$ as (Bosonic or
Fermionic) variables. Again any two ports which may be connected
by a link must be represented by a pair of variables which are
mutually complex conjugate, up to a sign. Since rule
(\ref{btildetobtilde}) allows for contractions between two $B$'s
differing in both their subscripts $\mu$ and $\nu$ we thus need to
have
\begin{equation}
B_{\nu k,\mu j}=\pm B_{\bar\nu k,\bar\mu j}^*\,.
\end{equation}

To write the matrices $B$ in a more elegant way, it is helpful to
divide $B$ into four $2\times 2$ blocks, each block having fixed
direction indices $\mu,\nu$ and matrix elements indexed by $k,j$.
The diagonally opposite blocks must be mutually complex conjugate
up to a sign (which may even vary with  $j$ and $k$). We may write
\begin{equation}\label{primitiveb}
 B\stackrel{?}=\left(\begin{array}{cc}b_p&b_a^*\\b_a&b_p^*\end{array}\right)\,,
\end{equation}
with the question mark on the equality a reminder of the as yet
unspecified signs. The index $p$ on $b_p,b_p^*$ signals $\mu=\nu$,
i.e., the ports associated with these subblocks are traversed in
{\it
  p}arallel directions by the original and the partner orbits. In
contrast, the ports associated with $b_a,b_a^*$ have $\mu\neq \nu$
and their traversal directions by original and partner orbits are
{\it
  a}ntiparallel.

In addition, rule (\ref{btobtilde}) allows contractions between
matrix elements of $B$ and $\tilde{B}$ with identical subscripts.
Hence these matrix elements, too, must be chosen as mutually
complex conjugate (up to a sign). Since the subscripts in $B$ and
$\tilde{B}$ are ordered differently, this means that $\tilde{B}$
must coincide with the adjoint of $B$ (again up to signs), i.e.,
\begin{equation}\label{uptosign_dagger_vs_tilde}
\tilde{B}\stackrel{?}=B^\dagger\;.
\end{equation}

A Gaussian-weight integral of a product of mutually conjugate
matrix elements will yield a sum over all permissible ways of
drawing contraction lines. To obtain the correct contribution from
each link the prefactor of $B_{\mu k,\nu j}B^*_{\mu k,\nu j}$ in
the exponent of the Gaussian again has to be chosen as
$\I(\epsilon_j+\epsilon_k')$. Furthermore, due to the internal
structure of $B$ the sum over all $B_{\mu k,\nu j}B^*_{\mu k,\nu
j}$ will include each pair of complex conjugate elements twice,
which means that we have to divide the exponent by $2$. Using the
composite indices $J,K$ and defining $\epsilon_J\equiv \epsilon_j$
and $\epsilon_K'\equiv \epsilon_k'$ independent of the Greek part
of $J,K$ we can write the exponent in the Gaussian as
$\sum_{JK}\pm\frac{\I}{2}(\epsilon_J+\epsilon'_K)B_{KJ}B_{KJ}^*$.
Incorporating also the factor
$\I(\epsilon_{J_{\sigma1}}+\epsilon'_{K_{\sigma1}})$ for each
encounter we obtain
\begin{eqnarray}
\label{uptosignorth} \pm \int d[B,\tilde{B}]\e^{\sum_{J,K}\pm
\frac{\I}{2}(\epsilon_J+\epsilon_K')B_{KJ}B_{KJ}^*}\nonumber\\
\times\prod_{\sigma=1}^V\left(\pm
\I(\epsilon_{J_{\sigma1}}+\epsilon'_{K_{\sigma1}})
B_{K_{\sigma1},J_{\sigma1}}\tilde{B}_{J_{\sigma1},K{\sigma2}}
B_{K_{\sigma2},J_{\sigma2}} \ldots \tilde{B}_{J_{\sigma
l(\sigma)},K_{\sigma1}}\right)\;;
\end{eqnarray}
each (convergent) integral of this type includes the correct
encounter and link factors for all structures with $V$ encounters
involving $l(1),l(2)\ldots l(V)$ encounter stretches.

\subsubsection{Signs}

In order to incorporate the sign factor $(-1)^{n_C+n_D}$ and the
correct prefactor of the diagonal approximation, we need to fully
specify the matrices $B$ and $\tilde{B}$: The off-diagonal entries
in all four blocks of $B$ are chosen Fermionic and the diagonal
entries Bosonic, and the signs in (\ref{uptosign_dagger_vs_tilde})
are picked as
\begin{eqnarray}\label{Borth}
B=\left(\begin{array}{cc}b_p&-b_a^*\\b_a&-\sigma_zb_p^*\end{array}\right),\qquad
\tilde{B}=\sigma_z B^\dagger\,;
\end{eqnarray}
 here the multiplication with $\sigma_z$ in $\tilde B$ means that each
$2\times 2$ block of $B^\dagger$ is multiplied with $\sigma_z$;
taking the adjoint of our $4\times 4$ supermatrices involves,
apart from complex conjugation and transposition in the sense of
ordinary matrices, a sign flip in the lower left (Fermi-Bose)
entries of all $2\times 2$ blocks.

The signs in the Gaussian weight and the matrix product are again
chosen such that both quantities can be written in terms of
supertraces.  The supertrace is now defined as the sum of the two
upper left (Bose-Bose) entries minus the sum of the lower right
(Fermi-Fermi) entries of the two diagonal blocks; in other words,
the supertrace includes a negative sign for all diagonal elements
associated to a Latin index 2 (regardless of the Greek index). The
integral (\ref{uptosignorth}) then acquires the form
\begin{eqnarray}
\label{orthoaver} \int
d[B,\tilde{B}]\exp\left(\frac{\I}{2}\str\hat\epsilon'B\tilde{B}
+\frac{\I}{2}\str\hat\epsilon\tilde{B}B\right)(\ldots)\equiv\langle\!
\langle\ldots\rangle\!\rangle\,,
\end{eqnarray}
with $\hat\epsilon,\hat{\epsilon'}$  4$\times$4 diagonal matrices
whose diagonal elements are $\hat\epsilon_{J}=\epsilon_{j},\;
\hat\epsilon'_{K}=\epsilon_{k}'$. After eliminating all
contraction lines from any integral of this type we are left with
the Gaussian integral
\begin{eqnarray}
\langle \!\langle 1\rangle\!\rangle=\int d[B,\tilde{B}]\exp\left(
\frac{\I}{2}\str\hat\epsilon'B\tilde{B}+\frac{\I}{2}\str\hat\epsilon\tilde{B}B\right)
\nonumber\\=\frac{(\epsilon_1+\epsilon'_2)^2(\epsilon_2+\epsilon'_1)^2}
{(\epsilon_1+\epsilon'_1)^2(\epsilon_2+\epsilon'_1)^2}
=\frac{(\epsilon_A+\epsilon_D)^2(\epsilon_C+\epsilon_B)^2}
{(\epsilon_A+\epsilon_B)^2(\epsilon_C+\epsilon_D)^2}\;.
\end{eqnarray}
as in the diagonal factor for time-reversal invariant systems.
Therefore, upon adding to (\ref{orthoaver}) the Weyl factor and
the factor $1/(2^VV!)$ from (\ref{masterorth}) we obtain the
contribution to the generating function (incorporating the
diagonal term) as
\begin{eqnarray}
\label{projorth} Z^{(1)}(V,\{l(\sigma),J_{\sigma i}, K_{\sigma
i}\})
=\frac{\e^{\I(\epsilon_A+\epsilon_B-\epsilon_C-\epsilon_D)/2}}{V!2^V}\nonumber\\
\times\left\langle \!\!\!\left\langle\prod_{\sigma=1}^V
\I(\epsilon_{J_{\sigma1}}+\epsilon'_{K_{\sigma1}}) {\rm Str}\;
P_{K_{\sigma1}}B P_{J_{\sigma1}}\tilde{B}P_{K_{\sigma2}} \ldots
P_{J_{\sigma
l(\sigma)}}\tilde{B}\right\rangle\!\!\!\right\rangle\;,
\end{eqnarray} \\
where $P_J$ ($P_K$) is a  projection matrix whose diagonal element
associated to $J$ ($K$) equals 1 while all other elements vanish.

 We still have to show  that with the choices made above
the sign factor $(-1)^{n_C+n_D}$ for each structure is obtained
correctly. The contraction rules (\ref{rule1}) and (\ref{rule2})
 we used to prove the corresponding result in the unitary
case carry over. In addition it is now possible to draw
contraction lines between two $B$'s or two $\tilde{B}$'s. These
lines represent links connecting two left ports or two right
ports.   The corresponding contraction rules are established in
Appendix \ref{sec:contraction_rules}; they read
\begin{eqnarray}
 \label{rule3}
 \left\langle\!\!\!\left\langle%
\contraction{\str (P_K}{B}{P_JY) \str (}{B}%
\str (P_K B P_JY) \str (P_{K'}BP_{J'}Z )%
\ldots\right\rangle\!\!\!\right\rangle
=-\frac{1}{\I(\epsilon_J+\epsilon'_K)}\delta_{\bar{K}K'}\delta_{\bar{J}J'}
\left\langle\!\!\left\langle\str( P_J Y P_K
\tilde{Z})\ldots\right\rangle\!\!\right\rangle
\\
 \label{rule4} \left\langle\!\!\!\left\langle%
\contraction{ \str\,(P_k}{B}{P_J YP_{K'}}{B}%
 \str\,(P_K B P_J YP_{K'} BP_{J'}Z)%
  \ldots\right\rangle\!\!\!\right\rangle
=-\frac{1}{\I(\epsilon_J+\epsilon'_K)}\delta_{\bar{K}K'}\delta_{\bar{J}J'}
\left\langle\!\!\left\langle\str(P_K \tilde{Y} P_{\bar{J}}
Z)\ldots\right\rangle\!\!\right\rangle;
\end{eqnarray}
analogously for $\tilde{B}$. Here $Y$ and $Z$ are matrix products
of the type required in (\ref{projorth}) starting and ending with
$\tilde{B}$.   The sequences $\tilde{Y}$, $\tilde{Z}$ are
 obtained from $Y$, $Z$ by  (i) reverting the order of
matrices; (ii) replacing all $B$'s  by $\tilde{B}$'s and vice
versa; (iii) replacing all projectors $P_K$ by $P_{\bar K}$ and
$P_{J}$ by $P_{\bar J}$. Eqs. (\ref{rule3}) and (\ref{rule4})
again contain the familiar factor for each link. We just chose the
more compact notation
\begin{equation}\label{barredJ}
J=(\mu,j)\,,\quad \bar J\equiv(\bar\mu,j)\,,\qquad
K=(\nu,k)\,,\quad \bar K\equiv(\bar\nu,k)
\end{equation}
The Kronecker deltas in the contraction rules
(\ref{rule3},\ref{rule4}) imply that the connected ports must
belong to the same pseudo-orbits and have opposite direction of
motion.

We can now use the contraction rules (\ref{rule1}), (\ref{rule2}),
(\ref{rule3}) and (\ref{rule4}) and their counterparts for
contractions of two $\tilde{B}$'s to step by step remove the
contraction lines corresponding to all links of a structure. In
doing so we do not meet any sign factors apart from the steps
where the final link belonging to an orbit is removed. This final
link must always connect a left to a right port, since otherwise
it would not be possible to close it to an orbit just by drawing
an encounter stretch. Hence in this step we only have to use the
rules (\ref{rule1}) and (\ref{rule2}) for removing links between
left ports (denoted by $B$) and right ports (denoted by
$\tilde{B}$). By the same argument as in Section
\ref{sec:build_sigma} we can show that the final step yields a
factor $\str(P_{J})$ for each orbit  in $A\cup C$ and a factor
$\str(P_{K})$ for each orbit in $B\cup D$.  According to our
definition of the supertrace these factors are equal to $-1$ if
the corresponding small indices are $j=2$ (corresponding to the
pseudo-orbit $C$) or $k=2$ (corresponding to the pseudo-orbit
$D$). We thus obtain a negative sign factor for every orbit
included in $C$ or $D$, and these factors combine to the desired
term $(-1)^{n_C+n_D}$.

\subsubsection{Towards a sigma model}

Summing over structures as in Section \ref{sec:build_sigma} and
again using the assumed largeness of the imaginary part $\eta$ of
all energy offsets, see (\ref{large_eta_prec}), we get the
generating function
\begin{eqnarray}
  Z^{(1)}
\sim\e^{\I\,(\epsilon_A+\epsilon_B-\epsilon_C-\epsilon_D)/2} \int
d[B,\tilde{B}]\exp\left[
\frac{\I}{2}\str\hat{\epsilon}\left(\sum_{l=1}^\infty(\tilde{B}B)^l\right)
+\frac{\I}{2}\str\hat{\epsilon}'\left(\sum_{l=1}^\infty
(B\tilde{B})^l\right)\right]\nonumber\\
= \int d[B,\tilde{B}]\exp\left[\frac{\I}{4}\;\str\! \left(
\hat{\epsilon}\,\frac{1+\tilde{B}B}{1-\tilde{B}B}
+\hat{\epsilon}'\,\frac{1+B\tilde{B}}{1-B\tilde{B}}
\right)\right]\,,\quad \eta\gg 1\,,\label{SUSYZ_orth}
\end{eqnarray}
equal in appearance (up to the factor $\frac{1}{2}$) to
(\ref{SUSYZ}) for the unitary symmetry class. The matrix integral
again sums up the contributions from all structures which form an
asymptotic expansion in inverse powers of energy offsets. The
integration domain in (\ref{SUSYZ_orth}) must include the vicinity
of the saddle point $B=\tilde B=0$, and be such that the
eigenvalues of $1-B\tilde B,\quad 1-\tilde B B$ be positive,
otherwise it is arbitrary.

To see that this asymptotic series agrees with the Gaussian
Orthogonal Ensemble (GOE) of RMT, we can argue like in the unitary
case: The sigma model for the GOE leads to a generating function
as (\ref{SUSYZ_orth}) but with a specified integration domain and
without the restriction $\eta\gg 1$. The large-$\epsilon$
asymptotics of the integral is determined by two saddle points of
the exponent  both of which have to be taken into account unless
$\eta\gg 1$. The so called standard saddle at $B=0$
 leads to a  contribution
dominated by small $B$ and thus coinciding with
(\ref{SUSYZ_orth}). The contribution of the Andreev-Altshuler
saddle  is related to $Z^{(1)}$ by
$Z^{(2)}(\epsilon_A,\epsilon_B,\epsilon_C,\epsilon_D)
=Z^{(1)}(\epsilon_A,\epsilon_B,-\epsilon_D,-\epsilon_C)$, again as
in semiclassics. Our semiclassical results thus recover the
asymptotics of the generating function in agreement with the sigma
model.

The same must then be true for $C(\epsilon)$. The two-point
correlation function of fully chaotic time-reversal invariant
systems is thus asymptotically given by the sum of two divergent
series in powers of $\epsilon^{-1}$ in Eq. (\ref{RMTseries}).

Now we proceed to the full $C(\epsilon)$. As is well known, an
analytic function can, under rather general conditions, be
uniquely restored from its asymptotic series by Borel summation
\cite{Sokal}. That method involves term-by-term Fourier transform
of the asymptotic expansion leading to a converging series,
followed by the inverse Fourier transform of the resulting
analytic function. The first stage of the Borel summation applied
to the two components of the correlation function gives the
spectral form factor $K(\tau)$ both for $\tau<1$ and $\tau>1$. The
inverse Fourier transform recovers the closed RMT expression for
the correlation function in Eq. (\ref{R}), for all energies
$\epsilon$, thus completing the semiclassical explanation of
universality.

\section{Conclusions}

We have shown that chaotic systems without any symmetries (beyond
time reversal invariance) display universal spectral statistics.
To obtain both the non-oscillatory and oscillatory contributions
to the spectral correlator $R(\epsilon)$ we had to (i) represent
the correlator through derivatives of a generating function $Z$,
(ii) semiclassically expand $Z$ as a sum over quadruplets of
pseudo-orbits (sets of classical periodic orbits), and (iii)
uncover a Weyl symmetry of $Z$. The non-oscillatory and the
oscillatory part of the correlation function are due to two
summands of $Z$ related by the Weyl symmetry. The oscillatory one
could only be recovered in a semiclassical approximation that
explicitly preserves unitarity; in this approximation the
oscillations are connected to a phase factor which accounts for
the average level density.

Constructively interfering contributions to the sum over
pseudo-orbit quadruplets originate from orbits that are either
identical or differ in close encounters in phase space. They
define asymptotic $\frac{1}{\epsilon}$-series. For dynamics
without time reversal invariance those series terminate after the
first term, both for the non-oscillatory and oscillatory part. For
time-reversal invariant systems, the series are infinite ones but
can be summed to give $R(\epsilon)$ in closed form.

 To implement the summation we write the semiclassical series for the
generating function $Z$ as a matrix integral. The sum over all
possible topologies then turns into an integral well known from
the zero dimensional sigma model of random-matrix theory.

For both the unitary and the orthogonal symmetry class, the
correlator $R(\epsilon)$ comes out as the closed-form expression
familiar from random-matrix theory. In contrast to the primary
$\frac{1}{\epsilon}$-series which makes sense only for
$|\epsilon|\gg1$, the final result holds for any real value of the
energy offset and yields, for $\epsilon\ll1$, the familiar power
law $R(\epsilon)\propto |\epsilon|^\beta$, with the exponent
$\beta$ revealing the degree of level repulsion characterizing
each symmetry class.  Our results thus imply the semiclassical
explanation of universal level repulsion. Likewise, all other
characteristics of universal spectral fluctuations based on the
two-point correlator $R(\epsilon)$, like the so called spectral
rigidity \cite{Mehta,Stoeckmann}, are now semiclassically
understood.

The techniques here elaborated (in particular the analogy between
orbit-based semiclassics and the sigma model) will be, and in fact
are already helpful in tackling further challenges in the field.
Examples are (i) localization effects in long thin wires
\cite{PBAAA}, (ii) non-universal effects related to the Ehrenfest
time (which has no counterpart in RMT)  \cite{Ehrenfest}, (iii) a
semiclassical treatment of arithmetic billiards \cite{Arithmetic}
(where intrinsically quantum Hecke symmetries have strong action
degeneracies as classical counterparts), (iv) spectral
correlations of higher order \cite{NagaoMueller} and the density
of nearest-neighbor spacings, (v) transport through ballistic
chaotic conductors \cite{RichterSieber,Transport} including the
``full counting statistics''  \cite{Novaes}, (vi) further symmetry
classes, (vii) transitions between symmetry classes
\cite{Japanese}, and (viii) last but not least, the relative
status of the ballistic sigma model and periodic-orbit theory
\cite{KoelleAlaaf}.

\begin{acknowledgments} We gratefully acknowledge helpful discussions with Jon
Keating, Taro Nagao, Keiji Saito, Martin Sieber, Ben Simons,
Hans-J\"urgen Sommers, and Martin Zirnbauer, as well as funding
from the Sonderforschungsbereich TR 12 of the Deutsche
Forschungsgemeinschaft.
\end{acknowledgments}


\appendix

\section{Contraction rules}
\label{sec:contraction_rules}

In this Appendix we want to derive the contraction rules
(\ref{rule1}), (\ref{rule2}), (\ref{rule3}), and (\ref{rule4})
 from Eqs. (\ref{simplecontr}) and (\ref{fermicontr}).
We start with {\bf systems without time-reversal invariance}.
First of all, contractions will be nonzero if the contracted
matrix elements are mutually complex conjugate. Using the
definition of the exponent according to Eq. (\ref{signgauss}) we
obtain for contractions between $B_{kj}$ and $\tilde{B}_{j'k'}$
(the latter agreeing up to the sign with $B_{k'j'}^*$),
\begin{eqnarray}
\label{medium}
 \left\langle\!\!\left\langle
\contraction {} {\!\!\!B_{kj}} {} {\!\!\!\tilde{B}_{j'k'}} {}
{B_{kj}} {} {\tilde{B}_{j'k'}}
g(B,\tilde{B})\right\rangle\!\!\right\rangle=-\frac{s_j}{\I(\epsilon_j
+\epsilon'_k)}\delta_{jj'}\delta_{kk'}\left\langle\!\!\left\langle
g(B,\tilde{B})\right\rangle\!\!\right\rangle\;.
\end{eqnarray}
Here  $s_j$ is a  sign factor  equal to 1 if $j=1$ and  to $-1$ if
$j=2$. To understand its origin one has to take into account the
contraction rule (\ref{simplecontr}) for the Bosonic variables
($j=k$) yielding $-\frac{1}{\I\epsilon}$  and (\ref{fermicontr})
for the Fermionic variables ($j\ne k$) yielding
$\frac{1}{\I\epsilon}$. An additional factor $-1$ emerges when
$j=k=2$  due to the negative sign in $\tilde{B}_{22}=-B_{22}^*$
and when $j=1,k=2$ because of minus at the term proportional to
$B_{21}\tilde{B}_{12}$ in the exponent of (\ref{signgauss}).

Using Eq. (\ref{medium}) we obtain the rule (\ref{rule1}) for
contractions between $B$ and $\tilde{B}$ in different supertraces
(inter-encounter contractions) which comes about as
\begin{eqnarray}
\left\langle\!\!\!\left\langle \contraction {\str(P_k} {B}
{P_jY)\str(P_{k'}XP_{j'}} {\tilde{B}} {\str(P_k} {B}
{P_jY)\str(P_{k'}XP_{j'}} {\tilde{B}}
)\ldots\right\rangle\!\!\!\right\rangle &=&
\left\langle\!\!\!\left\langle \contraction {\str(P_jYP_k} {B}
{)\str(P_{j'}} {\tilde{B}} {\str(P_jYP_k} {B} {)\str(P_{j'}}
{\tilde{B}} P_{k'}X)\ldots\right\rangle\!\!\!\right\rangle
\nonumber\\
&=&\left\langle\!\!\!\left\langle \contraction {s_jY_{jk}}
{\!\!\!B_{kj}} {s_{j'}} {\!\!\!\tilde{B}_{j'k'}} {s_jY_{jk}}
{B_{kj}} {s_{j'}} {\tilde{B}_{j'k'}}
X_{k'j'}\ldots\right\rangle\!\!\!\right\rangle
\nonumber\\
&=&
-\frac{s_j}{\I(\epsilon_j+\epsilon'_k)}\delta_{jj'}\delta_{kk'}\left\langle\!\left\langle
s_jY_{jk}s_j X_{kj}\ldots\right\rangle\!\right\rangle
\nonumber\\
&=&
-\frac{1}{\I(\epsilon_j+\epsilon'_k)}\delta_{jj'}\delta_{kk'}\left\langle\!\left\langle
s_jY_{jk}X_{kj}\ldots\right\rangle\!\right\rangle
\nonumber\\%
&=&
-\frac{1}{\I(\epsilon_j+\epsilon'_k)}\delta_{jj'}\delta_{kk'}\left\langle\!\left\langle
\str (P_j Y P_k X)\ldots\right\rangle\!\right\rangle\;.
\end{eqnarray}
Here we used the cyclic invariance of the  supertrace, then wrote
out the supertrace in components, with the help of $\str(P_k
Z)=s_kZ_{kk}$ for any supermatrix $Z$, and finally used Eq.
(\ref{medium}).

Similarly Eq. (\ref{rule2}) for intra-encounter contractions
between  $B$ and $\tilde{B}$  follows from
\begin{eqnarray}
\left\langle\!\!\!\left\langle \contraction {\str(P_k} {B}
{P_jUP_{j'}} {\tilde{B}} {\str(P_k} {B} {P_jUP_{j'}} {\tilde{B}}
P_{k'}V)\ldots\right\rangle\!\!\!\right\rangle &=&
\left\langle\!\!\!\left\langle \contraction {s_k} {\!\!\!B_{kj}}
{U_{jj'}} {\!\!\!\tilde{B}_{j'k'}} {s_k} {B_{kj}} {U_{jj'}}
{\tilde{B}_{j'k'}} {V_{k'k}}
\ldots\right\rangle\!\!\!\right\rangle
\nonumber\\
&=&\left\langle\! \!\!\left\langle \contraction {s_k}
{\!\!\!B_{kj}} {} {\!\!\!B_{j'k'}} {s_k} {B_{kj}} {}
{\tilde{B}_{j'k'}}
U_{jj'}V_{k'k}\ldots\right\rangle\!\!\!\right\rangle
\nonumber\\
&=&-\frac{s_j}{\I(\epsilon_j+\epsilon'_{k})}\delta_{jj'}\delta_{kk'}\left\langle\!\left\langle
s_k U_{jj}V_{kk}\ldots\right\rangle\!\right\rangle
\nonumber\\
&=&-\frac{1}{\I(\epsilon_j+\epsilon'_{k})}\delta_{jj'}\delta_{kk'}\left\langle\!\left\langle\str(P_j
U)\str(P_k V)\ldots\right\rangle\!\right\rangle\,;
\end{eqnarray}
here in the second line we could interchange $U_{jj'}$ and
$\tilde{B}_{j'k'}$, since a nonzero result arises only for $j=j'$
in which case $U_{jj}$ is Bosonic and thus commutes with all other
variables.

We now turn to {\bf time-reversal invariant systems}. For these
systems the derivation of rules (\ref{rule1}) and (\ref{rule2})
carries over directly, the only difference being that $J,K$ are
now double indices consisting of the Greek and Latin parts,
$J=(\mu,j)$, $K=(\nu,k)$; the sign factor $s_J$ has the values $1$
for $j=1$ and $-1$ for $j=2$ regardless of the Greek index, and
$s_K$ is defined analogously. In addition the fact that the matrix
elements of $B$ and $\tilde{B}$ are pairwise mutually conjugate
allows to draw contraction lines between two $B$'s or two
$\tilde{B}$'s. We thus obtain further contraction rules
(\ref{rule3}) and (\ref{rule4}) for contractions between two
$B$'s, and analogous rules for contractions between two
$\tilde{B}$'s. To derive these contraction rules we use the
important symmetry of $B$ as defined in (\ref{Borth}),
\begin{equation}
\label{conjugB} \Sigma B^t \Sigma^t=-\tilde{B}\,\qquad \Sigma
\tilde{B}^t \Sigma^t=-B
\end{equation}
where $B^t$ is the transpose of $B$, and we have introduced
$\Sigma=\left({0\atop 1}{\sigma_z\atop 0}\right)$ and
$\Sigma^t=\Sigma^{-1}=\left({0\atop \sigma_z}{1\atop 0}\right)$.
(Readers who want to check (\ref{conjugB}) are reminded that
transposition of a supermatrix involves interchanging the indices of
the matrix elements and afterwards flipping the sign of the lower left
elements in each $2\times 2$ block.)
Similarly conjugating the
projection matrices $P_K, P_J$ we get
\begin{equation}
\label{conjugP} \Sigma P_K\Sigma^t=P_{\bar{K}}\,,\qquad \Sigma P_J
\Sigma^t=P_{\bar{J}}\;;
\end{equation}
 Now let us consider the
matrix products $Y$ and $Z$ on the l.h.s. of
(\ref{rule3}) and (\ref{rule4}), involving
alternating sequences of $B$'s and $\tilde{B}$'s and interspersed
projection matrices.
Under
transposition and conjugation with $\Sigma$
these products have
(i) the order of matrices inverted (due to the transposition), (ii) $B$
replaced by $\tilde{B}$ and vice versa (due to (\ref{conjugB})), and (iii) the subscripts $J$ and
$K$ of all projection matrices replaced by $\bar J$
and $\bar K$ (due to (\ref{conjugP})). Since the numbers of $B$'s and $\tilde{B}$'s are equal the sign factors
from (\ref{conjugB}) all mutually compensate.

Thus equipped we can prove rule (\ref{rule3}) for inter-encounter
contractions between two $B$'s. By using the invariance of the
supertrace  both under transposition of its argument and under
conjugation with $\Sigma$ we write
\begin{eqnarray}
\left\langle\!\!\!\left\langle\str(P_K \contraction {} {B}
{P_JY)\str(P_{K'}} {B} {} {B} {P_JY)\str(P_{K'}} {B}
P_{J'}Z)\ldots\right\rangle\!\!\!\right\rangle &=&
\left\langle\!\!\!\left\langle\str(P_K \contraction {} {B}
{P_JY)\str(\Sigma  Y^t P_J} {B^t} {} {B} {P_JY)\str(\Sigma  Z^t
P_{J'}} {B^t}
 P_{K'}\Sigma^t)\ldots\right\rangle\!\!\!\right\rangle\nonumber\\
&=&\left\langle\!\!\! \left\langle\str(P_K \contraction {} {B}
{P_JY)\str(\tilde{Z}P_{\bar{J'}}} {\tilde{B}} {} {B}
{P_JY)\str(\tilde{Z}P_{\bar{J'}}} {\tilde{B}}
P_{\bar{K'}})\ldots\right\rangle\!\!\!\right\rangle\nonumber\\
&=&-\frac{1}{\I(\epsilon_J+\epsilon'_K)}\delta_{\bar J,J'
}\delta_{\bar{K},K' }\left\langle\!\!\left\langle\str(P_J Y P_K
\tilde{Z})\ldots\right\rangle\!\!\right\rangle\;.
\end{eqnarray}
Here in the third line $\tilde{Z}$ denotes the product obtained
from $Z$ by steps (i)-(iii) above.  We have thus proven Eq.
(\ref{rule3}).

We now move on to rule (\ref{rule4}) for intra-encounter contractions  between
two $B$'s, as in
\begin{equation}
\label{case4} \left\langle\!\!\!\left\langle\str(P_K \contraction {}
{B} {P_J Y P_{K'}} {B} {} {B} {P_J Y P_{K'}} {B}
P_{J'}Z)\ldots\right\rangle\!\!\!\right\rangle\;.
\end{equation}
First of all note that the contracted matrix elements are $B_{KJ}$
and $B_{K'J'}$, the latter coinciding up to a sign with
$B_{\bar{K},\bar{J}}^*$ (cf. Eq. (\ref{Borth})). Hence the only
relevant case is that  $J'=\bar{J},\,K'=\bar{K}$. For this case we
now write
\begin{eqnarray}
\label{case4again} \left\langle\!\!\!\left\langle\str(P_K
\contraction {} {B} {P_J Y P_{K}} {B} {} {B} {P_J Y P_{\bar K}}
{B} P_{\bar J}Z)\ldots\right\rangle\!\!\!\right\rangle =
\left\langle\!\!\!\left\langle s_K
 \contraction {}{B} {{}_{KJ}YP_{K'}} {B}
 {B_{KJ}} {(YP_{\bar K}}
{B)_{J\bar{J}}} Z_{\bar{J}K}\ldots\right\rangle\!\!\!\right\rangle
\end{eqnarray}
and evaluate the bracket in (\ref{case4again}) as
\begin{eqnarray}
\label{simplify}
(YP_{\bar{K}}B)_{J\bar{J}}&=&\left[\Sigma^t(\Sigma
B^t\Sigma^t)(\Sigma P_{\bar{K}}\Sigma^t)(\Sigma
Y^t\Sigma^t)\Sigma\right]^t_{J\bar{J}}\nonumber\\
&=&\left(\Sigma^t\tilde{B}P_{K}\tilde{Y}\Sigma\right)_{\bar{J}J}\nonumber\\
&=&\Sigma^t_{\bar{J},J}\tilde{B}_{J,K}
\tilde{Y}_{K,\bar{J}}\Sigma_{\bar{J},J}\nonumber\\
&=&s_J\tilde{B}_{J,K}\tilde{Y}_{K,\bar{J}}\;.
\end{eqnarray}
Here we first replaced the matrix sequence by itself double
transposed;  note that double transposition of a supermatrix (not
an identity operation!) leaves its Bosonic elements like the one
with the subscripts $J\bar J$ unchanged. Then we invoked Eqs.
(\ref{conjugB},\ref{conjugP}) and noted that for the matrix
elements at hand transposition just amounts to interchanging the
two subscripts.  In the third line we used that only those
elements of $\Sigma,\Sigma^t$ are nonzero for which one subscript
is equal to another one barred, i.e.,
 the port directions of motion are opposite. Finally, we  used
 $\Sigma^t_{\bar J,J}\Sigma_{\bar J,J}=s_J$
Inserting (\ref{simplify}) into (\ref{case4again}) we now obtain
\begin{eqnarray}
\left\langle\!\!\!\left\langle\str(P_K \contraction {} {B} {P_J Y
P_{K'}} {B} {} {B} {P_J Y P_{K'}} {B}
P_{J'}Z)\ldots\right\rangle\!\!\!\right\rangle &=&\delta_{\bar J,J'
}\delta_{\bar{K},K' }
 \left\langle\!\!\!\left\langle s_K
\contraction {} {B} {{}_{KJ}s_J} {\tilde{B}} {} {B_{KJ}} {s_J}
{\tilde{B}_{J,K}}
\tilde{Y}_{K,\bar{J}}Z_{\bar J,K}\ldots\right\rangle\!\!\!\right\rangle\nonumber\\
&=& -\frac{1}{\I(\epsilon_J+\epsilon'_K)}\delta_{\bar J,J'
}\delta_{\bar{K},K' }\left\langle\!\!\left\langle
s_K\tilde{Y}_{K,\bar{J}}Z_{\bar{J},K}\ldots\right\rangle\!\!\right\rangle
\nonumber\\
&=& -\frac{1}{\I(\epsilon_J+\epsilon'_K)}\delta_{\bar J,J'
}\delta_{\bar{K},K' }\left\langle\!\!\left\langle\str(
P_K\tilde{Y}P_{\bar{J}}Z)\ldots\right\rangle\!\!\right\rangle
\end{eqnarray}
which is (\ref{rule4}). Here we used (\ref{medium}),
defined $\tilde{Y}$ in analogy to $\tilde{Z}$, and in the
end rearranged the result as a supertrace.

One easily shows that rules analogous to (\ref{rule3}) and
(\ref{rule4}) also hold for contractions between two $\tilde{B}$'s.


\section{Supersymmetric sigma  model}\label{susy_sigma}

\label{sec:supersymmetric_sigma}

We propose to briefly review the basics of the sigma model;
details can be found in
\cite{Efetov,Haake,YurkevichLerner,KamenevMezard,MuellerPhD,
  TauSmallLong}. The sigma model is a way to implement averages over
random matrices. The appropriate random-matrix ensemble for
systems without time-reversal invariance is the Gaussian Unitary
Ensemble (GUE), consisting of complex Hermitian $N\times N$
matrices $H$ with a Gaussian weight proportional to
$\e^{-\frac{N}{2}\tr H^2}$; the matrix dimension $N$ is
sent to infinity. The pertinent generating function $Z$ looks like
(\ref{Zdef}), but with the energy average $\langle\ldots\rangle$
replaced by the GUE average. To perform that average, three
variants of the sigma model are available, all based on
representing determinants by suitable Gaussian integrals. The
supersymmetric variant deals with superintegrals over both
ordinary and Grassmannian variables while two replica variants
exclusively use either Grassmann or ordinary Gaussian integrals.

\subsection{From GUE to supermatrix integral over saddle-point manifold}

In the supersymmetric approach each of the determinants in the
generating function (\ref{Zdef}) is represented by a Gaussian
integral, Fermionic for the numerator and Bosonic for the denominator,
\begin{eqnarray}\label{Zsupint}
Z(\epsilon_A,\epsilon_B,\epsilon_C,\epsilon_D)&=&(-1)^N
\left(\prod_{m=1}^N\int \frac{d^2z_{Am}}{\pi}
\frac{d^2z_{Bm}}{\pi} d\eta^*_{Cm}d\eta_{Cm}d\eta^*_{Dm}d\eta_{Dm}
\right)\\ \nonumber &&\times\,\left\langle\e^{\,{-\I}\sum_{m,n}
\big\{
 -z_{Am}^*(E+\epsilon_{A}-H)_{mn}z_{An}
+z_{Bm}^*(E-\epsilon_{B}-H)_{mn}z_{Bn} \big\}}\right.
\\ \nonumber &&\quad\left.
\times\,\e^{\,-{\I}\sum_{m,n} \big\{
 \eta^*_{Cm}(E+\epsilon_{C}-H)_{mn}\eta_{Cn}
+\eta^*_{Dm}(E-\epsilon_{D}-H)_{mn}\eta_{Dn}
\big\}}\right\rangle\,;
\end{eqnarray}
at the expense of the overall factor $(-1)^N$ the signs of the
``Bosonic'' quadratic forms $z_A^*(\ldots)z_A$ and
$z_B^*(\ldots)z_B$ have been chosen so as to make the Bosonic
Gaussian integrals convergent, in consistency with ${\rm
Im}\,\epsilon_{A,B,C,D}>0$.
$\eta_C,\eta_C^*,\eta_D,\eta_D^*$ are
anticommuting Grassmann variables. As usual, the asterisk on commuting variables means
complex conjugation.  On anticommuting variables the asterisk may,
for the purposes of the present paper, be given the same
interpretation (complemented by $(\eta^*)^*=-\eta$, see
\cite{Haake}).

At this point hooking onto the presentation in Chapt.~10 of
Ref.~\cite{Haake} \footnote{We follow the notation of \cite{Haake}
apart from reordering the elements of $\Phi$ and setting the
parameter $\lambda$ in the random-matrix integrals equal to 1.},
we introduce the supervector
\begin{equation}\label{supvec}
\Phi=\left (\begin{array}{c}\Phi_1\\ \Phi_2\\ \Phi_3\\
\Phi_4\end{array}\right) =\left(\begin{array}{c}z_B\\ \eta_D\\
z_A\\ \eta_C\end{array}\right)
\end{equation}
where each of the four $\Phi_\alpha$'s has $N$ components
$\Phi_{\alpha m}$; the ordering of the $\Phi_\alpha$ in
(\ref{supvec}) is known as the advanced-retarded notation.
 To
accommodate the relative sign of the Bosonic quadratic forms
mentioned and the different energy arguments above we employ the
$(4\times 4)$ matrices,
\begin{eqnarray}
L={\rm diag}(1,1,-1,1)=\left({1\atop 0}{0\atop
-\sigma_z}\right)\,,\qquad
\nonumber\\
\hat{E}=E-\frac{1}{2\pi\bar{\rho}}\, {\rm
diag}\big(\epsilon_B,\epsilon_D,-\epsilon_A,-\epsilon_C
\big)\equiv E-\frac{1}{2\pi\bar{\rho}}\left({\hat{\epsilon}'\atop
0}{0 \atop -\hat{\epsilon}}\right)
\end{eqnarray}
whose block form will be convenient in the future. Our generating function may then be succinctly written as
\begin{equation}
Z=(-1)^N\int d\Phi^*d\Phi\, \left\langle \exp\Big\{\I\Phi^\dagger
L\big[H-\hat{E}\big]\Phi\Big\} \right\rangle\,,
\end{equation}
with $d\Phi^*d\Phi$ denoting the integration measure written
explicitly in (\ref{Zsupint}). The Gaussian average over
Hamiltonian matrices $H$
 with a normalized Gaussian weight $\propto e^{-\frac{N}{2}{\rm tr}H^2}$
leads to $Z=(-1)^N\exp\big(
-\frac{1}{2N}\sum_{\alpha,\beta}^{1\ldots 4}L_\alpha
L_\beta \sum_{m,n}^{1\ldots N} \Phi^*_{\alpha m}\Phi_{\alpha
n}\Phi^*_{\beta n}\Phi_{\beta m} \big)$. The Hermitian $4\times4$
supermatrix
\begin{equation}
\tilde{Q}_{\alpha \beta}= \frac{1}{N}\sum_{m=1}^N\Phi_{\alpha
m}\Phi^*_{\beta m}
\end{equation}
allows to further compact $Z$ as
\begin{equation}\label{supintZ}
Z=(-1)^N\int d\Phi^* d\Phi\, \exp\Big\{N\,{\rm Str}
\big[-\I\hat{E}\tilde{Q}L-\frac{1}{2}(\tilde{Q}L)^2\big]\Big\}
\end{equation}
where the symbol Str denotes the supertrace. The integrand would
be invariant under the transformation $\tilde{Q}\to
T\tilde{Q}T^\dagger$ with pseudounitary matrices $T$ defined by
\begin{equation}\label{pseudounitarity}
 T^\dagger L T=L\,,
\end{equation}
if the (real part of) the energy matrix were proportional to the
unit matrix. Due to
\begin{equation}
\bar{\rho}(E=0)=N/\pi
\end{equation}
the energy offsets $\epsilon_{A,B,C,D}/2\pi\bar{\rho}$
are of only order $1/N$, which means that in any case this symmetry
persists to leading order in $1/N$.

The exponent in (\ref{supintZ}) is quartic in the integration
variables $\Phi,\Phi^*$. Upon employing a suitable
Hubbard-Stratonovich transformation \cite{Haake} we render the
$\Phi$-integral Gaussian and end up with an integral over $4\times
4$ supermatrices $Q$,
\begin{eqnarray}
Z=-\int dQ\,\exp{\left\{-\frac{N}{2}{\rm Str}\left[
Q^2+2\ln(\hat{E}-Q)\right] \right\}}\,,
\label{susyZ1}
\end{eqnarray}
with a flat integration measure $dQ$.  For a details of the
transformation, in particular the overall minus sign and the
integration range in (\ref{susyZ1}), we refer the reader to
Ref.~\cite{Haake}. Important for what follows is that the integration range is restricted to
matrices $Q$ diagonalizable by pseudounitary transformations
$T$.

The large-$N$ limit allows for a saddle-point approximation. The
saddle manifold is determined by
\begin{equation}
Q=\frac{1}{\hat{E}-Q}\,.
\end{equation}
We may confine the discussion of this equation to  $E=0$, which
means that we evaluate $Z$ in the middle of Wigner's semicircle.
As seen
 the energy offsets
$\hat{E}-E$ are of order $1/N$ and thus irrelevant for saddles, we may even drop $\hat{E}$ entirely.
The saddle-point equation then simplifies
to $Q^2=-1$.

Next, we expand the exponent in Eq. (\ref{susyZ1}) about the
saddle manifold $Q^2=-1$, to first order in the energy offsets and
still sticking to the center of the Wigner's semicircle, $E=0$,
such that
$ \hat{E}=-\frac{1}{2\pi\bar{\rho}}\left({\hat{\epsilon}'\atop
0}{0 \atop -\hat{\epsilon}}\right). $
   Since the
supertrace of the unit matrix vanishes we get ${\rm
  Str}\big[{Q^2}+2\ln(\hat{E}-Q)\big]
  ={\rm Str}[\ln(\hat{E}^2-2\hat{E}Q+Q^2)]
  \approx{\rm Str}[\ln(-1)+\ln(1+2\hat{E}Q)]
  \approx{\rm Str} 2\hat{E}Q$, where we again neglected
corrections of higher order in $1/N$. Going to the limit
$N\to\infty$ we get the integral
\begin{eqnarray}\label{intsaddlemf}
Z&=& -\int_{Q^2=-1}dQ\, \e^{\,\frac{1}{2}{\rm Str}\,Q \,
\left({\hat{\epsilon}'\atop 0}{0 \atop -\hat{\epsilon}}\right)}
\,.
\end{eqnarray}

The eigenvalues of the matrices $Q$ must be equal to $\pm \I$.
This means that the saddle-point manifold can be divided into 5
conjugacy classes differing by the numbers $n_+,n_-$ of the
eigenvalues $+\I$ and $-\I$. In fact, only one of them is of
relevance \cite{Haake}, with $n_+=n_-=2$. Since the matrices $Q$
must be diagonalizable by pseudo-unitary transformations the
relevant $Q$'s can thus be represented as
\begin{equation}\label{conjugclass}
Q=\I\, T\Lambda^{(1)} T^{-1}
\end{equation}
with the diagonal matrix
 \begin{equation}\Lambda^{(1)}={\rm
diag}(1,1,-1,-1).
\end{equation}

\subsection{Rational parametrization}

There is an ambiguity in the
choice of the matrices $T=\left({T_{aa}\atop T_{ra}}{T_{ar}\atop
    T_{rr}}\right)$. Due to the degeneracy of the eigenvalues of $
\Lambda^{(1)}$, $Q$ remains unchanged if we replace $T$ by the
product $TK$ with $K$ any invertible block diagonal matrix.
Choosing $K=\left({T_{aa}^{-1}\atop 0}{0\atop T_{rr}^{-1}}\right)$
we have
\begin{equation}\label{ratpar}
  TK=1+\left({0 \atop \tilde{B}}{B \atop 0}\right)\,,\qquad
(TK)^{-1}=\left({(1-B\tilde{B})^{-1} \atop
-\tilde{B}(1-B\tilde{B})^{-1}}
    \;\,{-B(1-\tilde{B}B)^{-1} \atop (1-\tilde{B} B)^{-1}}\right)
\end{equation}
with $B=T_{ar}T_{rr}^{-1}$ and $\tilde{B}=T_{ra}T_{aa}^{-1}$. The
pseudounitarity of $T$ entails the matrix $(TK)^\dagger
L(TK)=K^\dagger LK= \left({(T_{aa}^{-1})^\dagger
(T_{aa}^{-1})\atop 0}
  {0\atop -(T_{rr}^{-1})^\dagger\sigma_z (T_{rr}^{-1}} \right)$ to be
block diagonal. On the other hand
\begin{equation}
(TK)^\dagger L(TK)=\left( \begin{array}{cc}
1-\tilde{B}^\dagger\sigma_z \tilde{B}&B-\tilde{B}^\dagger\sigma_z
\\B^\dagger-\sigma_z\tilde{B}&B^\dagger B-\sigma_z
\end{array}\right)\,,
\end{equation}
such that block diagonality yields $\tilde{B}=\sigma_z B^\dagger$.
With the latter relation between $B$ and $\tilde{B}$ established
we can write these $2\times2$ supermatrices as
\begin{equation}\label{superB}
B=\left({B_{11} \atop B_{21}}{B_{12} \atop B_{22}}\right), \qquad
B^\dagger=\left({B_{11}^* \atop -B_{12}^*}{B_{21}^* \atop
B_{22}^*}\right), \qquad \tilde{B}=\sigma_z
B^\dagger=\left({B_{11}^* \atop B_{12}*}{B_{21}^* \atop
-B_{22}^*}\right)\,;
\end{equation}
 note that the diagonal elements
$B_{11},B_{22}$ are Bosonic (commuting) variables whereas the
off-diagonals $B_{12},B_{21}$ are Fermionic, in accordance with
the advanced-retarded representation we are using. From this point
on we rename $TK$ as $T$.

The saddle-point manifold  results as
\begin{equation}
T\Lambda^{(1)} T^{-1}=\left( \begin{array}{cc}
\frac{1+B\tilde{B}}{1-B\tilde{B}}&-2B(1-\tilde{B}B)^{-1}
\\2\tilde{B}(1-B\tilde{B})^{-1}&-\frac{1+\tilde{B}B}{1-\tilde{B}B}
\end{array}\right)\,.
\end{equation}
The final integral over the  manifold becomes an integral over the
$2\times 2$ matrices $B,\tilde{B}$,
\begin{eqnarray}\label{Zsusy}
Z&\sim& -\int d[B,\tilde{B}]\,\exp\left[\frac{\I}{2}\,{\str}\,
T\Lambda^{(1)} T^{-1} \left({\hat{\epsilon}'\atop 0}{0 \atop
-\hat{\epsilon}}\right)\right]
\\ \nonumber
&=&-\int d[B,\tilde{B}]\exp\left[\frac{\I}{2}\,\str\! \left(
\hat{\epsilon}\,\frac{1+\tilde{B}B}{1-\tilde{B}B}
+\hat{\epsilon}'\,\frac{1+B\tilde{B}}{1-B\tilde{B}}
\right)\right]\\
&=& -\,{\rm
e}^{\frac{\I}{2}(\epsilon_A+\epsilon_B-\epsilon_C-\epsilon_D)/2}
\int d[B,\tilde{B}]\exp\left[\I\,\str\! \left(
\hat{\epsilon}\,\frac{\tilde{B}B}{1-\tilde{B}B}
+\hat{\epsilon}'\,\frac{B\tilde{B}}{1-B\tilde{B}} \right)\right]
\,.
\end{eqnarray}
with the flat measure
$d[B,B]=\frac{d^2 B_{11}}{\pi}\frac{d^2 B_{22}}{\pi}d B_{12}* dB_{12} dB_{21}^*dB_{21}$.
The integration ranges for the Bosonic integrals over $B_{11}$ and $B_{22}$
are respectively $|B_{11}|^2<1$ and $|B_{22}|^2<\infty$.  Note that the
foregoing integral representation of $Z^{(1)}$ equals the one
established semiclassically, see (\ref{SUSYZ}).

We now proceed to the asymptotics of the remaining integral
(\ref{Zsusy}) in the limit of large energy offsets,
$|\epsilon_{A,B,C,D}|\gg 1$. To that end we employ one more
saddle-point approximation\footnote{The previous saddle-point
approximation had the
  matrix dimension $N$ as its large parameter and we could take
  $\epsilon_{A,B,C,D}/N \to 0$ as $N\to\infty$. No conflict arises
  with subsequently considering the large-$\epsilon$ asymptotics.}.
The only finite stationary point of the exponent in (\ref{Zsusy})
occurs at $B=\tilde B=0$.  Let us assume, momentarily, that the
energy offsets are provided with a large positive imaginary
part, $\eta\gg 1$. Then this will be the only critical point of
the integrand responsible for the high energy asymptotics of the
integral (\ref{Zsusy}). To get an asymptotic series of $Z$ in
powers of $\epsilon^{-1}$ we expand the matrix
$(1-B\tilde{B})^{-1}$ and $(1-\tilde{B}B)^{-1}$ in the exponent in
powers of $B,\tilde B$ about the unit matrix and further expand
the resulting exponential about its Gaussian part. Scaling the
integration variables as $B=B'/\sqrt{|\epsilon|}$ and similarly
for $\tilde B$ with a typical dimensionless energy offset
$\epsilon$ close to $\epsilon_{A,B,C,D}$ we see that the
higher-order terms in the expansion indeed become small
perturbations in the high-energy limit.  The Gaussian
superintegral giving the leading-order contribution reads
\begin{eqnarray}\label{Z1FinalGauss}
Z^{(1)}&\sim&-\,
{\rm e}^{\frac{\I}{2}(\epsilon_A+\epsilon_B-\epsilon_C-\epsilon_D)/2}\\
&&\times\int \frac{d^2B_{11}}{\pi}\frac{d^2 B_{22}}{\pi}\, {\rm
e}^{\I\{B_{11}^*B_{11}(\epsilon_A+\epsilon_B)
            +B_{22}B_{22}^*(\epsilon_C+\epsilon_D)\}}\nonumber\\
&&\times\int dB_{12}^*B_{12}dB_{21}^* dB_{21}\,
  {\rm e}^{\I\{B_{12}^*B_{12}(\epsilon_B+\epsilon_C)
               +B_{21}^*B_{21}(\epsilon_A+\epsilon_D)\}}\,.\nonumber
\end{eqnarray}
The integration domain for the Bosonic variables can be extended
to the full complex plane; in view of $\eta\gg 1$ the error thus
introduced is exponentially small. Upon doing the remaining
elementary integrals we find
\begin{equation}\label{Z1sigma}
Z(\epsilon_A,\epsilon_B,\epsilon_C,\epsilon_D)\sim
Z^{(1)}(\epsilon_A,\epsilon_B,\epsilon_C,\epsilon_D)\sim {\rm
e}^{\I\,(\epsilon_A+\epsilon_B-\epsilon_C-\epsilon_D)/2}\,
\frac{(\epsilon_A+\epsilon_D)(\epsilon_B+\epsilon_C)}
{(\epsilon_A+\epsilon_B)(\epsilon_C+\epsilon_D)}, \quad \eta\gg
1\,.
\end{equation}

\subsection{Evaluation of the matrix integral}\label{subsec:integral}

In  fact the result (\ref{Z1sigma}) fully exhausts $Z^{(1)}$: In
agreement with the Duistermaat-Heckmann theorem \cite{DuisHeck},
all further terms of the perturbation series vanish for the
unitary symmetry class; in the framework of the semiclassical
theory this was already demonstrated in Section~\ref{sec:cancel}.

To prove the absence of perturbative corrections in the present RMT
context we transform integration variables such that the integral in
(\ref{Zsusy}) becomes Gaussian and similar to (\ref{Z1FinalGauss}). To
that end we introduce an analogue of the singular-value decomposition
of the matrix $B$,
\begin{eqnarray}\label{singval}
B=U\uB V^{-1},\quad \uB=\diag\,(x_B,x_F)
\end{eqnarray}
with unitary $U$ and pseudo-unitary $V$, and parametrize the
latter matrices as
\begin{equation}\label{UV}
U=\left({1+\eta\eta^*/2 \atop \eta^*}\,{\eta \atop
1-\eta\eta^*/2}\right) \left({\e^{-\I\phi_B}\atop 0}{0 \atop
\e^{-\I\phi_F}}\right)\,,\qquad V=\left({1-\tau\tau^*/2 \atop
-\tau^*}\,{\tau \atop 1+\tau\tau^*/2 }\right)=V^\dagger
=\sigma_zV^{-1}\sigma_z\,.
\end{equation}
The matrix $\uB $ is diagonal with positive (numerical parts of the)
diagonal entries, with the Bose-Bose entry obeying $0\le x_B<1$, due
to $|B_{11}|<1$. Obviously, $\tB=V\sigma_z\uB U^{-1}\equiv V\uBt
U^{-1}=\diag\,(x_B,-x_F) $, with the Fermi-Fermi entry negative. It
further follows that the matrix products $B\tB$ and $\tB B$ are
respectively diagonalized by $U$ and $V$,
\begin{equation}\label{bbtilde}
U^{-1}B\tilde{B}U=V^{-1}\tilde{B}B V=\uB\,\uBt=\left({x_{B}^2
\atop 0}{0\atop -x_{F}^2}\right)\equiv \diag\,(l_B,l_F),
\end{equation}
with the eigenvalues $0\le l_B<1,\, l_F\le 0$. The matrices $U,V$
also diagonalize the rational functions of $B\tB$ and $\tB B$
appearing in the sigma-model integral (\ref{SUSYZ}),
\begin{equation}
U^{-1}\frac{B\tilde{B}}{1-B\tB}U=V^{-1}\frac{\tilde{B}B}{1-\tB B}V
=\diag\left(\frac{l_{B}}{1-l_B}, \frac{l_{F}}{1-l_F}\right)\equiv
\diag(m_B,m_F)\,.
\end{equation}

The representation (\ref{singval}) can be inverted to express the
variables on the right-hand side in terms of the original elements
of $B,\tB$. The uniqueness of the decomposition of our $B$ is thus
ascertained. Moreover, the eight new variables
$l_B,l_F,\phi_B,\phi_F,\eta^*,\eta,\tau^*,\tau$ can be employed as
integration variables instead of the original elements of $B,\tB$.
A somewhat tedious calculation of the Berezinian yields the
measure \begin{equation}\label{measureB}
d[B,\tilde{B}]=\frac{dl_Bdl_Fd\phi_Bd\phi_Fd\eta^*d\eta d\tau^*
d\tau}{4\pi^2(l_B-l_F)^2}\,. \end{equation}

Our next step is yet another change of integration variables which
will be elements of matrices analogous to $B,\tB$ but with
$l_{B,F}$ replaced by $m_{B,F}$,
\begin{equation}
C=U\diag(\sqrt{m_B},\sqrt{-m_F})V^{-1}\,,\qquad \tC=\sigma_z
C^\dagger\,.
\end{equation}
We obviously have
\begin{equation}\label{CthroughB}
C\tC=\frac{B\tB}{1-B\tB}\,,\qquad \tC C=\frac{\tB B}{1-\tB B}\,,
\end{equation}
such that the integrand in the sigma-model integral becomes
Gaussian in terms of the matrices $C$ and $\tC$. Moreover, in
complete analogy with (\ref{measureB}) the integration measure can
be written as
\begin{equation}\label{measureC}
d[C,\tC]=\frac{dm_Bdm_Fd\phi_Bd\phi_Fd\eta^*d\eta d\tau^*
d\tau}{4\pi^2(m_B-m_F)^2}\,.
\end{equation}
Comparing the measures (\ref{measureB}) and (\ref{measureC}) we
get
\begin{equation}
d[C,\tC]=d[B,\tB]\frac{(l_B-l_F)^2}{(m_B-m_F)^2}\frac{dm_B}{dl_B}\frac{dm_F}{dl_F}
=d[B,\tB]\,.
\end{equation}
Therefore, in terms of the integration variables $C_{ik},C_{ik}^*$
the sigma-model integral becomes Gaussian. Again assuming that the
energy offsets are all provided with the large positive imaginary
part $\eta$ we can extend the integration domain for the Bosonic
variables to the full complex plane, at a risk only of corrections
of the order $\exp(-\eta)$. Then the resulting $C$-integral will
become identical with the leading-order integral
(\ref{Z1FinalGauss}). We have thus proven that contribution of all
higher-order terms in the expansion of the integrand about the
saddle point $B=0$ cancels, and the asymptotics of the generating
function, in all orders of $\frac {1}{\epsilon}$, is exhausted
(for $\eta\gg1$) by (\ref{Z1sigma}).

\subsection{ Andreev-Altshuler saddle point}\label{AndrAlt}

If the imaginary part $\eta$ of the energy arguments is small or
zero $Z^{(1)}$ does not capture the full high-energy asymptotics
of the generating function: an additive component $Z^{(2)}$
appears due to another, the Andreev-Altshuler or ``anomalous''
stationary point $B^{\rm AA}$. To show it let us use the
parametrization (\ref{singval})) introduced in the previous
subsection; note that the elements of the diagonal matrix $\uB$
satisfy $0\le x_B< 1,\ 0\le x_F<\infty$ .  Using
(\ref{bbtilde}),(\ref{CthroughB}) we can rewrite the supertrace in
the exponent of the sigma-model integral like
 \begin{eqnarray}
\str\!\left( \hat{\epsilon}\,\frac{1+\tilde{B}B}{1-\tilde{B}B}
+\hat{\epsilon}'\,\frac{1+B\tilde{B}}{1-B\tilde{B}} \right)
=\left(V^{-1}\hat{\epsilon}V\,+U^{-1}\hat{\epsilon}'U
\right)_{BB}\frac{1+x_B^2}{1-x_B^2}-
\left(V^{-1}\hat{\epsilon}\,V+U^{-1}\hat{\epsilon}'U
\right)_{FF}\frac{1-x_F^2}{1+x_F^2}\,.
\end{eqnarray}
At a stationary point the derivatives of this expression by $x_B$
and $x_F$ must be zero, from which it follows that $x_B$ must be
zero, whereas $x_F$ can be zero or infinity. Consequently there
exist two stationary points in the matrix space, (a) the standard
saddle $B=0$ giving birth to $Z^{(1)}$, and (b) the so called
Andreev-Altshuler saddle $B^{\rm AA}=\mbox{diag
}\left(0,\infty_C\right)$ where $\infty_C$ stands for the
infinitely remote point of the complex plane. The anomalous saddle
is responsible for the part $Z^{(2)}$ of the generating function
exponentially small when $\eta\gg 1$ but of the same order as
$Z^{(1)}$ if the energy offsets are real.

An elegant way to find the contribution of the anomalous stationary
point involves retracing the steps leading to the rational
parametrization and finding the point $Q^\mathrm{AA}$ of the
manifold (\ref{conjugclass}) associated with $B^{\rm AA}$. Let us
represent $B^\mathrm{AA}$ and $\tilde{B}^\mathrm{AA}$ as limits $
B^\mathrm{AA}=\lim_{ a\to 0}PK^{-1}_ a ,\quad \tilde
{B}^\mathrm{AA}=\lim_{ a\to 0}PK^{-1}_{- a} $ with the diagonal $2\times
2$ matrices
\begin{eqnarray}
P= \left( \begin{array}{cc} 0&0
\\0&1\end{array}\right),
\quad K_{\pm a}= \left( \begin{array}{cc} 1& 0
\\0&\pm a\end{array}\right)\,;
\end{eqnarray}
for simplicity we take $ a$ real, i.e., we approach the infinite point
of the complex plane along the real axis. When $ a$ is finite the
respective $4\times 4$ matrix $ T=\left({1 \atop \tilde{B}}{B \atop
    1}\right)$ factorizes,
\begin{eqnarray}
  T_ a=\left(
\begin{array}{cc} K_{-a}& P
\\P& K_{ a}\end{array}\right)\left(\begin{array}{cc} K_{-a}^{-1}&0
\\0&K_{a}^{-1}\end{array}\right).
\end{eqnarray}
Since the second factor in $T_{ a}$ commutes with $\Lambda^{(1)}$ we
find
\begin{eqnarray}
Q^\mathrm{AA}=\I \lim_{ a\to 0}T_ a \Lambda^{(1)}T_ a^{-1}=\I R
\Lambda^{(1)}R^{-1},\nonumber\\
R=R^{-1}=\left(
\begin{array}{cc} K_0& P
\\P& K_0\end{array}\right)=\left(\begin{array}{cccc}
1&0&0&0\\
0&0&0&1\\
0&0&1&0\\
0&1&0&0
\end{array}\right)\,.
\end{eqnarray}
The matrix $R$ describes a permutation transposing the second and fourth
entries, i.~e., the Fermi-Fermi ones. We thus have
\begin{equation}
\Lambda^{(2)}\equiv R\Lambda^{(1)}R^{-1}={\rm diag}(1,-1,-1,1)\,.
\end{equation}

To get the part of the high-energy asymptotics connected with the
Andreev-Altshuler stationary point we  reparameterize the
integration variables in (\ref{intsaddlemf}) as $Q'=RQR$. The
Jacobian of the transformation $Q\to Q'$ is unity. Denoting by
$\hat{E}'$ the offset matrix with the Fermi entries transposed,
$\hat{E}'\equiv R\hat{E}R={\rm
  diag}(\epsilon_B,-\epsilon_C,-\epsilon_A,\epsilon_D)$ we bring the
$Q$-integral to an equivalent form $Z= -\int_{Q'^2=-1} dQ'\,
\e^{\,\frac{\I}{2}{\rm Str}\, Q'\hat{E}'}$\,. Now we can set
$Q'=T\Lambda^{(1)}T^{-1}$ and rationally parameterize $T$ which
leads to a  superintegral differing from (\ref{Zsusy}) by the
interchange $\epsilon_D\leftrightarrow -\epsilon_C$. Expanding the
exponent around the stationary point $B=\tilde B=0$ corresponding
to $Q'=\I\Lambda^{(1)}$, or $Q=Q^\mathrm{AA}=\I\Lambda^{(2)}$ we
obtain an asymptotic series in $\epsilon^{-1}$ which truncates on
its first term,
$Z^{(2)}=Z^{(1)}|_{\epsilon_D\leftrightarrow-\epsilon_C}$. Adding
the contributions of both stationary points  we get the full GUE
generating function $Z=Z^{(1)}+Z^{(2)}$.

\section{Bosonic replicas}
\label{sec:bosonic_sigma}

The main technical difficulty in establishing a one-to-one correspondence between
semiclassics and the sigmal model of RMT are the spectral determinants in the numerator
of the generating functions. These determinants lead to unwelcome sign factors $(-1)^{n_C+n_D}$
in the semiclassical treatment, and in RMT they cannot be expressed through Gaussian integrals
over complex numbers. In the supersymmetric treatment the way around these difficulties is
to introduce anticommuting Fermionic variables. These variables incorporate the appropriate signs
in semiclassics, and allow for an integral representation of non-inverted determinants
in RMT. In the present appendix we want to present an alternative solution based
on the so-called replica trick. (In Appendix \ref{sec:fermionic_sigma} we will consider a combination of the
replica trick with Fermionic variables.)

While
the replica trick offers
considerable technical convenience, there is a price to pay: First, the replica trick is
hard to justify mathematically; second, at least in the present (Bosonic) variant only the part $Z^{(1)}$ is
accessible. That price has let
replicas fall into some disrepute.  From a semiclassical
point of view, the price looks more affordable. We have seen
$Z^{(1)}$ to yield the full two-point correlator with the help of
the ``crosswise'' procedure of Sect.~\ref{subsec:genfuncdef} or,
equivalently, $Z^{(2)}$ to be accessible from $Z^{(1)}$ by the
Riemann-Siegel lookalike of Sect.~\ref{subsec:RSlookalike}.

\subsection{Semiclassical construction of a sigma model}

To construct a replica sigma model from semiclassics, we start from formula (\ref{master2}).
We the assume that the pseudo-orbits $C$ and $D$ are each divided into $r-1$ subsets
yielding altogether $2r$ pseudo-orbits.
We denote the pseudo-orbit $A$ by $j=1$, and the $r-1$ parts of $C$ by
$j=2,\ldots,r$. Similarly $B$ is indicated by $k=1$, whereas the $r-1$ parts of
$D$ are labelled by $k=2,\ldots,r$.
Each of the $n_C$ orbits in $C$ can be included in any of the $r-1$ subsets of $C$,
and similarly for $D$.
This leaves altogether $(r-1)^{n_C+n_D}$ ways for distributing
the orbits of $C$ and $D$ among subsets. If we now formally take the limit $r\to 0$,
this number of possibilities turns into the factor $(-1)^{n_C+n_D}$ needed in our semiclassical
treatment.
We can thus drop the sign factor if
we write $Z_{\rm off}^{(1)}$ as
\begin{eqnarray}
\label{Bosemasterreplica1}
Z_{\rm off}^{(1)}&=&\lim_{r\to 0}Z_{\rm off}^{(1)}(r)\nonumber\\
Z_{\rm off}^{(1)}(r)&=&\sum_{{\rm structures}\atop{\rm of}\;2r\;{\rm-tuples}}\frac{1}{V!}\frac{\prod_{\rm enc}i(\epsilon_j+\epsilon'_k)}
{\prod_{\rm links}(-i(\epsilon_j+\epsilon'_k))}\;.
\end{eqnarray}
where we sum over structures of $2r$-tuples rather than quadruplets of pseudo-orbits. This means that now
every way of distributing the orbits in $C$ and $D$ among the subsets is considered to
give rise to a separate structure.
The indices of $\epsilon_j$, $\epsilon'_k$ in each link factor are chosen according to the pseudo-orbits $j,k$ that the link
belongs to; the corresponding indices in the encounter factor are chosen according to the pseudo-orbits the first entrance port
belongs to. In line with the above definitions of $j,k$ we take
\begin{eqnarray}
\epsilon_1=\epsilon_A,&&
\epsilon_2=\ldots=\epsilon_r=\epsilon_C, \nonumber\\ \epsilon'_1=\epsilon_B,&&
\epsilon'_2=\ldots=\epsilon'_r=\epsilon_D\;.
\end{eqnarray}
For the diagonal approximation we write
\begin{eqnarray}
\label{Bosemasterreplica2}
Z_{\rm diag}^{(1)}&=&\lim_{r\to 0}Z_{\rm off}^{(1)}(r)\nonumber\\
Z_{\rm diag}^{(1)}(r)&=&e^{i(\epsilon_A+\epsilon_B+(r-1)\epsilon_C+(r-1)\epsilon_D)/2}
(-i(\epsilon_A+\epsilon_B))^{-1}(-i(\epsilon_A+\epsilon_D))^{-(r-1)}(-i(\epsilon_C+\epsilon_B))^{-(r-1)}(-i(\epsilon_C+\epsilon_D))^{-(r-1)^2}\nonumber\\
&=&e^{i\sum_j \epsilon_j/2+i\sum_k \epsilon'_k/2}\prod_{j,k=1\ldots r}(-i(\epsilon_j+\epsilon'_k))^{-1}
\end{eqnarray}
where the factors $-i$ are just inserted for later convenience; they cancel mutually in the replica limit $r\to 0$.
Now can now construct a sigma model similarly as in the main part: Each entrance port belonging to an original pseudo-orbit $j$ (either $A$ or a part of $C$)
and a partner pseudo-orbit $k$ (either $B$ or a part of $D$) is indicated by a symbol $B_{kj}$. Correspondingly each exit
port is denoted by a symbol $\tilde{B}_{jk}$. The indices of $B$'s and $\tilde{B}$'s belonging to
the same encounter have to coincide like the subscripts in the trace of a matrix product.
We can again sum over all ways of connecting these ports by links with the help of a Gaussian integral over $B_{kj},\tilde B_{jk}$.
This is done in the same way as in the supersymmetric approach, with only one difference:
Since the sign factor $(-1)^{n_C+n_D}$ has been eliminated by the replica trick, all
sign factors in the previous treatment can be omitted including those in the commutation relation of $B,\tilde{B}$. The coefficients $B_{jk},\tilde{B}_{kj}$ can thus be taken as ordinary complex variables with $\tilde{B}_{jk}=B_{kj}^*$ (i.e. $\tilde{B}=B^\dagger$). All supermatrices are thus replaced by
ordinary matrices which are now larger due to the $r$ different values of $j$ and $k$.

To go through the argument in detail, we split the sum over structures into sums over
the possible numbers and sizes of encounters,
assignments of ports to pseudo-orbits, and ways of connecting the
ports through links:
\begin{equation}
\sum_{\rm{structures}}= \sum_{\#\,\rm{enc}}\; \sum_{\rm{ports}}\;
\sum_{\rm{links}}\;.
\end{equation}
The three subsums are done step by step.

{\bf Link sum}: We start by summing over link connections for
fixed encounters and assignment of ports to pseudo-orbits. We
denote by $L_{kj}^{\IN}$, $L_{kj}^{\OUT}$ the numbers of entrance
and exit ports forming part of the original pseudo-orbits
$j=1,2,\ldots r$ and the partner pseudo-orbits $k=1,2,\ldots r$.
The links now have to connect each of the $L_{kj}^{\OUT}$ exit
ports of $j,k$ to one of the $L_{kj}^{\IN}$ corresponding entrance
ports, and we can conclude $L_{kj}^{\IN}=L_{kj}^{\OUT}$ for all
$j$ and $k$. With that condition satisfied, all connections
between entrance and exit ports associated with the same $j$ and
$k$ are permissible, i.e., there are
$\prod_{jk}L_{kj}^{\IN}!=\prod_{jk}L_{kj}^{\OUT}!$ possibilities.
For each $j,k$ we obtain a combinatorial factor
$\delta_{L_{kj}^{\IN},L_{kj}^{\OUT}}L_{kj}^{\rm in}!$.

We now combine the latter combinatorial factor with  the
$L_{kj}+1$ contributions $\frac{1}{-\I(\epsilon_j+\epsilon'_k)}$,
one from each link within  $j$ and $k$ and one from the diagonal
approximation. We can represent the product as an integral over
complex parameters $B_{kj}$, where we let
$\tilde{B}_{jk}=B_{kj}^*$,
\begin{equation}
\label{Bcompl} (-\I(\epsilon_j+\epsilon'_k))^{-(L_{kj}^{\IN}+1)}
\delta_{L_{kj}^{\IN},L_{kj}^{\OUT}}{L_{kj}^{\IN}!} = \int
d^2B_{kj}{\rm e}^{\I(\epsilon_j+\epsilon'_k)|B_{kj}|^2}
(B_{kj})^{L_{kj}^{\IN}}({\tilde B}_{jk})^{L_{kj}^{\OUT}}
\end{equation}
with $d^2B_{kj}=\frac{d{\rm Re}B_{kj}d{\rm Im}B_{kj}}{\pi}$, for
each pair  $j$ and $k$. The integral converges due to ${\rm
Im}\,\epsilon_j,{\rm Im}\,\epsilon'_k>0$.

The only terms not yet taken into account are the encounter
contributions $\I(\epsilon_j+\epsilon'_k)$, and the factor ${\rm
  e}^{\,\I(\sum_j \epsilon_j+\sum_k
  \epsilon'_k)/2}\frac{1}{V!}$ from Eqs.
(\ref{Bosemasterreplica1}) and (\ref{Bosemasterreplica2}). With these terms included the link sum
yields the generating function
\begin{eqnarray}
\label{linkssummed} Z^{(1)}(r)&=& \sum_{\#\,\rm{enc}}\;
\sum_{\rm{ports}} \frac{1}{V!}\left(\prod_{\rm
enc}\I(\epsilon_j+\epsilon'_k)\right) \prod_{jk}\int d^2B_{kj}
{\rm
e}^{\,\I(\epsilon_j+\epsilon'_k)\left(\frac{1}{2}+|B_{kj}|^2\right)}
(B_{kj})^{L_{kj}^{\IN}}(B_{kj}^*)^{L_{kj}^{\OUT}}\;.
\end{eqnarray}
We have thus re-expressed $Z^{(1)}_r$ as an integral over
arbitrary complex $r\times r$ matrices $B$.


{\bf Port sum}: Next, we sum over all possibilities to assign
entrance and exit ports of encounters to pseudo-orbits $j$ and
$k$. These assignments will determine $L_{kj}^{\IN}$ and
$L_{kj}^{\OUT}$ and thus the numbers of factors $B_{kj}$,
$\tilde{B}_{jk}$ for the summands in Eq. (\ref{linkssummed}). To perform
the sum in a compact manner, we arrange the factors $B_{kj}$,
$\tilde{B}_{jk}$ such that all factors originating from ports of the
same encounter are grouped together. Each of these groups starts
with the factor associated to the first entrance port, followed by
the first exit port, the second entrance port, etc., up to the
last exit port. This leads to an alternating sequence of
$B_{kj}$'s and $\tilde{B}_{jk}^*$'s.
As we had seen, the subscripts in this sequence must coincide as for the trace
of a matrix product.
We can also absorb the encounter factor from (\ref{master2}) into
the latter sequence. The indices in that factor represent the
original and partner pseudo-orbits to which the first entrance
port belongs and thus must coincide with the subscripts of the
first $B$ in the sequence above.
If we sum over all assignments of ports to
pseudo-orbits by summing over all independent indices,
each $l$-encounter then gives rise to a factor
\begin{equation}
\sum_{j_1,\ldots,j_l,k_1,\ldots,k_l}i(\epsilon_{j_1}+\epsilon'_{k_1})
B_{k_1,j_1}\tilde{B}_{j_1,k_2}B_{k_2,j_2}\ldots B_{k_l,j_l}\tilde{B}_{j_l,k_1}
=
i\,{\rm tr} \,\hat{\epsilon}(\tilde{B} B)^2 +i\,{\rm tr}\,
\hat{\epsilon'} (B\tilde{B})^2
\end{equation}
with
\begin{eqnarray}
\hat{\epsilon}&=&\diag(\epsilon_1,\epsilon_2,\ldots,\epsilon_r)
=\diag(\epsilon_A,\epsilon_C,\ldots,\epsilon_C)\nonumber\\
\hat{\epsilon}'&=&\diag(\epsilon'_1,\epsilon'_2,\ldots,\epsilon'_r)
=\diag(\epsilon_B,\epsilon_D,\ldots,\epsilon_D)\;.
\end{eqnarray}
If we proceed in this way, and also write the
exponentials of Eq.  (\ref{linkssummed}) as
$\prod_{jk}{\e}^{\,\I\,(\epsilon_j+\epsilon'_k)
  (\frac{1}{2}+|B_{kj}|^2)} ={\e}^{\I\,{\rm tr}\,\hat{\epsilon}
  (\frac{1}{2}+\tilde{B} B) + \I\,{\rm tr}\, \hat{\epsilon}'\,
  (\frac{1}{2}+B\tilde{B})}$, Eq. (\ref{linkssummed}) turns into
\begin{equation}
\label{almost} Z^{(1)}_r= \int d[B,\tilde{B}]{\rm e}^{\I\,{\rm tr}\,
\hat{\epsilon}\, (\frac{1}{2}+B^\dagger B) + \I\,{\rm tr}
\,\hat{\epsilon}'\,(\frac{1}{2}+BB^\dagger)} \sum_{\#\,\rm{enc}}
\frac{1}{V!}\prod_\sigma \left( \I\,{\rm tr}\,\hat{\epsilon}\,
(\tilde{B} B)^{l(\sigma)} +\I\,{\rm tr}\,\hat{\epsilon}'\,
(B\tilde{B})^{l(\sigma)}\right)\,.
\end{equation}
%


{\bf {Encounter sum}}: The remaining summation over the number $V$
of encounters and over their sizes $l(\sigma)$,
$\sigma=1,2,\ldots,V$, is trivial. We get
\begin{eqnarray}\label{encounter_sum}
Z^{(1)}(r)&=&\int d[B,\tilde{B}]\, {\e}^{\,\I\,{\rm tr}\,\hat{\epsilon}\,
(\frac{1}{2}+B^\dagger B) + \I\,{\rm
tr}\,\hat{\epsilon}'\,(\frac{1}{2}+BB^\dagger)}
\sum_{V}\frac{1}{V!}\left(\sum_{l=2}^\infty\left( \I\,{\rm
tr}\,\hat{\epsilon} (B\tilde{B})^{l}+\I\,{\rm tr}
\hat{\epsilon}' (B\tilde{B})^{l}\right) \right)^V\nonumber\\
&=& \int d[B,\tilde{B}]\exp\left[\frac{\I}{2}\;{\rm tr}\! \left(
\hat{\epsilon}\,\frac{1+\tilde{B}B}{1-\tilde{B}B}
+\hat{\epsilon}'\,\frac{1+B\tilde{B}}{1-B\tilde{B}}
\right)\right]\;.
\end{eqnarray}

\subsection{The sigma model of RMT}

Eq. (\ref{encounter_sum}) agrees with the random-matrix result, which we shall derive in the following.
In random-matrix theory the replica trick amounts to replacing
the spectral
determinants in the numerator of the generating function
by powers $\Delta(E)^{-(r-1)}$ with
the number of replicas $r$ an integer larger than unity. One
evaluates the thus generalized generating function for $r\geq1$,
but in the very end takes the limit $r\to 0$.  This leads to the following representation of the
generating function
\begin{eqnarray}\label{ZBose}
Z&=&\lim_{r\to 0}Z(r)\nonumber\\
Z(r)&=&\left\langle
{\Delta\left(E+\frac{\epsilon_A}{2\pi\overline{\rho}}\right)}^{-1}
{\Delta\left(E-\frac{\epsilon_B}{2\pi\overline{\rho}}\right)}^{-1}
{\Delta\left(E+\frac{\epsilon_C}{2\pi\overline{\rho}}\right)}^{-(r-1)}
{\Delta\left(E-\frac{\epsilon_D}{2\pi\overline{\rho}}\right)}^{-(r-1)}
\right\rangle\nonumber\\
&=&\left\langle\prod_{j=1}^r
\Delta\left(E+\frac{\epsilon_j}{2\pi\overline{\rho}}\right)^{-1}
\Delta\left(E-\frac{\epsilon_k'}{2\pi\overline{\rho}}\right)^{-1}
\right\rangle\;.
\end{eqnarray}
The advantage of the replica trick is that now only inverted
spectral determinants show up, which can be represented by
Gaussian integrals over ordinary complex variables. If we go
through the same steps as for the supersymmetric sigma model
(average over matrices representing the Hamiltonian,
Hubbart-Stratonivich transformation, stationary-phase
approximation) we obtain the following integral representation for
$Z(r)$,
\begin{equation}
\label{BoseQint}
 Z(r)\sim\int_{Q^2=-1} d[Q]\exp\left(\frac{1}{2}\mathrm{tr}\,
 \left({\hat{\epsilon}\atop 0}{0\atop\hat{\epsilon}'}\right)Q\right)\;.
\end{equation}
where we dropped a proportionality factor converging to 1 in the replica limit $r\to 0$.
In contrast to the supersymmetric case, the matrices $Q$ are now complex $2r\times 2r$
matrices.
They have to be diagonalizable bt pseudo-unitary transformations $T$ which are now defined by
\begin{equation}
 T\Lambda^{(1)}T^\dagger=\Lambda^{(1)}.
\end{equation}
where $\Lambda^{(1)}=(1,1,\ldots,-1,-1,\ldots)$.
The saddle point manifold $Q^2=-1$ has the shape of a hyperboloid.
In the replica case the relevant solutions of $Q^2=-1$
are matrices that can be obtained through conjugation from $i\Lambda^{(1)}$ where
,
\begin{equation}
Q=T\Lambda^{(1)}T^{-1}\;.
\end{equation}
Multiplication with a block diagonal matrix as in the
supersymmetric now leads to the rational parametrization
\begin{equation}
\label{Boseparam} T=1-\left(
\begin{array}{cc}
0&B
\\
\tilde{B}&0
\end{array}\right)
\end{equation}
where $B$ is an arbitrary $r\times r$ matrix and $\tilde{B}=B^\dagger$. The fact that the
off-diagonal blocks of this matrix are Hermitian-conjugate to one
another follows from the pseudo-unitarity condition.

In the limit of large $\epsilon$ the integral (\ref{BoseQint}) can
be calculated in a saddle point approximation (not to be mixed with
the already implemented limit $N\to\infty$).  Using
the parametrization (\ref{Boseparam}) in $Q$ and expanding the
exponent in the vicinity of the saddle point $B=0$ we obtain the
integral
\begin{eqnarray}
\label{Bosesigma} Z^{(1)}(r)&=& \int d[B,\tilde{B}]\exp\left[ \I\,{\rm tr}\, \hat\epsilon
\left(\frac{1}{2}+\sum_{l=1}^\infty(\tilde{B} B)^{l}\right)
+\I\,{\rm tr}\, \hat\epsilon'
\left(\frac{1}{2}+\sum_{l=1}^\infty(B\tilde{B})^{l}\right)
\right]\nonumber\\
&=&
\int d[B,\tilde{B}]\exp\left[\frac{\I}{2}\;{\rm tr}\! \left(
\hat{\epsilon}\,\frac{1+\tilde{B}B}{1-\tilde{B}B}
+\hat{\epsilon}'\,\frac{1+B\tilde{B}}{1-B\tilde{B}}
\right)\right]\;
\end{eqnarray}
agreeing with our semiclassical formula.

Both the semiclassical and the RMT results can be generalized to
the orthogonal case. One then obtains a formula as in
(\ref{Bosesigma}) but with $B$ as a $2r\times 2r$ matrix subject
to symmetries analogous to the supersymmetric treatment. In fact,
even the crossover between the unitary and the orthogonal symmetry
class is accessible in this way \cite{Japanese}.

\section{Fermionic replicas}
 \label{sec:fermionic_sigma}

\subsection{Semiclassical construction of a sigma model}

A further variant of the sigma model makes use both replicas and Fermionic variables (the latter appearing only
in the intermediate stages of the derivation from RMT).
To derive this variant from semiclassics, we write the sign factor in Eq. (\ref{master2}) as
\begin{equation}
(-1)^{n_C+n_D}=(-1)^{n_A+n_B}(-1)^{(n_B+n_D)-(n_A+n_C)}
\end{equation}
Here the factor $(-1)^{n_A+n_B}$ can be dealt with in a similar way as the factor $(-1)^{n_C+n_D}$ in
the Bosonic approach. This time we split the pseudo-orbits $A$ and $B$ into $r-1$ parts. There are
 $(r-1)^{n_A+n_B}$ ways to do this. If we take the limit $r\to 0$ this number of possibilities
 converges to $(-1)^{n_A+n_B}$.

\begin{figure}
\begin{center}
\includegraphics[scale=0.3]{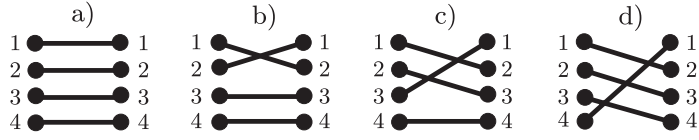}
\end{center}
\caption{Reconnections in a 4-encounter in three steps}
\label{fig:steps}
\end{figure}

The remaining sign factor $(-1)^{(n_B+n_D)-(n_A+n_C)}$
depends on the difference between the overall
numbers of partner orbits and original orbits.  To
keep track of this factor we envisage the reconnections inside
encounters as a step-by-step process. In each step the connections
of just two stretches are changed, as in Fig. \ref{fig:ARS}.
This changes the number of orbits by one and thus flips the above sign.
If we only have one 2-encounter
we need only one such step and the sign factor is negative.  For the
example of a 4-encounter, Fig.~\ref{fig:steps} shows how we can
get from connections in (a) to the final connections in (d) in
three steps, each time only reshuffling two stretches. In general for each $l$-encounter we
need $l-1$ steps.
Let us denote by $v_l$ the number of $l$-encounters.
 Then the
overall number of steps, and thus sign flips, is given by the
number $L=\sum_l v_l l$ of encounter stretches or links minus the
number $V=\sum_l v_l$ of encounters.  We are thus free to replace the
sign factor by $(-1)^{L-V}$.

Altogether we then obtain
\begin{eqnarray}
\label{Z1fermi}
Z^{(1)}_{\rm off}&=&\lim_{r\to 0}Z^{(1)}_{\rm off}(r)\nonumber\\
Z^{(1)}_{\rm off}(r)&=&\sum_{{\rm
structures}\atop {\rm of}\; 2r\;{\rm-tuples}}\frac{1}{V!}(-1)^{L-V}\frac{\prod_{\rm enc}\I
(\epsilon_j+\epsilon'_k)} {\prod_{\rm
links}(-\I(\epsilon_j+\epsilon'_j))}\;.
\end{eqnarray}
Here the definition of structures is adapted to consider separately
situations where the orbits in $A$ or $B$ are distributed differently among the $r-1$
subsets of $A$ and $B$. Furthermore $\epsilon_j$, $\epsilon'_k$ (and the corresponing diagonal matrices)
are now defined as
\begin{eqnarray}
\label{fermiepsilon}
\hat{\epsilon}&=&\diag(\epsilon_1,\epsilon_2,\ldots,\epsilon_r)=\diag(\epsilon_A,\ldots,\epsilon_A,\epsilon_C)\nonumber\\
\hat{\epsilon}'&=&\diag(\epsilon'_1,\epsilon'_2,\ldots,\epsilon'_r)=\diag(\epsilon_B,\ldots,\epsilon_B,\epsilon_D)\;.
\end{eqnarray}
For the diagonal term we have
\begin{eqnarray}
Z_{\rm diag}^{(1)}&=&\lim_{r\to 0}Z_{\rm diag}^{(1)}(r)\nonumber\\
Z_{\rm diag}^{(1)}(r)&=&e^{i(-(r-1)\epsilon_A-(r-1)\epsilon_B-\epsilon_C-\epsilon_D)/2}
(\epsilon_A+\epsilon_B)^{-(r-1)^2}(\epsilon_A+\epsilon_D)^{-(r-1)}(\epsilon_C+\epsilon_B)^{-(r-1)}(\epsilon_D+\epsilon_C)^{-1}\nonumber\\
&=&e^{-i\sum_j \epsilon_j/2-i\sum_k \epsilon'_k/2}\prod_{j,k=1\ldots r}(-i(\epsilon_j+\epsilon'_k))^{-1}\;.
\end{eqnarray}
If we use (\ref{Z1fermi}) as a starting point we can repeat the semiclassical construction of the sigma
model from the previous appendix. We just have to use the new definitions of $\hat{\epsilon}$, $\hat{\epsilon}'$
and insert sign factors in the oscillatory prefactor, for each of the $V$ encounters, and for each of the $L$ encounter stretch represented by $\tilde{B}B$, $B\tilde{B}$.
This yields
\begin{eqnarray}\label{fermiencounter_sum}
Z^{(1)}(r)&=&\int d[B,\tilde{B}]\, {\e}^{-\,\I\,{\rm tr}\,\hat{\epsilon}\,
(\frac{1}{2}-\tilde{B} B) - \I\,{\rm
tr}\,\hat{\epsilon}'\,(\frac{1}{2}-B\tilde{B})}
\sum_{V}\frac{1}{V!}\left(-\sum_{l=2}^\infty\left( \I\,{\rm
tr}\,\hat{\epsilon} (-B\tilde{B})^{l}+\I\,{\rm tr}
\hat{\epsilon}' (-B\tilde{B})^{l}\right) \right)^V\nonumber\\
&=& \int d[B,\tilde{B}]\exp\left[-\frac{\I}{2}\;{\rm tr}\! \left(
\hat{\epsilon}\,\frac{1-\tilde{B}B}{1+\tilde{B}B}
+\hat{\epsilon}'\,\frac{1-B\tilde{B}}{1+B\tilde{B}}
\right)\right]\;.
\end{eqnarray}

\subsection{The sigma model of RMT}

In RMT, the Fermionic replica sigma model is obtained if we
represent {\it all} spectral
determinants the generating function (\ref{Zdef}) by Gaussian
Grassmann integrals. To that end one brings the determinants of
the denominator ``upstairs'' as \cite{KamenevMezard}
$\det^{-1}=\det^{r-1}$, with the replica parameter $r$ first taken
as an integer larger than unity; only in the end, $r$ is assumed
real and sent to zero.
The spectral determinants can the be represented through Gaussian integrals
over Fermionic variables. By going through the same steps as in the supersymmetric
and the Bosonic replica variant, these integrals can be traded for
and integral over complex $2r\times 2r$ matrices which are diagonalizable by unitary transformations
and satisfy $Q^2=-1$,
\begin{eqnarray}\label{intmf}
Z(r)&\sim& \int_{Q^2=-1} dQ\, \e^{\,\frac{1}{2}{\rm tr}\,Q
\left({\hat\epsilon\atop0}{0\atop-\hat\epsilon'}\right)}
\end{eqnarray}
with $\hat{\epsilon}$, $\hat{\epsilon}'$ as in (\ref{fermiepsilon}).
The matrices $Q$ relevant for the non-oscillatory terms
in the correlator are again of the type
\begin{equation}
Q=iT\Lambda^{(1)}T^{-1}\;.
\end{equation}
The rational parametrization then amounts to
\begin{equation}\label{ratparfermi}
T=1+\left({0 \atop \tilde{B}}{-B \atop 0}\right)
\end{equation}
which inserted
into the integral (\ref{intmf}) yields
\begin{eqnarray}
Z^{(1)}(r)
&=& \int d[B,\tilde{B}]\exp\left[-\frac{\I}{2}\;{\rm tr}\! \left(
\hat{\epsilon}\,\frac{1-\tilde{B}B}{1+\tilde{B}B}
+\hat{\epsilon}'\,\frac{1-B\tilde{B}}{1+B\tilde{B}}
\right)\right]\;.
\end{eqnarray}
just as in semiclassics.

Again both the semiclassical and the RMT calculation can be generalized to
the orthogonal case.

The contribution of the saddle $B=\tilde{B}=0$ does not exhaust
the high-$\epsilon$ asymptotics of the generating function.
Another contribution is connected with the vicinity of the so
called anomalous or Andreev-Altshuler saddle, $\Lambda^{(2)}={\rm
  diag}(-1,1,\ldots,1;1,-1,\ldots,-1)$ which can be obtained from
$\Lambda^{(1)}$ by swapping the first and the $(r+1)$-th entry,
i.~e.,
  \begin{eqnarray}
\Lambda^{(2)}=R\Lambda^{(1)}R^\dagger\,,\qquad R=\left(
\begin{array}{c|c|c|c}
&&1&\\ \hline &1_{r-1}&&\\ \hline 1&&&\\ \hline &&&1_{r-1}
\end{array}
\right)
\end{eqnarray}
where all missing entries are zero. $\Lambda^{(2)}$ cannot be
reached using the rational parametrization (\ref{ratparfermi});
its vicinity corresponds to $B,\tilde{B}$ tending to infinity.

To appreciate the contribution of the Andreev-Altshuler saddle, we
substitute $Q=RQ'R^{\dagger}$ in the integral (\ref{intmf}) and
adopt $Q'$ as the new integration variable. The Jacobian of this
transformation is unity such that we get $Z_r= \int_{Q'^2=1} dQ'\,
\e^{\,\frac{1}{2}{\rm tr}\,Q'R^{\dagger}XR}$ an expression still
equivalent to (\ref{intmf}), the replacement of $X$ by
$R^\dagger XR=\mathrm{diag}\left(
  -\epsilon_1',\epsilon_2,\ldots,\epsilon_r;
  \epsilon_1,-\epsilon_2',\ldots,-\epsilon_r'\right)$
notwithstanding. We then employ the rational parametrization for
$Q'$ and obtain the asymptotic series in $\epsilon^{-1}$ expanding
the integrand around the stationary point $B=\tilde{B}=0$
corresponding to $Q'=\Lambda^{(1)}$ or $Q=\Lambda^{(2)}$. Again,
the Duistermaat-Heckmann theorem guarantees that the series breaks
up after the leading term.  The final result for $Z^{(2)}$ differs
from $Z^{(1)}$ only by the appearance of the permuted offset
matrix $R^\dagger X R$ instead of $X$, i.~e.
\begin{equation}\label{Weyl2}
  Z^{(2)}(\epsilon_A,\epsilon_B,\epsilon_C,\epsilon_D)=
  Z^{(1)}(\epsilon_A,\epsilon_B,-\epsilon_D,-\epsilon_C)\,.
\end{equation}
The Weyl symmetry of $Z=Z^{(1)}+Z^{(2)}$ thus arises again.


\section{Structures, permutations and symmetry}
\label{app_permutations}%

Here we formalize the definition of a structure in terms of
permutations, in order to clarify the relation of structures to
diagrams and investigate the symmetry aspects of that relation. A
``diagram'' is understood as the set of all physically distinct
pseudo-orbit quadruplets with a given topology, i.e, with a fixed
set of encounters whose ports are connected by links in a
specified way. Such a diagram can have many equivalent structures
associated. For simplicity we limit the following to the unitary
symmetry class.

We imagine that the encounters are numbered by $\sigma=1,\ldots,V$
and the stretches and ports in the $\sigma$-th encounter by
$j=1,\ldots,l(\sigma)$. Each port is thus denoted by a label
$\sigma.j$. According to the convention of section
\ref{sec:structures} in the original pseudo-orbits $A,C$ each
entrance port is connected with the exit port of the same number.
After the reconnection leading to the pseudo-orbits $B,D$ the
entrance port $\sigma.j$ will be connected to the exit port
$\sigma.(j-1)$, with the exception of the first entrance port of
each encounter which will be connected to the last exit port. The
port reconnections can be depicted by a table in which the lower
and the upper row list the entrance and exit ports, respectively,
and each encounter is represented by a  block of width
$l(\sigma),$
\begin{equation*}
P_{\mathrm{enc}}=%
\begin{pmatrix}
1.1 & \ldots  & 1.l(1) & 2.1 & \ldots  & 2.l(2) & \ldots & V. 1&
\ldots  &
V.l(V) \\
1.l(1) & \ldots  & 1.(l(1)-1) & 2.l(2) & \ldots  & 2.(l(2)-1) &
\ldots  & V.l(V) & \ldots
& V.(l(V)-1)%
\end{pmatrix}%
.
\end{equation*}%
This table can be viewed as a permutation of $L$ numbers
consisting of $V$ independent cycles
$P_{\mathrm{enc}}=(1.l(1),\ldots ,1.2,1.1)\ldots (V.l(V), \ldots
V.2,V.1)$, each cycle associated with an encounter. The port
connections by the encounter stretches in the original
pseudo-orbits $A,C$ are described in this language by the trivial
permutation $I$.

 The connection of exit to entrance ports by links can also be
represented by a similar permutation $P_{\rm link}$ where the
upper row refers to exit ports and the lower one to entrance
ports. Unlike $P_{\mathrm{enc}}$ the link permutation is totally
arbitrary.

The permutation products $P_{\mathrm{link}}I=P_{\mathrm{link}}$
and
$P_{%
  \mathrm{link}}P_{\mathrm{enc}}$ define connections between the
consecutive entrance ports before and after reconnection,
respectively. Each cycle of $P_{\mathrm{link}}$ represents the
sequence of the entrance ports of a pre-reconnection orbit in
$A,C$; similarly, each cycle of
$P_{\mathrm{link}}P_{\mathrm{enc}}$ defines such a sequence for a
post-reconnection orbit in $B$ or $D$. The numbers $n,n^{\prime }$
of cycles in these permutations give the numbers of pre- and
post-reconnection orbits.

 Our definition of a structure can thus
be formalized as a particular choice of (S1) the numbers $V$ and
$l(\sigma)$ determining the permutation $P_{\mathrm{enc}}$, (S2) a
link permutation $P_{\mathrm{link}}$,  and  (S3) a division of the
$n$ cycles of $P_{\mathrm{link}}$ between the
pseudo-orbits $A$ and $C,$ and of the $n^{\prime }$ cycles of $P_{\mathrm{link}%
}P_{\mathrm{enc}}$ between the pseudo-orbits $B$ and $D$.

 The order in which the encounters are numbered 
 and the
 choice of the first stretch in an encounter is
irrelevant. Their change amounts to  port renumbering, i.e., also
to a permutation (an arbitrary permutation of the encounter
numbers, and a cyclic permutation of the labels for each
encounter). The possible renumberings corresponding to a given
$V$, $l(\sigma)$ form a group $\cal{C}$ with
$N({\cal{C}})=V!\prod_{\sigma =1}^{V}l(\sigma)$ elements. The
renumbering $C\in {\cal C}$  transforms the encounter and link
permutations as $P\to CPC^{-1}$. For a structure devoid of any
symmetry, none of the non-trivial $C\in {\cal C}$ commute
simultaneously with the link and encounter permutation, hence
renumberings will produce $V!\prod_{\sigma =1}^{V}l_{\sigma }$
structures described by $P_{\mathrm{enc}}^{\prime
}=CP_{\mathrm{enc}}C^{-1},\quad P_{\mathrm{link}}^{\prime
}=CP_{\mathrm{link}}C^{-1}$, all of them corresponding to the same
diagram  and making the same contribution to the generating
function. This overcounting is cancelled by the denominator
$1/N({\cal C})$ in (\ref{master3}) such that we are left with the
contribution of the respective diagram in the form of the ratio of
the encounter and link factors.

In a symmetric topology $C$ may simultaneously commute with
$P_{\rm link}$ and $P_{\rm enc}$, and hence the number of
equivalent structures is smaller than $N({\cal C})$. However
renumbering the encounters and stretches will still change the
variables $s_{\sigma j}, u_{\sigma j}$ parametrizing the
separations inside the encounters; the indexing of the link
durations is also changed. Since we integrate over all values of
$s_{\sigma j}, u_{\sigma j}$ as well as the link durations,
relabelling thus leads to a separate contribution even in the
symmetric case. To count each quadruplet only once we thus always
have to divide by the number of possible relabellings $N({\cal
C})$.

Alternatively, we could formulate our approach in terms of a to
sum over diagrams rather than structures. However, then the
diagrams with symmetries would require special treatment. This was
done in \cite{taucube} for the diagrams with $L-V=2$, but it
becomes too cumbersome for a generalization to arbitrary orders.

\begin{figure}
\begin{center}
\includegraphics[ scale=0.4]{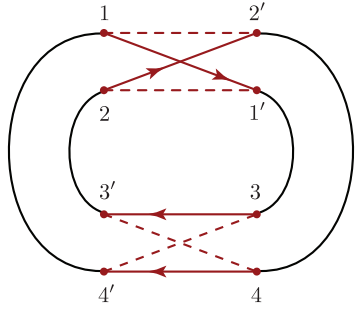}
\caption{A symmetric orbit pair with two encounters, already shown
in the gallery of Fig.~\ref{fig:gallery} as the uppermost one in
the left column for $L-V=2$; here, one of the possible port
numberings is indicated. }
 \label{fig:symm}
\end{center}
\end{figure}

For an example we consider a quadruplet containing, before and
after the reconnection, one periodic orbit  with  two 2-encounters
connected as shown in Fig.~\ref{fig:symm}.
 The encounter and link permutations with the
port numbering shown in Fig.~\ref{fig:symm} read
\begin{equation*}
P_{\mathrm{enc}}=%
\begin{pmatrix}
1.1 & 1.2 & 2.1 & 2.2 \\
1.2' & 1.1' & 2.2' & 2.1'%
\end{pmatrix}%
,\quad P_{\mathrm{link}}=%
\begin{pmatrix}
1.1' & 1.2' & 2.1' & 2.2' \\
2.1 & 2.2 & 1.2 & 1.1%
\end{pmatrix}%
\end{equation*}%
where we put primes over the indices of the exit ports, to
distinguish them from the entrance ports. We have the following
possibilities to relabel encounters and stretches (i) interchange
all first indices 1 and 2, (ii) interchange 1.1 and 1.2, and/or
(iii) interchange 2.1 and 2.2. As (i) and simultaneous application
of (ii) and (iii) both leave $P_{\rm enc}$ and $P_{\rm link}$
invariant there are only two equivalent structures associated with
the diagram. However application of (i) and simultaneous
application of (ii) and (iii) still changes the parameters
$s_{\sigma j}$ and $u_{\sigma j}$. Hence when calculating the
contribution of each structure by integrating over all parameters
each quadruplet will be counted four times per structure. To avoid
avoid overcounting we thus have to divide out $N({\cal C})=8$, as
we would without any symmetries.

\section{Complex vs real correlator}\label{sec:complex_corr}

 Let us prove the relation $R(\epsilon)=\lim_{\;\mbox{Im}\,
  \epsilon\to+0}\mbox{Re}\,C(\epsilon)$ with the complex correlator
defined by (\ref{defC}), in particular the presence of the
constant $1/2$ in (\ref{defC}). Averaging in the definition of the
spectral correlator, be it real or complex, is carried out over an
interval of its both arguments, namely the central energy $E$ and
the energy offset $\epsilon/\pi\bar\rho$; to obtain a smooth
function  the product of the two averaging intervals must be large
compared with the squared mean level spacing,
\begin{equation}\label{avercond}
\delta E\; \delta\epsilon/\pi\bar\rho\gg
\left(1/\bar\rho\right)^2\,. \end{equation} If we want
oscillations of the correlator on the scale of the mean level
spacing to be resolved the averaging interval of $\epsilon$ has to
be taken small, $\delta\epsilon\ll 1$.  In view of
(\ref{avercond}) this means that the averaging interval $\left[
a,b\right]$ of the central energy with $\delta E=b-a$ must be
large compared with the mean level spacing multiplied  by
$1/\delta\epsilon$. On the other hand the averaging interval must
be small compared with the range populated by the consecutive
levels $E_{1}<\ldots <E_{N}$.
 (The generalization to unbounded spectra or bounded spectra with $N=\infty$ is
straightforward.)
We define the average over the central energy as $\left\langle
\ldots \right\rangle _{ E}=\left( b-a\right)
^{-1}\int_{a}^{b}dE\left( \ldots \right) $. In the energy range
$E_1\leq E\leq E_N$ the mean level spacing $1/\bar{\rho}$ can be
taken as constant. Without loss of generality we choose to discuss
spectral correlations in the middle of the range considered,
i.~e.~at $E=\frac{E_N+E_1}{2}$ and can therefore take that energy
as the center of the averaging interval $\left[
  a,b\right]$.
 By explicitly doing the average over $E$ in the definition of the correlator (\ref{defC})
we obtain
\begin{eqnarray*}
C(\epsilon)+\frac{1}{2}=\left\langle\frac{1}{2\pi^{2}\bar{\rho}^2(b-a)}
\sum_{i,k=1}^{N}\frac{1}{ \frac{\epsilon^+}{\pi\bar{\rho} }
+E_{i}-E_{k} }\left[ \ln \frac{ x-\frac{\epsilon^+}{2\pi\bar{\rho}
} -E_{i}}{x+\frac{\epsilon^+}{2\pi\bar{\rho} }-E_{k}} \right]
_{x=a}^{x=b}\;\right\rangle_{\epsilon}\;. \end{eqnarray*}
 where the remaining average is over $\epsilon$ only.
Using $\left( x+\I\eta \right) ^{-1}\leftrightarrow\mathrm{{\cal
    P}\,}x^{-1}-\I\,\pi \delta \left( x\right)$ for $\eta\to 0^+$
and $\ln x=\ln \left\vert x\right\vert +\I\,\mbox{Arg }x$, we can
write the real part of $C$ as
\begin{equation}
\mbox{Re }C\left( \epsilon \right) =-\frac{1}{2}+R_{I,I}+R_{R,R}
\label{compco3}
\end{equation}%
where $R_{I,I}$ ($R_{R,R}$) stem from products of the imaginary
(real) parts of the fraction and the square bracket.

We begin with $R_{I,I}$ assuming $\mbox{Im }\epsilon ^{+}=\eta
\rightarrow +0.$ In the summand $\left( i,k\right) $\ the phase of
the expression in the square brackets can be (i) $2\pi $ (both
$E_{i}$ and $E_{k}$ lie within the averaging interval of the
central energy $\left[ a,b\right] $); (ii) $\pi $ (only one
eigenvalue lies within \ $\left[ a,b\right] $); (iii) zero (both
eigenvalues are outside $\left[ a,b\right] $). Let us assume that
the energy offset $\epsilon/\pi\bar \rho$ is small compared with
$b-a$; then considering the factor $ \delta \left( \frac{\epsilon
  }{\pi\bar{\rho} }+E_{i}-E_{k}\right) $, we may neglect the
possibility (ii). Up to an additive constant, $R_{I,I}$ must then
coincide with the real correlation function,
\begin{eqnarray}
R_{I,I}
=\frac{1}{(b-a)\bar{\rho}^2}\left\langle\;\sum_{a<E_{i},E_{k}<b}\delta
\left( \frac{\epsilon }{\pi\bar{\rho}
}+E_{i}-E_{k}\right)\;\right\rangle_{ \epsilon} \label{RII}
=R\left( \epsilon\right) +1.  \nonumber
\end{eqnarray}

In the sum of products of real parts
\begin{equation}
R_{R,R}=\frac{1}{2\pi^2\bar{\rho}^2(
b-a)}\left\langle\;\sum_{i,k=1}^{N} \frac{1}{\frac{\epsilon
}{\pi\bar{\rho} }+E_{i}-E_{k} } \ln \left\vert \frac{\left(
b-\frac{\epsilon }{2\pi\bar{\rho} }-E_{i}\right) } {\left(
a-\frac{\epsilon }{2\pi\bar{\rho} }-E_{i}\right) } \frac{\left(
a+\frac{\epsilon }{2\pi\bar{\rho} }-E_{k}\right) }
{\left(b+\frac{\epsilon}{2\pi\bar{\rho} }-E_{k}\right) }
\right\vert  \;\right\rangle_{ \epsilon}\label{RRsum}
\end{equation}%
the poles of $\left( \frac{\epsilon }{\pi\bar{\rho}
}+E_{i}-E_{k}\right) ^{-1}$ are cancelled by the zero of the
respective logarithms; the  mild logarithmic singularities of
$R_{R,R}$ at $E_{i}=a,b\pm \frac{\epsilon }{2\pi\bar{\rho} }$ are
unconnected with the level correlation and thus of no physical
relevance. (They are caused by the rectangular cut-off of the
integrand at  the borders of
the averaging interval in $%
\left\langle \ldots \right\rangle_{ E}$ and disappear if this
cut-off is smoothed out). Hence the summand has are no relevant
singularities or other fluctuations on the scale of $\epsilon$. We
are thus free to replace the double sum over a discrete eigenvalue
spectrum by a double integral over a uniformly distributed
density, as in
\[
R_{R,R}=\frac{1}{2\pi ^{2}\left( b-a\right) }\int_{E_{1}}^{E_{N}}dxdy%
\frac{1}{\left( \frac{\epsilon }{\pi\bar{\rho} }+x-y\right) }\ln
\left\vert \frac{\left( b-\frac{\epsilon }{2\pi\bar{\rho}
}-x\right) } {\left( a-\frac{\epsilon }{2\pi\bar{\rho} }-x\right)
} \frac{\left( a+\frac{\epsilon }{2\pi\bar{\rho} }-y\right) }
{\left(b+\frac{\epsilon }{2\pi\bar{\rho} }-y\right) } \right\vert
.
\]
 Note that the smoothing w.r.t. the offset variable is thus
rendered superfluous.

Consider now the semiclassical limit $\hbar\to \infty$  which
means that $\bar \rho\to\infty$; we also assume the spectral span
 constant (in classical units) which means that the number $N$ of levels taken into account
 goes to infinity. Let the averaging interval $[a,b]$ simultaneously shrink
compared with the spectral span in such a way that the hierarchy
 $E_{N}-E_{1}\gg b-a\gg \epsilon/\pi\bar \rho $ is observed.
Neglecting $\frac{\epsilon }{\pi\bar{\rho}}$ and letting the
parameter $A\equiv \frac{E_N-E_1}{b-a}$ go to infinity we get
\begin{eqnarray}
R_{R,R} &=&\frac{1}{4\pi^2}\lim_{A\rightarrow \infty }
\int_{-A}^{A}dudv\frac{1}{u-v} \ln \left\vert \frac{\left(
1-u\right) }{\left( 1+u\right) } \frac{\left( 1+v\right) }{\left(
1-v\right) } \right\vert \label{RRR} \\ \nonumber
&=&-\frac{2}{\pi^2}\int_{0}^{1}\frac{dx}{x}\,\ln \left\vert
\frac{1+x}{1-x}\right\vert =-\frac{1}{2}.
\end{eqnarray}
Collecting (\ref{compco3}), (\ref{RII}), and (\ref{RRR}) we come
to the desired relation, $\lim_{\eta \rightarrow 0}\mbox{Re
}C\left( \epsilon ^{+}\right) =R\left( \epsilon \right) $.

\end{document}